\documentclass[acmsmall,screen,nonacm]{acmart}

\setcopyright{none}
\renewcommand{\footnotetextcopyrightpermission}[1]{}

\usepackage{mathtools,stmaryrd,bm}
\usepackage{booktabs,tabularx,enumitem,microtype}
\usepackage{placeins,flafter}
\usepackage{algorithm,algpseudocode}
\usepackage{tikz}
\usetikzlibrary{positioning,arrows.meta}
\usepackage{cleveref}

\AtEndPreamble{%
  \theoremstyle{acmdefinition}%
  \newtheorem{remark}[theorem]{Remark}}

\newcommand{\N}{\mathbb N}
\newcommand{\C}{\mathbb C}

\newcommand{\Hil}{\mathcal H}
\newcommand{\Lin}{\mathcal L}
\newcommand{\Pos}{\operatorname{Pos}}
\newcommand{\tr}{\operatorname{tr}}
\newcommand{\id}{\mathrm{id}}
\newcommand{\supp}{\operatorname{supp}}
\newcommand{\csupp}{\operatorname{supp}_{\mathrm{cl}}}
\newcommand{\spr}{\operatorname{spr}}
\newcommand{\spanop}{\operatorname{span}}

\newcommand{\vecop}{\operatorname{vec}}
\newcommand{\Choi}{\operatorname{Choi}}
\newcommand{\OFPS}[2]{\mathsf{OFPS}(#1,#2)}
\newcommand{\TF}{\mathfrak T}
\newcommand{\bX}{\mathbf X}
\newcommand{\bY}{\mathbf Y}
\newcommand{\bZ}{\mathbf Z}
\newcommand{\bz}{\mathbf z}
\newcommand{\by}{\mathbf y}
\newcommand{\br}{\mathbf r}
\newcommand{\bh}{\mathbf h}
\newcommand{\zero}{\mathbf 0}
\newcommand{\one}{\mathbf 1}
\newcommand{\mass}[1]{\lvert #1\rvert}
\newcommand{\sem}[1]{\llbracket #1\rrbracket}
\newcommand{\fy}[1]{\llbracket #1\rrbracket_{\mathrm{FY}}}
\newcommand{\ket}[1]{\lvert #1\rangle}
\newcommand{\bra}[1]{\langle #1\rvert}
\newcommand{\proj}[1]{\ket{#1}\!\bra{#1}}
\newcommand{\kw}[1]{\mathbf{#1}}
\newcommand{\cqwhile}{\textnormal{\textsf{cqWhile}}}
\newcommand{\sample}{\mathrel{\overset{\scriptscriptstyle\$}{\leftarrow}}}
\newcommand{\Cmap}{\mathcal C}
\newcommand{\Emap}{\mathcal X}
\newcommand{\Cone}{\overline{\Cmap}}
\newcommand{\Xone}{\overline{\Emap}}
\newcommand{\Qop}{\mathcal E}
\newcommand{\Rmap}{\mathcal R}
\newcommand{\Vsp}{\mathbb V}
\newcommand{\Act}{\mathsf A}
\newcommand{\Ex}{\mathsf E}
\newcommand{\Enc}{\operatorname{Enc}}
\newcommand{\E}{\mathbb E}
\newcommand{\Prb}{\mathbb P}
\newcommand{\Var}{\operatorname{Var}}
\newcommand{\cvar}{\mathsf{cVar}}
\newcommand{\qvar}{\mathsf{qVar}}
\newcommand{\iv}[1]{[\,#1\,]}
\newcommand{\NOT}{\mathsf{NOT}}
\newcommand{\ops}{\textnormal{\textit{ops}}}
\renewcommand{\wp}{\operatorname{wp}}

\newcommand{\lz}[1]{}

\author{Minbo Gao}
\affiliation{%
  \institution{Institute of Software, Chinese Academy of Sciences}
  \city{Beijing}
  \country{China}}
\affiliation{%
  \institution{University of Chinese Academy of Sciences}
  \city{Beijing}
  \country{China}}
\email{gmb17@tsinghua.org.cn}

\author{Bohan Ma}
\affiliation{%
  \institution{Zhejiang University}
  \city{Hangzhou}
  \country{China}}
\email{bohanma@zju.edu.cn}

\author{Tianshi Yu}
\affiliation{%
  \institution{Institute of Software, Chinese Academy of Sciences}
  \city{Beijing}
  \country{China}}
\affiliation{%
  \institution{University of Chinese Academy of Sciences}
  \city{Beijing}
  \country{China}}
\email{yuts@ios.ac.cn}

\author{Mingshuai Chen}
\affiliation{%
  \institution{Zhejiang University}
  \city{Hangzhou}
  \country{China}}
\email{m.chen@zju.edu.cn}

\author{Jianwei Yin}
\affiliation{%
  \institution{Zhejiang University}
  \city{Hangzhou}
  \country{China}}
\email{zjuyjw@zju.edu.cn}

\author{Li Zhou}
\affiliation{%
  \institution{Institute of Software, Chinese Academy of Sciences}
  \city{Beijing}
  \country{China}}
\email{zhouli@iscas.ac.cn}

\title{Operator-Valued Generating Functions for Classical-Quantum Programs}

\begin{abstract}
Quantum loops may execute arbitrarily many iterations, and quantitative properties such as almost-sure termination and expected running time are often analyzed using different techniques. We ask whether a single finite symbolic representation of a quantum loop can support multiple quantitative analyses. For probabilistic programs, generating functions provide a way to analyze distributions over classical variables, including counters that record loop iterations. In quantum programs, however, classical outcomes may be correlated with quantum states, so classical output distributions alone are insufficient.

We introduce operator-valued formal power series (OFPSs) to represent joint classical-quantum outputs and develop a compositional semantics. Although an OFPS may contain infinitely many coefficients, certain loops have finite control structures that allow exact symbolic analysis. We identify classical-quantum homogeneous-bounded loops and prove that their outputs admit rational representations, computable exactly using finite linear algebra without user-supplied invariants or assuming almost-sure termination. A single representation determines probabilities, moments, and conditional quantum states. For more general loops, we develop sound superinvariant bounds.
\end{abstract}


\begin{document}
\maketitle

\section{Introduction}\label{sec:intro}

Quantum loops may execute arbitrarily many iterations, making their quantitative properties difficult to analyze. Important questions include whether a loop terminates almost surely, its expected running time, and higher-order moments of its running time, such as variance. Previous work has developed methods for analyzing termination~\citep{LiYing2018,YingFeng2010,Ying2013} and expected running time~\citep{Avanzini2022,Liu2025,Yu2023}. These properties are often analyzed using different techniques, each focusing on specific quantitative questions. This raises a natural question: 

{
\begin{quote}  
    \emph{Can we compute a finite, exact symbolic representation of a quantum loop, from which multiple quantitative properties can be derived?}
\end{quote}
}

One way to analyze loop properties through program outputs is to record execution information in classical variables. For example, a classical counter that increments on each iteration records the number of loop iterations. The output distribution of this counter determines the iteration-count distribution and its moments for terminating executions. For probabilistic programs, generating-function methods provide symbolic representations of such output distributions, allowing multiple quantitative properties to be computed from a single representation~\citep{Klinkenberg2021,Klinkenberg2024}. The same idea can be applied to quantum loops by introducing classical variables to record iteration counts or accumulated costs. These variables may be unbounded, even when the quantum computation operates on a fixed number of qubits. Classical-quantum (CQ) programming languages~\citep{DengFeng2022,FengYing2021} provide a natural setting for this analysis, where classical variables are represented separately from finite-dimensional quantum memory.

However, the generating-function approach for probabilistic programs cannot be directly applied to CQ programs. Although classical and quantum memory are represented separately, their outputs may be correlated. A quantum measurement produces a classical outcome and updates the quantum state, while subsequent quantum operations may depend on the measurement outcome. Consequently, a loop may return different quantum states after different numbers of iterations. A classical generating function captures the distribution of iteration counts, but not the quantum state associated with each count. Even knowing both the classical output distribution and the marginal quantum state is insufficient to determine their joint output. Therefore, analyzing CQ loops requires a symbolic representation that preserves the correlations between classical outcomes and quantum states. Moreover, to analyze loops with infinitely many possible classical outputs, we need to compute this joint representation exactly without enumerating infinitely many classical states.

To represent joint classical-quantum outputs, we introduce \emph{operator-valued formal power series (OFPSs)}. Each monomial represents a classical state, and its coefficient is the corresponding unnormalized quantum state. The trace of each coefficient gives the probability of the classical outcome, while normalization gives the corresponding quantum state when the probability is nonzero. We define a compositional transformer semantics for CQ programs using OFPSs, prove its adequacy with respect to existing CQ semantics~\citep{FengYing2021}, and show that it conservatively extends probabilistic generating-function semantics~\citep{Klinkenberg2021}. A single OFPS determines termination probabilities, expected running times, higher-order count moments, and conditional quantum states. However, an OFPS may still have infinitely many nonzero coefficients. The question is therefore when such an output admits a finite representation that can be computed exactly.

The key observation is that infinitely many classical states do not necessarily require infinitely many distinct transitions. In many loops, the control flow depends on finitely many classical configurations, while unbounded counters are incremented without affecting subsequent computation. Homogeneous-bounded loops exploit this structure to obtain finite summaries for probabilistic programs~\citep{Klinkenberg2021}. Extending this approach to CQ programs requires preserving quantum-state transformations. Even when classical transition probabilities remain unchanged, quantum operations may depend on counter values and produce different quantum outputs. Thus, homogeneity must hold for the joint CQ output, not just its classical distribution. We define \emph{classical-quantum homogeneous-bounded (CQ-HB) loops} and construct finite summaries whose transitions record both classical increments and quantum operations. These summaries give finite systems of linear equations with polynomial coefficients.

Our main result establishes that CQ-HB loops have rational OFPS outputs for finitely supported inputs. Thus, infinitely many classical outcomes and their associated quantum states can be represented by finitely many rational functions. Given a finite loop summary and algebraic data, these representations can be computed exactly by solving a finite linear system, without user-supplied loop invariants. The rationality result does not require almost-sure termination: the joint output of terminating executions remains rational even when some executions diverge.

Our exact analysis applies to CQ-HB loops, but loops with unbounded classical control may fall outside this class and have non-rational outputs. For such programs, we develop a \emph{superinvariant} rule for deriving sound upper bounds on joint CQ outputs. We reduce the verification conditions to positivity of linear maps and provide sufficient conditions based on positive-semidefinite Choi matrices. These bounds can be used to analyze multiple quantitative properties when exact rational representations are unavailable.

\paragraph{Summary of Contributions.} We make three main contributions:

\begin{enumerate}
\item Compositional joint-output semantics. We develop an OFPS semantics for CQ programs that preserves correlations between classical outcomes and quantum states. We establish its adequacy with respect to existing CQ semantics, show that it conservatively extends probabilistic generating-function semantics, and characterize its observational expressiveness (Sections~\ref{sec:ofps}--\ref{sec:semantics}).

\item Exact symbolic analysis of infinite-state CQ loops. We identify CQ-HB loops and establish that their complete joint outputs admit rational representations for finitely supported inputs, without assuming almost-sure termination. We further show how these representations can be computed exactly from finite summaries with algebraic data, without user-supplied loop invariants (Section~\ref{sec:hb}).

\item Sound analysis beyond rational outputs. We develop superinvariant-based reasoning for bounding joint CQ outputs and establish positivity-based verification conditions, including sufficient conditions based on Choi matrices (Section~\ref{sec:superinv}).
\end{enumerate}

We demonstrate these capabilities through repeat-until-success circuits, sequential quantum readout, a quantum Bernoulli-factory subroutine, and absorbing quantum walks (Sections~\ref{sec:overview}, \ref{sec:hb}--\ref{sec:case-studies}). These case studies illustrate how joint-output analysis provides information beyond classical output distributions and individual quantitative properties.

\paragraph{Relation to Closest Work.} Existing work provides two important foundations for our approach. Classical-quantum semantics~\citep{Barbosa2021,DengFeng2022,FengYing2021} faithfully represents joint classical-quantum outputs, but these representations may be infinite when classical memory is unbounded. Generating-function methods~\citep{Klinkenberg2021,Klinkenberg2024}, on the other hand, enable exact symbolic analysis of infinite classical distributions, but their scalar coefficients do not capture the associated quantum states. Operator-valued generating functions also have precedents in quantum recurrence~\citep{Bourgain2014,Grunbaum2013} and graph-series semantics~\citep{Magnot2026}. Our main contribution is a program-level exact-analysis guarantee: we identify a class of infinite-state CQ loops whose complete joint outputs admit finite rational representations, computable exactly without assuming almost-sure termination.

A complementary line of work studies quantitative properties of quantum programs. \citet{Avanzini2022} develop expectation transformers for CQ programs, including expected-cost reasoning and a characterization of CQ denotational semantics. \citet{Liu2025} and \citet{Yu2023} provide finite-dimensional methods for computing expected runtimes of quantum programs. In comparison, our analysis of CQ-HB loops computes a finite symbolic representation of the joint output of terminating executions. From this representation, we can derive the iteration-count subdistribution over terminating executions, its submoments and tail probabilities, and quantum states conditioned on iteration counts.

Hybrid path-sum methods also support symbolic reasoning about CQ programs~\citep{Chareton2026hps}, with invariant-based reasoning for unbounded loops~\citep{Chareton2026}. Our exact analysis instead identifies CQ-HB loops whose joint outputs can be computed as rational OFPSs without supplied loop invariants.

\FloatBarrier
\section{A Running Example}\label{sec:overview}\label{sec:overview-retry}
This section illustrates our approach on a small loop, before any formal development. Readers unfamiliar with density operators and the ket notation $\ket\psi$ may first consult Section~\ref{sec:background}.

Many quantum procedures retry an operation on the same qubit. When a measurement indicates that another attempt is needed, the qubit is retained for the next round. The program~\eqref{prog:retry} is the simplest such loop. It repeatedly applies a two-outcome measurement $M$ to a qubit $q$ until outcome $0$ occurs. The counter $n$ records the number of measurement attempts.

Starting with $b=1$ and $n=0$, the loop is
\begin{equation}\label{prog:retry}
 \mathrm{Retry}\equiv
 \kw{while}\ b=1\ \kw{do}\ n:=n+1;\ b\leftarrow\kw{meas}\ M[q]\ \kw{od}.
\end{equation}
Here, the measurement $M=\{M_0,M_1\}$ has operators
\[
 M_0=\frac1{\sqrt5}\mathrm{diag}(1,2),\qquad
 M_1=\frac1{\sqrt5}\mathrm{diag}(2,1).
\]
The loop exits on outcome $0$ and repeats on outcome $1$.
For an input density operator $\rho$, we seek the joint output of the attempt count $n$ and the final qubit state.

Tracking probabilities alone does not capture this joint output. For example, with initial state $\ket+=(\ket0+\ket1)/\sqrt2$, the first measurement yields outcome $1$ with probability $\frac12$. If the probability of continuing were $\frac12$ at every attempt, the loop would take $2$ attempts on average, but it actually takes $3.125$. The reason is that outcome $1$ changes the qubit state. This outcome is four times as likely on $\ket0$ as on $\ket1$, so observing it shifts the state toward $\ket0$ and increases the probability of continuing at the next attempt. The loop therefore acts on quantum states rather than on a fixed continuation probability, and the returned state depends on the number of attempts.

We now derive the joint output for a general input $\rho$. The loop exits after $m$ attempts with probability $p_m$ and returns the corresponding normalized qubit state $\rho_m$. We encode these joint outputs in the operator-valued formal power series (OFPS)
\[
 F_\rho=p_1\rho_1N+p_2\rho_2N^2+\cdots
       =\sum_{m\ge1}p_m\rho_mN^m.
\]
The formal indeterminate $N$ records the classical count $n$. The coefficient of $N^m$ is the unnormalized state $p_m\rho_m$ for the event $n=m$: its trace gives the probability $p_m$, and dividing by $p_m$ recovers the conditional output state $\rho_m$. The final flag $b$ is always $0$ and is omitted.

For an unnormalized input $p\rho$, we write $F_{p\rho}=pF_\rho$ to retain the branch probability $p$. Splitting on the first measurement gives
\begin{equation}\label{eq:retry-equation}
 F_\rho=N\,M_0\rho M_0^\dagger+N\,F_{M_1\rho M_1^\dagger}.
\end{equation}
Here $F_\rho$ describes all terminating executions that start in $\rho$. If the first measurement gives $0$, the loop exits with unnormalized state $M_0\rho M_0^\dagger$. This contributes $N M_0\rho M_0^\dagger$, since one attempt has been made. If it gives $1$, the loop continues from $M_1\rho M_1^\dagger$, with the probability of that outcome retained in its trace. The series $F_{M_1\rho M_1^\dagger}$ describes all subsequent attempts from this state. Because $n$ affects neither the guard nor the measurement, the first attempt is accounted for by multiplying this series by $N$. These two cases partition the terminating executions, so their contributions add.

Write $\rho_{ij}=\bra i\rho\ket j$ in the basis $(\ket0,\ket1)$. Since $M_0$ and $M_1$ are diagonal, we can write $(F_\rho)_{ij}=c_{ij}\rho_{ij}$, where $c_{ij}\in\C[[N]]$ is a scalar series independent of the input $\rho$. Substituting this form into Equation~\eqref{eq:retry-equation} gives, for example,
\[
 c_{00}=\frac N5+\frac{4N}5c_{00}.
\]
Solving this and the corresponding equations for the remaining entries gives
\begin{equation}\label{eq:retry-solution}
 F_\rho=
 \begin{pmatrix}
  \dfrac{N}{5-4N}\rho_{00} & \dfrac{2N}{5-2N}\rho_{01}\\[9pt]
  \dfrac{2N}{5-2N}\rho_{10} & \dfrac{4N}{5-N}\rho_{11}
 \end{pmatrix}.
\end{equation}
Specializing Equation~\eqref{eq:retry-solution} to $\rho=\proj+$ and expanding $F_\rho$ gives the count probabilities and the corresponding output states in the table below. The last column concerns an additional measurement of the output qubit in the basis $\ket\pm=(\ket0\pm\ket1)/\sqrt2$ after the loop has exited. The probability of outcome $+$, conditioned on $n=m$, is $\Prb(+\mid n=m)=\bra+\rho_m\ket+$.
\begin{center}
\small
\begin{tabular}{@{}cccc@{}}
\toprule
attempts $m$ & $\Prb(n=m)$ & state after exit & $\Prb(+\mid n=m)$\\
\midrule
$1$ & $0.5$ & $\sqrt{0.2}\ket0+\sqrt{0.8}\ket1$ & $0.9$\\
$2$ & $0.16$ & $\ket+$ & $1$\\
$3$ & $0.08$ & $\frac1{\sqrt5}(2\ket0+\ket1)$ & $0.9$\\
$4$ & $0.054$ & $\frac1{\sqrt{17}}(4\ket0+\ket1)$ & $0.74$\\
\bottomrule
\end{tabular}
\end{center}

The table illustrates two features of the joint output. The conditional state $\rho_m$ depends on $m$: a measurement in the basis $\{\ket0,\ket1\}$ yields $0$ with probability $0.2$ for $m=1$ and $0.8$ for $m=3$. The output also retains quantum coherence. For $m=2$, $\rho_2=\proj+$ yields outcome $+$ with probability $1$ in the additional measurement. The classical mixture $\frac12\proj0+\frac12\proj1$ has the same probabilities in the $\{\ket0,\ket1\}$ basis but yields $+$ with probability $\frac12$. Equation~\eqref{eq:retry-solution} records both the dependence on $m$ and the coherence of the conditional states, together with the count distribution and its mean (Section~\ref{sec:ofps-queries}).

The rest of the paper makes this calculation precise and general. Sections~\ref{sec:ofps} and~\ref{sec:semantics} define OFPSs and the semantics that justifies unfolding equations such as~\eqref{eq:retry-equation}. Section~\ref{sec:hb} computes the exact outputs of a class of loops that includes this one, and Section~\ref{sec:superinv} provides a superinvariant rule for bounding the outputs of arbitrary loops.

\FloatBarrier
\section{Background}\label{sec:background}
This section introduces the quantum notions used in the paper, following \citet{NielsenChuang2010} and \citet{Watrous2018}. The classical-quantum model and cq-state definitions in Section~\ref{sec:ofps} follow \citet{FengYing2021}.

\paragraph{Conventions.}
Throughout, $\N=\{0,1,\ldots\}$. Program variables such as $x,n,b$ are distinct from the formal indeterminates $X,N,B$ that encode them. We use $\bX,\bY,\bZ$ for tuples of indeterminates and $\bz,\by,t$ for numerical arguments of generating functions. We distinguish a formal series $F$ from its generating function $f_F$ (Definition~\ref{def:gf}). A pointed $\omega$-complete partial order ($\omega$-cpo) has a least element and least upper bounds of increasing sequences; Kleene's theorem computes least fixed points of continuous maps as limits of iterates~\citep{AbramskyJung1994}. More details can be found in the appendices.

\paragraph{Linear algebra notions and quantum states}\label{sec:quantum-states}

We model a quantum system by a finite-dimensional \emph{Hilbert space} $\Hil\cong\C^D$, a complex vector space equipped with an inner product. We use Dirac notation: a unit vector $\ket\psi\in\Hil$ represents a \emph{pure quantum state}, and $\bra\psi$ denotes its conjugate transpose. The matrix $\ket i\bra j$ has a single $1$ in row $i$ and column $j$, and $\ket\psi\ket\varphi \coloneq \ket\psi \otimes \ket\phi$ stands for the joint state of two subsystems. We write $\Lin(\Hil)$ for the set of linear operators on $\Hil$, $A^\dagger$ for the conjugate transpose, and $I$ for the identity.

An operator $A\in\Lin(\Hil)$ is \emph{positive (semidefinite)}, written $A\succeq0$, if $\bra\psi A\ket\psi\ge0$ for every $\ket\psi\in\Hil$, and $\Pos(\Hil)$ denotes the set of positive operators. The \emph{trace} of $A$ is $\tr (A)\coloneq \sum_i \bra{i} A \ket{i}$, i.e., the sum of its diagonal entries in any orthonormal basis. A \emph{density operator} is a positive operator with trace $1$, and a \emph{partial density operator} is a positive operator with trace at most $1$. The \emph{L\"owner order} is defined by $A\preceq B$ iff $B-A\succeq0$. Intuitively speaking, positivity generalizes nonnegativity of probabilities, the trace generalizes total probability mass, and the L\"owner order generalizes the pointwise order on subdistributions.

We use $\rho$ for quantum states, including partial density operators, and specify normalization when needed.

Following \citet{Selinger2004}, programs never renormalize. A branch reached with probability $p$ in normalized state $\rho$ is represented by the single operator $p\rho$: its trace $p$ is the probability, and its normalization $\rho$ is the state. This makes branching and merging linear, just as for subdistributions.

\begin{table}[!htbp]
\caption{Probabilistic, quantum, and classical-quantum notions, extending the comparison of \citet[Table~1]{FengYing2021}. The last column uses the notation of Sections~\ref{sec:ofps} and~\ref{sec:semantics}.}\label{tab:analogy}
\small
\begin{tabular}{@{}llll@{}}
\toprule
 & probabilistic & quantum & classical-quantum\\
\midrule
state & subdistribution $\mu$ & partial density operator $\rho$ & cq-state $\Delta:\Sigma\to\Pos(\Hil)$\\
mass & $\sum_\sigma\mu(\sigma)$ & $\tr\rho$ & $\sum_\sigma\tr\Delta(\sigma)$\\
order & pointwise $\le$ & L\"owner $\preceq$ & $\preceq$ at each classical state\\
one step & substochastic matrix & quantum operation $\Qop$ & kernels $\kappa^S_{\tau,\sigma}$\\
observation & expected value & $\tr(O\rho)$ for an effect $O$ & $\sum_\sigma\tr\bigl(O(\sigma)\Delta(\sigma)\bigr)$\\
generating function & $\sum_\sigma\mu(\sigma)\bX^\sigma$ & --- & OFPS $\sum_\sigma \rho_\sigma\bX^\sigma$\\
\bottomrule
\end{tabular}
\end{table}

\paragraph{Quantum operations.}\label{sec:quantum-ops}
Programs change quantum states in three ways. A \emph{unitary} $U$ maps $\rho$ to $U\rho U^\dagger$; it is the quantum analogue of a bijection on states and preserves the trace. A \emph{reset} $\Rmap_q$ discards the state of subsystem $q$ and replaces it by $\ket0$. A \emph{measurement} produces a classical outcome and updates the state.
All three are \emph{quantum operations}.

Formally, a quantum operation is a linear map $\Qop$ on $\Lin(\Hil)$ of the form $\Qop(A)=\sum_iK_iAK_i^\dagger$ for finitely many operators $K_i$ with $\sum_iK_i^\dagger K_i\preceq I$, and $\mathrm{QO}(\Hil)$ denotes the set of quantum operations. Equivalently, $\Qop$ does not increase the trace of positive operators and is \emph{completely positive} (CP): it maps positive operators to positive operators even when it acts on one part of a larger system, that is, $\Qop\otimes\id$ is positive for every extension. 

A \emph{measurement} is a family $M=\{M_i\}_{i\in I_M}$ of operators, indexed by a set $I_M\subseteq\N$ of outcome labels, with $\sum_{i\in I_M}M_i^\dagger M_i=I$. On a partial density operator $\rho$, outcome $i$ occurs with probability $\tr(M_i\rho M_i^\dagger)$ and leaves the unnormalized state $M_i\rho M_i^\dagger$. The condition $\sum_iM_i^\dagger M_i=I$ says that the outcome probabilities add up to $\tr \rho$. In the retry loop of Section~\ref{sec:overview-retry}, $M_0$ and $M_1$ correspond to exit and continuation, respectively, and $M_0\rho M_0^\dagger$ and $M_1\rho M_1^\dagger$ are the corresponding unnormalized states. An \emph{observable} is a Hermitian operator $O$, and $\tr(O\rho)$ is its expected value on $\rho$. An \emph{effect} is an observable with $0\preceq O\preceq I$; then $\tr(O\rho)$ is the probability that a yes/no measurement accepts $\rho$. Effects are the quantum analogue of $[0,1]$-valued random variables. Table~\ref{tab:analogy} summarizes the analogy.

\paragraph{Choi matrices.}\label{sec:choi}
Fix an orthonormal basis $\ket0,\ldots,\ket{D-1}$ of $\Hil$ and write $E_{ij}=\ket i\bra j$. The \emph{Choi matrix} of a linear map $\Theta$ on $\Lin(\Hil)$ is the $D^2\times D^2$ block matrix whose $(i,j)$ block is $\Theta(E_{ij})$; equivalently, $\Choi(\Theta)=\sum_{i,j=0}^{D-1}E_{ij}\otimes\Theta(E_{ij})$.

\FloatBarrier
\section{Operator-Valued Formal Power Series}\label{sec:ofps}
This section develops the mathematical framework for OFPSs. 

Throughout, following \citet{FengYing2021}, fix a finite set $\cvar=\{x_1,\ldots,x_k\}$ of classical variables with values in $\N$ and a finite set $\qvar$ of quantum subsystems forming the quantum memory. A \emph{classical state} is a vector $\sigma\in\Sigma:=\N^k$ of values of the classical variables, and $\Hil$, a Hilbert space of dimension $D$, is the joint space of the subsystems in $\qvar$. There is one indeterminate $X_i$ per classical variable $x_i$, and $\bX^\sigma=X_1^{\sigma_1}\cdots X_k^{\sigma_k}$ is the monomial of $\sigma$.

\subsection{Classical-Quantum States and their Series}\label{sec:ofps-def}
We use two objects that carry the same information but play different roles. A \emph{classical-quantum state} is the semantic notion of a program state used by \citet{FengYing2021,DengFeng2022,Barbosa2021}: a function from classical states to quantum states. An \emph{operator-valued formal power series} is an algebraic representation of such a function. The two are related as a probability distribution is related to its generating function.

\begin{definition}[Classical-quantum state~\citep{FengYing2021}]\label{def:cqstate}
A \emph{classical-quantum state} (cq-state) is a function $\Delta:\Sigma\to\Pos(\Hil)$ with $\sum_{\sigma\in\Sigma}\tr\Delta(\sigma)\le1$.
\end{definition}

The operator $\Delta(\sigma)$ is the unnormalized quantum state in the event that the classical state is $\sigma$. Its trace is the probability of that event, and its normalization is the quantum state in that event. When $\Delta$ is the output from a normalized program input, its total $\sum_\sigma\tr\Delta(\sigma)$ is the probability of termination. For instance, measuring the qubit $\ket+$ in the basis $\ket0,\ket1$ and storing the outcome in $x$ produces the cq-state that maps $x=0$ to $\frac12\proj0$ and $x=1$ to $\frac12\proj1$.

\begin{definition}[Operator-valued formal power series]\label{def:ofps}
An \emph{operator-valued formal power series} (OFPS) over $\Hil$ in $\bX$ is a formal sum $F=\sum_{\sigma\in\Sigma}A_\sigma\bX^\sigma$ with coefficients $[\bX^\sigma]F:=A_\sigma\in\Lin(\Hil)$. We write $\Lin(\Hil)[[\bX]]$ for the set of OFPSs and $\Pos(\Hil)[[\bX]]$ for those with positive coefficients. Addition and scalar multiplication are coefficientwise, and for $\br\in\N^k$ the \emph{shift} is $\bX^\br F:=\sum_\sigma A_\sigma\bX^{\sigma+\br}$.
\end{definition}

This definition imposes no convergence requirement.

\begin{definition}[Admissible OFPS]\label{def:admissible}
The set of \emph{admissible} OFPSs is
\[
 \OFPS{\Hil}{\bX}:=\Bigl\{F\in\Pos(\Hil)[[\bX]]\;:\;\mass F\le1\Bigr\},
 \qquad\text{where}\quad \mass F:=\sum_{\sigma\in\Sigma}\tr\bigl([\bX^\sigma]F\bigr)
\]
is the \emph{mass} of $F$. The \emph{classical support} of $F$ is $\csupp F:=\{\sigma\in\Sigma:[\bX^\sigma]F\ne0\}$.
\end{definition}

When $\Hil=\C$, admissible OFPSs reduce to the scalar generating series for subdistributions used by \citet{Klinkenberg2021}.

The \emph{encoding} of a cq-state $\Delta$ is $\Enc\Delta:=\sum_{\sigma}\Delta(\sigma)\bX^\sigma\in\OFPS{\Hil}{\bX}$. For example, the cq-state produced by the measurement above is encoded as $\frac12\proj0+\frac12\proj1X$.

\begin{proposition}[Encoding]\label{prop:encoding}
The map $\Enc$ is a bijection between cq-states $\Delta:\Sigma\to\Pos(\Hil)$ and $\OFPS{\Hil}{\bX}$.
\end{proposition}
\begin{proof}
Its inverse sends $F\in\OFPS{\Hil}{\bX}$ to the cq-state $\sigma\mapsto[\bX^\sigma]F$; admissibility ensures positivity and total trace at most one.
\end{proof}

For an admissible OFPS $F=\sum_\sigma\rho_\sigma\bX^\sigma$, each coefficient $\rho_\sigma$ is a partial density operator, and $p_\sigma:=\tr\rho_\sigma$ is the probability of the classical outcome $\sigma$. When $p_\sigma>0$, let $\widehat\rho_\sigma:=\rho_\sigma/p_\sigma$ be its normalized conditional quantum state. Then
\[
 F=\sum_{\sigma:p_\sigma>0}p_\sigma\widehat\rho_\sigma\bX^\sigma.
\]

We retain cq-states to connect our semantics to the existing semantics of classical-quantum programs (Section~\ref{sec:adequacy}). We use OFPSs for symbolic analysis: increments correspond to multiplication by monomials, and loop unfolding yields equations between series. For admissible OFPSs, the associated generating functions also support marginalization and moment computation (Section~\ref{sec:ofps-queries}).

A special case is an OFPS of the form $G\rho$, where $G$ is a probability generating function and $\rho$ is a fixed normalized state. In such an output the classical result and the quantum state are independent: whatever classical state is observed, the quantum state is $\rho$.

\paragraph{Order and limits.}\label{sec:ofps-order}
Loops are interpreted as limits of their finite unrollings, so we need an order in which these unrollings increase.

\begin{definition}[Coefficientwise order]\label{def:ofps-order}
For $F,G\in\Pos(\Hil)[[\bX]]$, the \emph{coefficientwise L\"owner order} is defined by
\[
 F\sqsubseteq G
 \quad\Longleftrightarrow\quad
 [\bX^\sigma]F\preceq[\bX^\sigma]G
 \quad\text{for every }\sigma\in\Sigma.
\]
\end{definition}

A finite or countable family $(F_i)_{i\in I}$ in $\OFPS{\Hil}{\bX}$ is \emph{admissible} if $\sum_i\mass{F_i}\le1$, and its sum is taken coefficientwise. Addition is only partially defined within $\OFPS{\Hil}{\bX}$: the sum of two series of mass one is not admissible.

\begin{proposition}[Domain of admissible OFPSs]\label{prop:ofps-domain}
$(\OFPS{\Hil}{\bX},\sqsubseteq)$ is a pointed $\omega$-cpo with the zero series as its least element. For every increasing sequence $F_1\sqsubseteq F_2\sqsubseteq\cdots$ in $\OFPS{\Hil}{\bX}$, its supremum $F=\bigsqcup_{n\ge1}F_n$ belongs to $\OFPS{\Hil}{\bX}$ and satisfies
\[
 [\bX^\sigma]F=\lim_{n\to\infty}[\bX^\sigma]F_n
 \quad(\sigma\in\Sigma),
 \qquad
 \mass F=\sup_{n\ge1}\mass{F_n}.
\]
The coefficient limits are taken in trace norm. The sum of an admissible family is the supremum of its finite partial sums.
\end{proposition}
The proof is given in Appendix~\ref{app:domain}.

\paragraph{Restriction and shift.}\label{sec:ofps-ops}
A conditional or a loop guard splits an input according to its classical state. For $F=\sum_\sigma A_\sigma\bX^\sigma\in\Lin(\Hil)[[\bX]]$ and $\mathcal S\subseteq\Sigma$, the \emph{restriction} of $F$ to $\mathcal S$ is $F|_{\mathcal S}:=\sum_{\sigma\in\mathcal S}A_\sigma\bX^\sigma$. We identify a guard $e$ with the set of classical states that satisfy it, so that $F=F|_e+F|_{\neg e}$. The shift of Definition~\ref{def:ofps} is the effect of incrementing variables. Restriction and shift do not commute: for the guard $x=0$, shifting first gives $(XF)|_{x=0}=0$, whereas restricting first keeps $X(F|_{x=0})$.

\subsection{Computing Output Properties}\label{sec:ofps-queries}
\begin{definition}[Generating function; rational OFPS]\label{def:gf}
For an admissible OFPS $F=\sum_\sigma \rho_\sigma\bX^\sigma\in\OFPS{\Hil}{\bX}$, its \emph{generating function} is the operator-valued function $f_F(\bz)=\sum_\sigma \rho_\sigma\bz^\sigma$ on the closed unit polydisc $|z_i|\le1$.

A general OFPS $G=\sum_\sigma B_\sigma\bX^\sigma\in\Lin(\Hil)[[\bX]]$ is \emph{rational} if for every matrix position $(i,j)$ the scalar series $\sum_\sigma(B_\sigma)_{ij}\bX^\sigma$ is the Taylor expansion at $\zero$ of a quotient $P/Q$ of polynomials with $Q(\zero)\ne0$. For admissible $F$, this is equivalent to every matrix entry of its generating function $f_F$ agreeing with a rational function in a neighborhood of $\zero$.
\end{definition}

For $F\in\OFPS{\Hil}{\bX}$, the series defining $f_F$ converges absolutely and uniformly in trace norm on the closed unit polydisc. Its sum is analytic in the interior, with the coefficients of $F$ as its Taylor coefficients at $\zero$ (Proposition~\ref{prop:gf}). A rational OFPS is a finite description of infinitely many coefficients, as in~\eqref{eq:retry-solution}.

We now describe how to obtain output properties from an admissible OFPS $F=\sum_\sigma \rho_\sigma\bX^\sigma\in\OFPS{\Hil}{\bX}$. For an \emph{event} $\mathcal S\subseteq\Sigma$, the \emph{event operator} is $\rho_{\mathcal S}:=\sum_{\sigma\in\mathcal S}\rho_\sigma$, and $p_{\mathcal S}:=\tr \rho_{\mathcal S}$ is the \emph{probability} of $\mathcal S$. If $p_{\mathcal S}>0$, the normalized \emph{conditional state} on $\mathcal S$ is $\widehat\rho_{\mathcal S}:=\rho_{\mathcal S}/p_{\mathcal S}$. For $x_i\in\cvar$ and $r\in\N$, the \emph{$r$-th factorial submoment} of $x_i$ is $\sum_\sigma\sigma_i^{\underline{r}}\,\tr \rho_\sigma\in[0,\infty]$, where $m^{\underline{r}}:=m(m-1)\cdots(m-r+1)$ for $r\ge1$ and $m^{\underline{0}}:=1$.

For a normalized program input and a counter $n=x_i$ initialized at zero, let $p_m=\Prb(n=m,\mathrm{termination})$. Its marginal generating function is $g_i(z)=\sum_{m\ge0}p_mz^m$. For time-to-termination queries, assume the counter records elapsed steps, and define $T$ as the recorded count on termination and $\infty$ on divergence.

\begin{table}[!htbp]
\caption{Output properties computed from an admissible output OFPS $F$, for normalized program input. The last column illustrates these properties for the retry loop on input $\ket+$ (Section~\ref{sec:overview-retry}).}\label{tab:queries}
\small
\begin{tabular}{@{}lll@{}}
\toprule
Output property & Characterization & Retry loop\\
\midrule
Marginal generating function & $g_i(z_i):=\tr f_F(1,\ldots,z_i,\ldots,1)$ & $\frac12\bigl(\frac{z}{5-4z}+\frac{4z}{5-z}\bigr)$\\
Termination probability & $g_i(1)$ & $1$\\
First terminal submoment & $g_i'(1^-)$ & $25/8$\\
Conditional expected count & $g_i'(1^-)/g_i(1)$ & $25/8$\\
Expected time to termination & $\begin{cases}g_i'(1^-),&g_i(1)=1,\\\infty,&g_i(1)<1\end{cases}$ & $25/8$\\
Conditional state on $\mathcal S$ & $\widehat\rho_{\mathcal S}=\rho_{\mathcal S}/p_{\mathcal S}$ & $\widehat\rho_{\{n=2\}}=\proj+$\\
Conditional acceptance of $O$ & $\tr(O\rho_{\mathcal S})/p_{\mathcal S}$ & $\Prb(+\mid n=2)=1$\\
Independence & $F=G\rho$, $\tr\rho=1$ & Dependent\\
\bottomrule
\end{tabular}
\par\smallskip
{\footnotesize\raggedright\emph{Note.} Conditional expected counts require $g_i(1)>0$. Conditional moments on an event $\mathcal S$ are obtained by applying the derivative formula to $F|_{\mathcal S}$ and dividing by $p_{\mathcal S}>0$.\par}
\end{table}
\FloatBarrier

More generally, $\lim_{z\to1^-}g_i^{(r)}(z)$ is the $r$-th terminal factorial submoment, computed without normalization. Dividing by $g_i(1)>0$ gives moments conditional on termination. Under almost-sure termination, $g_i(1)=1$, these coincide with ordinary count moments, which may still be infinite. For example, $g_i(z)=z/2$ gives first terminal submoment $1/2$ and conditional expected count $1$, but $\E[T]=\infty$. Appendix~\ref{app:counts} gives the derivation and discusses accumulated costs on diverging executions.

\FloatBarrier
\section{\cqwhile{} and its OFPS Semantics}\label{sec:semantics}
We give the classical-quantum language \cqwhile{} an OFPS transformer semantics. We establish agreement with existing semantics and characterize semantic equivalence through classical observations in program contexts.

\subsection{Syntax}\label{sec:syntax}
In this paper, we restrict classical variables to $\N=\{0,1,\ldots\}$, as in the probabilistic generating-function semantics of \citet{Klinkenberg2021}. This is a natural choice for power-series representations, where classical values appear as nonnegative exponents. The integer-valued case can be approached along similar lines, but requires additional care in the encoding and analysis.

We follow the syntax of \citet[Section~4.1]{FengYing2021}, with $\cvar$ and $\qvar$ from Section~\ref{sec:ofps}. Classical expressions $e$ are total functions of the state, with natural-number values in assignments and Boolean values in guards. We write $\sigma[v/x]$ for $\sigma$ with $x$ updated to $v$.

\begin{definition}[Syntax of \cqwhile{}]\label{def:syntax}
Programs are generated by
\[
\begin{aligned}
 S::={}&\kw{skip}\mid x:=e\mid x\sample\mu\\
 &\mid q:=\ket0\mid\bar q:=U[\bar q]\mid x\leftarrow\kw{meas}\ M[\bar q]\\
 &\mid S_1;S_2\mid\kw{if}\ e\ \kw{then}\ S_1\ \kw{else}\ S_2\ \kw{fi}
 \mid\kw{while}\ e\ \kw{do}\ S\ \kw{od},
\end{aligned}
\]
where $x\in\cvar$, $q\in\qvar$, $\bar q$ is a tuple of distinct subsystems in $\qvar$, $\mu$ is a probability distribution on $\N$, $U$ is a unitary, and $M$ is a measurement on the subsystems $\bar q$ (Section~\ref{sec:quantum-ops}). Quantum operations act as the identity on subsystems outside their operands.
\end{definition}

Note that, measurement outcomes are stored in classical variables and can be used in guards.

\subsection{Transformer Semantics}\label{sec:transformers}
A program denotes a map $\OFPS{\Hil}{\bX}\to\OFPS{\Hil}{\bX}$. As in the probabilistic setting, these maps are continuous and do not create mass. Linearity is required on admissible combinations, since $\OFPS{\Hil}{\bX}$ is not closed under arbitrary sums.

\begin{definition}[Healthy transformer]\label{def:healthy}
A map $\psi:\OFPS{\Hil}{\bX}\to\OFPS{\Hil}{\bX}$ is \emph{healthy} if the following conditions hold for all $F,G\in\OFPS{\Hil}{\bX}$:
\begin{enumerate}
\item it does not create mass: $\mass{\psi(F)}\le\mass F$;
\item it is \emph{linear on admissible combinations}: $\psi(aF+bG)=a\psi(F)+b\psi(G)$ whenever $a,b\ge0$ and $a\mass F+b\mass G\le1$;
\item it is \emph{continuous}: it is monotone and preserves the limits of increasing sequences.
\end{enumerate}
$\TF$ is the set of healthy transformers, ordered by $\psi\sqsubseteq\psi'$ iff $\psi(F)\sqsubseteq\psi'(F)$ for every $F\in\OFPS{\Hil}{\bX}$.
\end{definition}

\begin{definition}[OFPS transformer semantics]\label{def:semantics}
The OFPS semantics assigns to each program $S$ a transformer $\sem S:\OFPS{\Hil}{\bX}\to\OFPS{\Hil}{\bX}$, defined by structural recursion according to Table~\ref{tab:semantics}. For a guard $e$ and loop body $S$, the \emph{unfolding functional} $\Phi_{e,S}:\TF\to\TF$ is defined by
\begin{equation}\label{eq:unfolding}
 \Phi_{e,S}(\psi)(F):=F|_{\neg e}+\psi\bigl(\sem S(F|_e)\bigr)
\end{equation}
for all $\psi\in\TF$ and $F\in\OFPS{\Hil}{\bX}$. In the loop clause, $\operatorname{lfp}_{\sqsubseteq}\Phi_{e,S}$ denotes the least fixed point of $\Phi_{e,S}$ in $\TF$ under the pointwise order of Definition~\ref{def:healthy}.
\end{definition}
Theorem~\ref{thm:well-defined} proves that these clauses are well defined and yield healthy transformers.

\begin{table}[!htbp]
\caption{OFPS semantics of \cqwhile{} for $F=\sum_\sigma\rho_\sigma\bX^\sigma\in\OFPS{\Hil}{\bX}$.}\label{tab:semantics}
\small
\renewcommand{\arraystretch}{1.25}
\begin{tabularx}{\linewidth}{@{}lX@{}}
\toprule
$S$ & $\sem S(F)$\\
\midrule
$\kw{skip}$ & $F$\\
$x:=e$ & $\sum_\sigma \rho_\sigma\,\bX^{\sigma[e(\sigma)/x]}$\\
$x\sample\mu$ & $\sum_\sigma\sum_{v}\mu(v)\,\rho_\sigma\,\bX^{\sigma[v/x]}$\\
$q:=\ket0$ & $\sum_\sigma\Rmap_q(\rho_\sigma)\,\bX^\sigma$\\
$\bar q:=U[\bar q]$ & $\sum_\sigma U\rho_\sigma U^\dagger\,\bX^\sigma$\\
$x\leftarrow\kw{meas}\ M[\bar q]$ & $\sum_\sigma\sum_{i}M_i\rho_\sigma M_i^\dagger\,\bX^{\sigma[i/x]}$\\
$S_1;S_2$ & $\sem{S_2}\bigl(\sem{S_1}(F)\bigr)$\\
$\kw{if}\ e\ \kw{then}\ S_1\ \kw{else}\ S_2\ \kw{fi}$ & $\sem{S_1}(F|_e)+\sem{S_2}(F|_{\neg e})$\\
$\kw{while}\ e\ \kw{do}\ S\ \kw{od}$ & $\bigl(\operatorname{lfp}_{\sqsubseteq}\Phi_{e,S}\bigr)(F)$\\
\bottomrule
\end{tabularx}
\end{table}

\FloatBarrier
\subsection{Properties of the Semantics}\label{sec:properties}
\paragraph{Well-definedness.}\label{sec:well-defined}
The Kleene iterates of the unfolding functional have a direct operational reading. We write $\kw{abort}$ for $\kw{while}\ \mathrm{true}\ \kw{do}\ \kw{skip}\ \kw{od}$. For a loop $W=\kw{while}\ e\ \kw{do}\ S\ \kw{od}$, let $W_0=\kw{abort}$ and
\[
 W_{m+1}=\kw{if}\ e\ \kw{then}\ S;W_m\ \kw{else}\ \kw{skip}\ \kw{fi},
\]
so that $W_m$ is the loop unrolled $m$ times.

\begin{theorem}[Healthiness and unfolding]\label{thm:well-defined}
\leavevmode
\begin{enumerate}
\item For every program $S$, $\sem S\in\TF$.
\item If $S$ is loop-free, then $\mass{\sem S(F)}=\mass F$ for every $F\in\OFPS{\Hil}{\bX}$.
\item For every loop $W=\kw{while}\ e\ \kw{do}\ S\ \kw{od}$, $\sem{W_m}=\Phi^m_{e,S}(0)$ for every $m\in\N$, where $0$ is the zero transformer, $\sem W=\bigsqcup_m\sem{W_m}$, and $\sem W(F)=F|_{\neg e}+\sem W\bigl(\sem S(F|_e)\bigr)$ for every $F\in\OFPS{\Hil}{\bX}$.
\end{enumerate}
\end{theorem}
\begin{proof}[Proof sketch]
We proceed by structural induction on $S$. Primitive commands are healthy and preserve mass. Sequencing and conditionals preserve healthiness, and preserve mass when their subprograms do. For loops, the induction hypothesis and the guard partition ensure that $\Phi_{e,S}:\TF\to\TF$ is well defined and $\omega$-continuous. Since $\TF$ is a pointed $\omega$-cpo with bottom $0$ (Lemma~\ref{lem:transformers}), Kleene's theorem yields the loop denotation. Induction on $m$ gives $\sem{W_m}=\Phi_{e,S}^m(0)$; the fixed-point identity gives the unfolding equation. The full proof is in Appendix~\ref{app:well-defined}.
\end{proof}

Linearity on admissible combinations and continuity allow the semantics on arbitrary inputs to be reconstructed from its action on elementary inputs $\rho\bX^\sigma$, as formalized by the coefficient kernels below. The unfolding equation supplies the loop equations used in the exact analysis of Section~\ref{sec:hb}. The characterization $\sem W=\bigsqcup_m\sem{W_m}$ allows upper bounds to be proved for each unrolling and then passed to the supremum, as in Section~\ref{sec:superinv}.

\paragraph{Coefficient kernels.}\label{sec:kernels}
For a program $S$ and fixed classical states $\sigma,\tau$, we extract from $\sem S$ the linear map sending a quantum input $\rho$ at $\sigma$ to the sum of the unnormalized outputs of all terminating executions ending in $\tau$. For normalized $\rho$, the trace of the resulting output operator is the probability that $S$ terminates in $\tau$.

\begin{definition}[Coefficient kernel]\label{def:kernel}
For a program $S$ and classical states $\sigma,\tau$, the \emph{kernel} $\kappa^S_{\tau,\sigma}$ is the linear map on $\Lin(\Hil)$ such that, for every partial density operator $\rho$, $\kappa^S_{\tau,\sigma}(\rho)$ is the coefficient of $\bX^\tau$ in $\sem S(\rho\,\bX^\sigma)$.
\end{definition}

For example, consider the retry program~\eqref{prog:retry}. In a single iteration, the kernel from $(b,n)=(1,h)$ to $(0,h+1)$ is $\rho\mapsto M_0\rho M_0^\dagger$, the kernel to $(1,h+1)$ is $\rho\mapsto M_1\rho M_1^\dagger$, and all other kernels from $(1,h)$ are zero. With $B$ marking $b$, the input $\rho B$ encodes $b=1$ and $n=0$, so $\sem{\mathrm{Retry}}(\rho B)=F_\rho$, the output OFPS of Section~\ref{sec:overview}. Inserting these kernels into~\eqref{eq:unfolding}, and shifting the counter $n$ out of the loop, gives Equation~\eqref{eq:retry-equation}.

Each kernel is a quantum operation: it is completely positive and does not increase the trace~\citep[Proposition~A.9]{Barbosa2021}. The kernels of $S$ determine its semantics, since the coefficient of $\bX^\tau$ in $\sem S(F)$ is $\sum_\sigma\kappa^S_{\tau,\sigma}\bigl([\bX^\sigma]F\bigr)$. 

\paragraph{Adequacy and conservativity.}\label{sec:adequacy}
\citet{FengYing2021} define the denotational semantics $\fy S$ of a classical-quantum program $S$ on cq-states by summing the outputs of terminating computations. The encoding $\Enc$ gives a bijection between cq-states and admissible OFPSs (Proposition~\ref{prop:encoding}). The following theorem establishes that, for \cqwhile{} programs, $\sem S$ and $\fy S$ agree under this encoding.

\begin{theorem}[Adequacy]\label{thm:adequacy}
For every \cqwhile{} program $S$ and every cq-state $\Delta$, $\sem S(\Enc\Delta)=\Enc\bigl(\fy S(\Delta)\bigr)$.
\end{theorem}

The adequacy proof is given in Appendix~\ref{app:adequacy}. For $\Hil=\C$, our semantics coincides with the probabilistic generating-function semantics of \citet{Klinkenberg2021} on the common language fragment and admissible inputs (Corollary~\ref{cor:conservative}).

\paragraph{Classical observations.}\label{sec:observations}
The quantum output of a program can affect the classical output of a surrounding computation. For example, two programs with identical classical outputs but qubit outputs $\ket0$ and $\ket1$ can be distinguished by a subsequent computational-basis measurement.

Let $\ket0$ denote the all-zero product state of the quantum memory. With all classical variables initially $0$, the input OFPS is $\proj0\bX^\zero$. For a program $R$, define its \emph{classical observation} by
\[
 \operatorname{obs}(R):=\sum_{\tau\in\Sigma}\tr\bigl([\bX^\tau]\sem R(\proj0\bX^\zero)\bigr)\,\bX^\tau .
\]
The coefficient of $\bX^\tau$ is the probability that $R$ terminates in classical state $\tau$ from this initial input.

A \emph{context} $C[\,\cdot\,]$ is a program with one hole; $C[S]$ fills it with $S$. A context $P;[\,\cdot\,];T$ can prepare inputs using $P$ and measure quantum outputs using $T$. The following proposition characterizes semantic equality through classical observations in contexts.

\begin{proposition}[Contextual characterization]\label{prop:observations}
Assume that $\cvar$ is nonempty. For all programs $S_1,S_2$,
\[
 \sem{S_1}=\sem{S_2}
 \quad\Longleftrightarrow\quad
 \operatorname{obs}(C[S_1])=\operatorname{obs}(C[S_2])
 \quad\text{for every context }C.
\]
Contexts of the form $P;[\,\cdot\,];T$ with loop-free programs $P$ and $T$ suffice.
\end{proposition}

\begin{proof}[Proof sketch]
Compositionality gives the forward implication. For the converse, a loop-free $P$ prepares an arbitrary classical state $\sigma$ and pure quantum state $\proj\psi$. A loop-free $T$ records in $x_1$ the outcome of the measurement with operators $M_0=\proj\varphi$ and $M_1=I-\proj\varphi$ when the classical output is $\tau$, and sets $x_1:=2$ otherwise. For $j=1,2$, the probability that $x_1=0$ after $P;S_j;T$ is $\bra\varphi\kappa^{S_j}_{\tau,\sigma}(\proj\psi)\ket\varphi$. Equality of these probabilities for all $\sigma,\tau$ and unit vectors $\ket\psi,\ket\varphi$ forces equality of every coefficient kernel, and hence of the denotations. When $\Hil=\C$, classical tests suffice. The full proof is in Appendix~\ref{app:observations}.
\end{proof}

With $\cvar$ nonempty, any compositional semantics that determines classical output distributions must therefore distinguish programs whose OFPS denotations differ.

\FloatBarrier
\section{Exact Analysis of Homogeneous-Bounded Loops}\label{sec:hb}
The retry loop in Section~\ref{sec:overview} has infinitely many reachable classical states because its counter is unbounded, yet its behavior is captured by a single equation. In this section, we explain why, identify the class of loops that admit such a description, and establish our main result: the output of every such loop is a rational OFPS, even without almost-sure termination. We use the retry loop as a running example throughout.

\subsection{Folding an Infinite State Space}\label{sec:hb-fold}
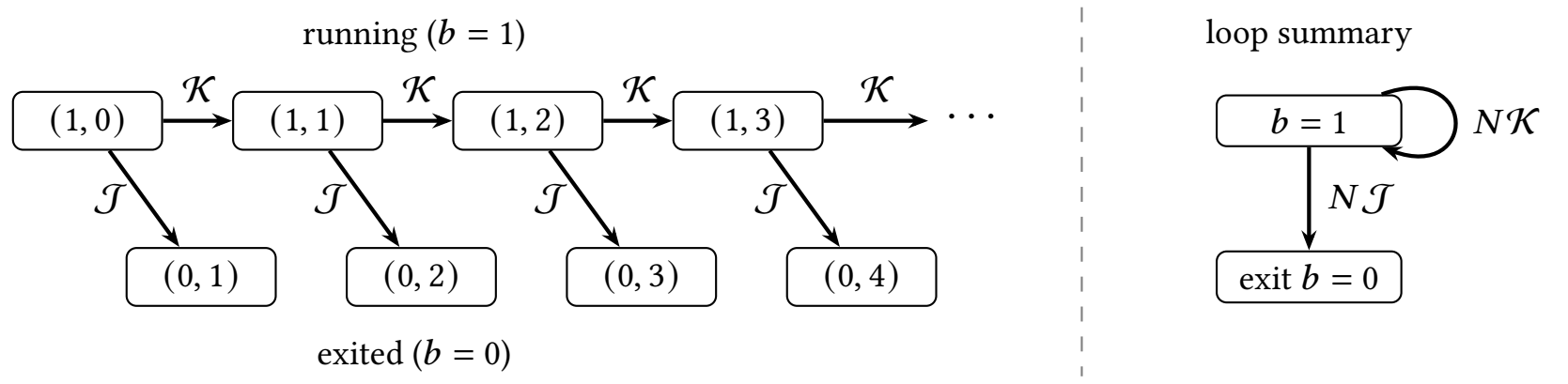
\begin{figure}[!htbp]
\centering
\begin{tikzpicture}[
  st/.style={draw,rounded corners=2pt,font=\scriptsize,inner sep=2.5pt,minimum width=1.05cm},
  arr/.style={-{Stealth[length=4pt]},thick},
  lab/.style={font=\scriptsize}]
\foreach \h in {0,1,2,3}{
  \node[st] (a\h) at (1.55*\h,0) {$(1,\h)$};
}
\foreach \h in {1,2,3,4}{
  \node[st] (e\h) at (1.55*\h-0.75,-1.1) {$(0,\h)$};
}
\node[font=\small] (dots) at (6.25,0) {$\cdots$};
\foreach \h/\n in {0/1,1/2,2/3}{
  \draw[arr] (a\h) -- node[above,lab]{$\mathcal K$} (a\n);
}
\draw[arr] (a3) -- node[above,lab]{$\mathcal K$} (dots);
\foreach \h/\n in {0/1,1/2,2/3,3/4}{
  \draw[arr] (a\h) -- node[left,lab]{$\mathcal J$} (e\n);
}
\node[font=\scriptsize,align=left] at (2.3,0.6) {running ($b=1$)};
\node[font=\scriptsize,align=left] at (2.3,-1.65) {exited ($b=0$)};
\draw[dashed,gray] (7.0,0.8) -- (7.0,-1.8);
\node[st,minimum width=1.3cm] (L) at (8.6,0) {$b=1$};
\node[st,minimum width=1.3cm] (X) at (8.6,-1.1) {exit $b=0$};
\draw[arr] (L) to[out=20,in=-20,looseness=5] node[right,lab]{$N\mathcal K$} (L);
\draw[arr] (L) -- node[right,lab]{$N\mathcal J$} (X);
\node[font=\scriptsize] at (8.6,0.6) {loop summary};
\end{tikzpicture}
\caption{Folding the retry loop. Left: its classical states and the quantum operations applied by one iteration. Right: the loop summary, in which the indeterminate $N$ records the increment of the counter $n$. An arrow labeled $N\mathcal K$ is the transition $1\xrightarrow{1,\mathcal K}1$ of Example~\ref{ex:retry-summary}, with increment $1$ and quantum operation $\mathcal K$.}
\label{fig:fold}
\end{figure}

Consider the classical states $(b,n)$ of the retry loop~\eqref{prog:retry}, shown on the left of Figure~\ref{fig:fold}. While the loop runs, $b=1$ and $n$ takes the values $0,1,2,\ldots$. One iteration moves the state $(1,h)$ either to $(1,h+1)$ on outcome $1$, applying the quantum operation $\mathcal K(\rho)=M_1\rho M_1^\dagger$ to the qubit, or to $(0,h+1)$ on outcome $0$, applying $\mathcal J(\rho)=M_0\rho M_0^\dagger$ and leaving the loop.  Although the state space is infinite, the transition structure is the same at every counter value \(h\), up to a shift. This follows from two properties.
\begin{itemize}
\item The counter $n$ affects neither control flow nor quantum operations: the guard does not read it, and the body only increments it. Starting from $n=h$ therefore shifts the final counter value by $h$ relative to starting from $n=0$. In the output OFPS, this shift corresponds to multiplication by $N^h$.
\item The flag $b$ ranges over a finite set of values while the loop is active; here, it is always $1$.
\end{itemize}
We can therefore fold the picture into the finite graph on the right of Figure~\ref{fig:fold}. It has one node for the running loop, a self-loop labeled $N\mathcal K$ for outcome $1$, and an exit labeled $N\mathcal J$ for outcome $0$. The indeterminate $N$ records that each arrow increases $n$ by one. Reading the folded graph as an equation, the output is the exit applied after any number of self-loops, which is exactly~\eqref{eq:retry-equation}.

This is the idea behind the homogeneous-bounded loops of \citet{Klinkenberg2021}. The variables that stay bounded while the loop runs, like $b$, form a finite control. The other variables, like $n$, only accumulate, as rewards in a Markov chain do. Two things are new in the quantum setting. The arrows carry quantum operations instead of probabilities, so the folded equations have matrices as unknowns. And the conditions that permit folding must hold for the quantum output, not only for its probabilities.

\FloatBarrier
\subsection{Homogeneous-Bounded Loops}\label{sec:hb-def}
\begin{definition}[Homogeneous variable]\label{def:homogeneous}
A classical variable $x_i$ is \emph{homogeneous} for a program $S$ if $\sem S(X_iF)=X_i\sem S(F)$ for every $F\in\OFPS{\Hil}{\bX}$.
\end{definition}

Homogeneity says that incrementing $x_i$ before running $S$ has the same effect as incrementing it afterward, for the joint classical-quantum output. Requiring this of the full output rather than of probabilities matters. Suppose the retry body additionally applied the phase flip $Z=\mathrm{diag}(1,-1)$ to the qubit whenever $n$ is even. A phase flip negates the coherence of the qubit and leaves its populations, and hence all probabilities, unchanged. A probabilistic analysis would therefore still treat $n$ as a counter. The returned state, however, now depends on the parity of $n$, so starting the loop at $n=1$ instead of $n=0$ changes the output, and $n$ is not homogeneous. If $x_i$ is homogeneous for the body and the guard is independent of $x_i$, then $x_i$ is also homogeneous for the whole loop. This homogeneity property justifies the folding construction.

We partition the $k=|\cvar|$ classical variables into $\ell$ \emph{retained} and $d$ \emph{homogeneous} variables, with $\ell+d=k$. We write classical states as $(s,\bh)\in\N^\ell\times\N^d$, with corresponding indeterminates $\bZ$ and $\bY$.

\begin{definition}[CQ homogeneous-bounded loop]\label{def:hb}
A loop $W=\kw{while}\ e\ \kw{do}\ S\ \kw{od}$ is \emph{classical-quantum homogeneous-bounded} (CQ-HB) for a given partition of $\cvar$ if
\begin{enumerate}
\item every homogeneous variable $x_i$ is homogeneous for $S$, and $e$ does not depend on it, that is, $\sigma\models e$ iff $\sigma[v/x_i]\models e$ for all $\sigma\in\Sigma$ and $v\in\N$;
\item the set of \emph{active} retained values $\Act:=\{s\in\N^\ell:(s,\zero)\models e\}$ is finite;
\item for every $s\in\Act$ there is a finite set $\Omega_s\subseteq\Sigma$ such that $\csupp\sem S(\rho\,\bZ^s\bY^\zero)\subseteq\Omega_s$ for every partial density operator $\rho$.
\end{enumerate}
\end{definition}

Condition~(1) allows the counters to be factored out; because the guard does not depend on them, $(s,\bh)\models e$ iff $s\in\Act$. Condition~(2) says that every retained variable is bounded by $e$, which makes the control finite. Condition~(3) says that one iteration from an active value can end in only finitely many classical states, whatever the quantum input, which gives each node of the folded graph finitely many arrows. Condition~(3) must hold uniformly over a continuum of quantum inputs. It suffices, however, to check a single input, the completely random state $I/D$. Every state $\rho$ lies below $D\cdot(I/D)$ in the L\"owner order, and kernels preserve that order, so every output that is possible from some state is possible from $I/D$.

For example, the retry loop is CQ-HB with $b$ retained and $n$ homogeneous: the body changes $n$ only by $n:=n+1$ and the guard does not mention it, the only active value is $b=1$, and one iteration from $(1,0)$ ends in $(0,1)$ or $(1,1)$.

Homogeneity is a semantic property, undecidable in general, but a simple syntactic condition implies it. If the body is loop-free with finitely many sampling and measurement outcomes, each homogeneous variable $h$ occurs only in increments $h:=h+e$ where $e$ reads only retained variables, and no guard or other expression reads a homogeneous variable, then conditions~(1) and~(3) hold. All case studies of this paper satisfy this condition, with the counter $n$ homogeneous.

\subsection{Loop Summaries}\label{sec:summaries}
The folded graph of Figure~\ref{fig:fold} is an instance of a simple structure, which we call a \emph{loop summary}. It is a finite graph whose edges carry quantum operations and monomials, the quantum counterpart of the Markov reward chains behind homogeneous-bounded probabilistic loops~\citep{Klinkenberg2021}. Throughout, $\bY=(Y_1,\ldots,Y_d)$ are indeterminates, and $\mathrm{QO}(\Hil)$ is the set of quantum operations on $\Lin(\Hil)$ (Section~\ref{sec:quantum-ops}).

In the summary constructed from a CQ-HB loop, locations represent active valuations of the retained variables, while exits represent retained valuations at which the guard is false. The homogeneous values are recorded by monomials in $\bY$.

\begin{definition}[Loop summary]\label{def:summary}
A \emph{loop summary} is a triple $(\Lambda,\Ex,{\to})$ in which $\Lambda$ is a finite set of \emph{locations}, $\Ex$ is a finite set of \emph{exits} disjoint from $\Lambda$, and
\[
{\to}\ \subseteq\ \Lambda\times\N^d\times\mathrm{QO}(\Hil)\times(\Lambda\cup\Ex)
\]
is a finite set of \emph{transitions}. We write $s\xrightarrow{\br,\Qop}t$ for $(s,\br,\Qop,t)\in{\to}$; the transition leads from $s$ to $t$, increments the homogeneous variables by $\br$, and applies $\Qop$. It is \emph{continuing} if $t\in\Lambda$ and \emph{exiting} if $t\in\Ex$. For every $s\in\Lambda$ and every $A\succeq0$,
\begin{equation}\label{eq:summary-mass}
 \textstyle\sum_{s\xrightarrow{\br,\Qop}t}\tr\Qop(A)\le\tr A ,
\end{equation}
where the sum ranges over all transitions leaving $s$.
\end{definition}

A transition $s\xrightarrow{\br,\Qop}t$ sends the unnormalized quantum state $A_s$ at $s$ to $\Qop(A_s)$ at $t$ and records the homogeneous increment by the factor $\bY^\br$. Contributions reaching the same location with the same homogeneous values are added. We therefore represent the quantum state and homogeneous values by an operator-valued polynomial or series at each location. Condition~\eqref{eq:summary-mass} ensures that the total trace over the outgoing transitions does not exceed the input trace. The following definitions express these updates as linear maps and collect the contributions of all finite runs.

\begin{definition}[Continuation and exit maps]\label{def:summary-maps}
Let $(\Lambda,\Ex,{\to})$ be a loop summary, and let $\Vsp=\bigoplus_{s\in\Lambda}\Lin(\Hil)$ be the space of families $A=(A_s)_{s\in\Lambda}$ of operators. The \emph{continuation} $\Cmap(\bY):\Vsp\to\Vsp[\bY]$ and, for $u\in\Ex$, the \emph{exit map} $\Emap_u(\bY):\Vsp\to\Lin(\Hil)[\bY]$ are
\[
 \bigl(\Cmap(\bY)A\bigr)_t=\sum_{s\xrightarrow{\br,\Qop}t}\bY^\br\,\Qop(A_s)\quad(t\in\Lambda),
 \qquad
 \Emap_u(\bY)A=\sum_{s\xrightarrow{\br,\Qop}u}\bY^\br\,\Qop(A_s),
\]
extended to families of polynomials and of series coefficientwise.
\end{definition}

The map $\Cmap(\bY)$ collects the continuing transitions and $\Emap_u(\bY)$ the transitions into $u$.

\begin{definition}[Input to a loop summary]\label{def:summary-input}
An \emph{input to the loop summary} is a pair $(\iota,\delta)$ of families $\iota=(\iota_s)_{s\in\Lambda}\in\Vsp[\bY]$ and $\delta=(\delta_u)_{u\in\Ex}$ with $\iota_s,\delta_u\in\Pos(\Hil)[\bY]$, such that $\sum_{s}\tr\iota_s(\one)+\sum_u\tr\delta_u(\one)\le1$.
\end{definition}

For a CQ-HB loop, $\iota$ collects the input satisfying the guard at loop entry, while $\delta_u$ collects the input already at exit $u$, where the guard is false. The latter is returned without executing the loop body.

\begin{remark}[Retained values as indices]
For a CQ-HB loop, the family $\iota=(\iota_s(\bY))_{s\in\Lambda}$ encodes the same active input as the OFPS $\sum_{s\in\Lambda}\bZ^s\iota_s(\bY)$. Since $\Lambda$ is finite, using retained values as component indices gives the continuation map a finite block matrix representation. The monomials in $\bY$ record initial homogeneous values and subsequent increments. The input is polynomial because its classical support is finite.
\end{remark}

\begin{definition}[Live and terminal series]\label{def:live-terminal}
For an input $(\iota,\delta)$ and a fresh indeterminate $T$, the \emph{graded live series} $\widehat L(T)$ and the \emph{terminal series} $F_u$, for $u\in\Ex$, are
\begin{equation}\label{eq:terminal-series}
 \widehat L(T)=\sum_{m\ge0}T^m\Cmap(\bY)^m\,\iota,
 \qquad
 F_u=\delta_u+\sum_{m\ge0}\Emap_u(\bY)\,\Cmap(\bY)^m\,\iota .
\end{equation}
We define the \emph{ungraded live series} $L=(L_s)_{s\in\Lambda}:=\sum_{m\ge0}\Cmap(\bY)^m\iota\in\Vsp[[\bY]]$ only when every coefficient sum converges in trace norm.
\end{definition}

The series $\widehat L(T)$ always belongs to $\Vsp[\bY][[T]]$, since each coefficient of $T^m$ is a polynomial in $\bY$. The terminal sums converge coefficientwise in trace norm by~\eqref{eq:summary-mass}.

The indeterminate $T$ records the number of continuing transitions: $[T^m]\widehat L(T)$ collects the contributions arriving at the locations after exactly $m$ continuing transitions. When $L$ exists, its component $L_s$ collects everything that \emph{ever} arrives at $s$. The $m$-th summand in the terminal sum is what leaves through $u$ after $m$ continuing transitions, so $F_u$ collects everything that \emph{eventually} leaves through $u$. The terminal series are the outputs, and each $F_u$ is an admissible OFPS in $\bY$. The ungraded live series behaves differently, because it counts repeated visits. Whenever $L$ exists, it satisfies
\[
 L=\iota+\Cmap(\bY)L,\qquad F_u=\delta_u+\Emap_u(\bY)L .
\]
What arrives is the input plus what one more transition sends back, and what leaves is the direct input plus one exit step applied to what arrives. If every continuing transition increments some homogeneous variable, each coefficient of $L$ is a finite sum, so $L$ exists. Without coefficient convergence, $L$ is undefined: for $\Cmap=\id$ and constant input $\iota=\rho\succeq0$, $\rho\ne0$, its putative constant coefficient is $\sum_{m\ge0}\rho$, and $L=\rho+L$ has no solution. Section~\ref{sec:rationality} computes the terminal series using $\widehat L(T)$ even in such cases.

\begin{example}[The summary of the retry loop]\label{ex:retry-summary}
With $\bY=(N)$, the summary on the right of Figure~\ref{fig:fold} has $\Lambda=\{1\}$, $\Ex=\{0\}$, and the two transitions
\[
 1\xrightarrow{1,\mathcal K}1\ \ \text{(continue)},\qquad 1\xrightarrow{1,\mathcal J}0\ \ \text{(exit)},
 \qquad\text{where }\mathcal K(\rho)=M_1\rho M_1^\dagger,\ \ \mathcal J(\rho)=M_0\rho M_0^\dagger .
\]
Thus $\Cmap(N)=N\mathcal K$ and $\Emap_0(N)=N\mathcal J$, and~\eqref{eq:summary-mass} follows from $M_1^\dagger M_1+M_0^\dagger M_0=I$. With $\iota=\rho$ and $\delta_0=0$,
\[
 L=\rho+N\mathcal K(L),\qquad F_0=N\mathcal J(L).
\]
Since $M_0$ and $M_1$ are diagonal, these equations decouple entrywise and yield $F_0=F_\rho$ from~\eqref{eq:retry-solution}.
\end{example}

Every CQ-HB loop folds into a summary. Its locations are the active retained values, and its transitions are the kernels of one iteration of the body.

\begin{definition}[Summary of a CQ-HB loop]\label{def:folded}
Let $W=\kw{while}\ e\ \kw{do}\ S\ \kw{od}$ be CQ-HB, and let $F_{\mathrm{in}}=\sum_{t}\bZ^t R_t(\bY)\in\OFPS{\Hil}{\bX}$ be finitely supported, with $R_t\in\Pos(\Hil)[\bY]$. The \emph{summary of $W$ for $F_{\mathrm{in}}$} consists of the data
\[
 (\Lambda,\Ex,{\to};\iota,\delta),
\]
where the first three components specify the graph and the last two its input:
\begin{itemize}
\item The \emph{locations} are $\Lambda=\Act$, the active retained values.
\item The \emph{exits} $\Ex$ are those retained values $u\notin\Act$ for which either a transition defined below enters $u$ or $R_u\ne0$.
\item The \emph{transition relation} ${\to}$ consists of exactly the edges
\[
 s\xrightarrow{\br,\Qop}t\qquad\text{for }s\in\Act,\ (t,\br)\in\Omega_s,\ \Qop=\kappa^S_{(t,\br),(s,\zero)}\ne0.
\]
\item The \emph{running input} is $\iota=(R_s)_{s\in\Lambda}$, the part initially at the locations.
\item The \emph{direct exit input} is $\delta=(R_u)_{u\in\Ex}$, the part that exits without executing the body.
\end{itemize}
The continuation and exit maps and the live and terminal series are then given by Definitions~\ref{def:summary-maps} and~\ref{def:live-terminal}.
\end{definition}

The transition $s\xrightarrow{\br,\Qop}t$ summarizes all terminating executions of $S$ from $(s,\zero)$ to $(t,\br)$: for quantum input $\rho$, their combined quantum output is $\Qop(\rho)$. By homogeneity, the same transition applies from $(s,\bh)$, with destination $(t,\bh+\br)$.

Conditions~(2)--(3) of Definition~\ref{def:hb} give finitely many source locations and finitely many outgoing transitions per source. The exit set is finite because each exit is reached by a transition or occurs in the finitely supported input. Condition~\eqref{eq:summary-mass} follows from the bound on the sum of traces over all outputs of one body execution (Proposition~\ref{prop:kernels}(2)). Since the guard depends only on retained values, splitting the input between $\Act$ and its complement yields $\iota$ and $\delta$, respectively, with unchanged total mass. Together with loop unfolding, these observations yield the following soundness result; the full proof is given in Appendix~\ref{app:path-expansion}.

\begin{proposition}[Soundness of summaries]\label{prop:summary}
For every CQ-HB loop $W$ and every finitely supported $F_{\mathrm{in}}\in\OFPS{\Hil}{\bX}$, the summary of $W$ for $F_{\mathrm{in}}$ is a loop summary with an input, and
\[
 \sem W(F_{\mathrm{in}})=\sum_{u\in\Ex}\bZ^u\,F_u ,
\]
where $F_u$ are its terminal series.
\end{proposition}

\paragraph{Further examples.}
Proposition~\ref{prop:summary} reduces loop analysis to computing the terminal series of a summary. We first give two further examples; Section~\ref{sec:rationality} presents the general solution method.

Summaries with several locations arise when a loop has a finite classical mode: each value of the mode becomes a location. For instance, repeating the retry measurement until outcome $0$ occurs twice in a row needs a variable $c$ that counts consecutive occurrences of that outcome. Its summary has the locations $c=0$ and $c=1$, the exit $c=2$, and four transitions,
\[
 0\xrightarrow{1,\mathcal K}0,\qquad 0\xrightarrow{1,\mathcal J}1,\qquad 1\xrightarrow{1,\mathcal K}0,\qquad 1\xrightarrow{1,\mathcal J}2 .
\]
Outcome $0$ moves $c=0$ to $c=1$ and exits from $c=1$, whereas outcome $1$ resets $c$ to $0$. The live series is then a pair of matrices, one for each location, and the equations of the summary couple them.

The next example illustrates how omitting a correction step can change the quantum output while leaving the distribution of iteration counts unchanged.

\begin{example}[A repeat-until-success circuit, correct and buggy]\label{ex:rus}
A repeat-until-success (RUS) circuit implements a target gate $V$ on a qubit $q$ with the help of an ancilla $a$~\citep{Paetznick2014,Bocharov2015}. For $0<p\le1$, a gadget unitary $G$ on $q$ and $a$ satisfies $G\ket\psi\ket0=\sqrt p\,V\ket\psi\ket0+\sqrt{1-p}\,W\ket\psi\ket1$ for every $\ket\psi$. Measuring $a$ therefore yields $0$ with probability $p$ and leaves $V\ket\psi$, or yields $1$ and leaves $W\ket\psi$, where $W$ is a gate that is easy to undo. The correct version in~\eqref{prog:rus} undoes $W$ after failure; the buggy version omits this correction. The boxed commands mark the difference.
\begin{equation}\label{prog:rus}
\begin{array}{@{}l@{\qquad}l@{}}
 \multicolumn{1}{c}{\text{\bfseries Correct}} &
 \multicolumn{1}{c}{\text{\bfseries Buggy}} \\[3pt]
 \kw{while}\ b=1\ \kw{do} & \kw{while}\ b=1\ \kw{do}\\
 \quad n:=n+1;\ a:=\ket0; & \quad n:=n+1;\ a:=\ket0;\\
 \quad q,a:=G[q,a]; & \quad q,a:=G[q,a];\\
 \quad b\leftarrow\kw{meas}\ M_{\mathrm{comp}}[a]; &
 \quad b\leftarrow\kw{meas}\ M_{\mathrm{comp}}[a];\\
 \quad\kw{if}\ b=1\ \kw{then} & \quad\kw{if}\ b=1\ \kw{then}\\
 \qquad\boxed{q:=W^\dagger[q]} & \qquad\boxed{\kw{skip}}\\
 \quad\kw{else}\ \kw{skip}\ \kw{fi} & \quad\kw{else}\ \kw{skip}\ \kw{fi}\\
 \kw{od} & \kw{od}.
\end{array}
\end{equation}
The ancilla is reset at the start of every round, so it suffices to track $q$. Let $\rho$ be its normalized input state. The summary has $\Lambda=\{1\}$, $\Ex=\{0\}$, and the transitions $1\xrightarrow{1,\Qop_{\mathrm f}}1$ and $1\xrightarrow{1,\Qop_{\mathrm s}}0$, where $\Qop_{\mathrm f}(\rho)=(1-p)\rho$ (fail, undo $W$, try again) and $\Qop_{\mathrm s}(\rho)=pV\rho V^\dagger$ (success). The equation $L=\rho+N(1-p)L$ gives the output
\[
 \frac{pN}{1-(1-p)N}\;V\rho V^\dagger :
\]
a geometric number of rounds with mean $1/p$, and the target state $V\rho V^\dagger$ regardless of the count. In the buggy version, the correction is replaced by $\kw{skip}$. The failure transition then applies $\rho\mapsto(1-p)\,W\rho W^\dagger$, and the state returned after $m$ rounds becomes $VW^{m-1}\rho W^{\dagger(m-1)}V^\dagger$. For the common failure gate $W=Z$, the phase flip of Section~\ref{sec:hb-def}, $Z^{m-1}$ is the identity for odd $m$ and $Z$ for even $m$, and the output splits into
\[
 \frac{pN}{1-(1-p)^2N^2}\;V\rho V^\dagger\;+\;\frac{p(1-p)N^2}{1-(1-p)^2N^2}\;VZ\rho ZV^\dagger .
\]
With probability $(1-p)/(2-p)$ ($1/3$ for $p=\frac12$), the buggy circuit returns a phase-flipped state. Both versions have the same geometric round count, so no analysis of the counter alone can distinguish them. The OFPS exposes the dependence on count parity. At $p=0$, both circuits diverge and their formal terminal series are zero.
\end{example}

\subsection{Main Result: Rational Outputs}\label{sec:rationality}
Our main result gives a finite construction of a rational representation of the loop's joint classical-quantum output, without assuming almost-sure termination. The explicit matrix construction follows the theorem.

\begin{theorem}[Rational outputs and exact computation]\label{thm:main-rational}
Let $W$ be a CQ-HB loop, let $F_{\mathrm{in}}\in\OFPS{\Hil}{\bX}$ be finitely supported, and let $(\Lambda,\Ex,{\to};\iota,\delta)$ be the summary of $W$ for $F_{\mathrm{in}}$.
\begin{enumerate}
\item \emph{Finite output representation.} The joint output $\sem W(F_{\mathrm{in}})$ is a rational OFPS, represented by finitely many rational functions. Each terminal series $F_u$ is rational in $\bY$. No almost-sure termination assumption is required.
\item \emph{Exact construction.} Given a summary and input with algebraic matrix entries, Algorithm~\ref{alg:summary-output} computes rational representations of all $F_u$, and hence of the joint output, by solving a linear system with $|\Lambda|D^2$ unknowns, where $D=\dim\Hil$ is the dimension of the quantum memory.
\end{enumerate}
\end{theorem}

\paragraph{Matrix notation.}
Fix an orthonormal basis $\ket0,\ldots,\ket{D-1}$ of $\Hil$ and write $E_{ij}=\ket i\bra j$. Define $\vecop(E_{ij}):=\ket i\otimes\ket j$, $\Choi(\Qop):=\sum_{i,j=0}^{D-1}E_{ij}\otimes\Qop(E_{ij})$, and $\mathfrak R(E_{ij}\otimes E_{ab}):=E_{ai}\otimes E_{bj}$, extending $\vecop$ and $\mathfrak R$ linearly. Then $\vecop(\Qop(A))=\mathfrak R(\Choi(\Qop))\vecop(A)$.

For a location family, $\vecop$ stacks its component vectors in a fixed order of $\Lambda$; $\vecop^{-1}$ reshapes an exit vector into a $D\times D$ matrix. In the algorithm, $\C(T,\bY)$ is the field of rational functions in $T,\bY$, $\boldsymbol\ell=\vecop(\widehat L(T))$, and $\operatorname{cancel}$ cancels common polynomial factors between the numerator and denominator of each entry.
\begin{algorithm}[H]
\caption{Exact output from a CQ-HB summary}\label{alg:summary-output}
\small
\begin{algorithmic}[1]
\Statex \textbf{Input:} Summary $(\Lambda,\Ex,{\to};\iota,\delta)$ with algebraic data
\Statex \textbf{Output:} A rational representation of $\sem W(F_{\mathrm{in}})$
\State $\mathbf M_{t,s}\gets\sum_{s\xrightarrow{\br,\Qop}t}\bY^\br\,\mathfrak R(\Choi(\Qop))$ \quad $(s,t\in\Lambda)$
\State $(\mathbf B_u)_s\gets\sum_{s\xrightarrow{\br,\Qop}u}\bY^\br\,\mathfrak R(\Choi(\Qop))$ \quad $(s\in\Lambda,\ u\in\Ex)$
\State Solve $(I-T\mathbf M)\boldsymbol\ell=\vecop(\iota)$ over $\C(T,\bY)$
\For{$u\in\Ex$}
\State $\mathbf h_u\gets\operatorname{cancel}\bigl(\vecop(\delta_u)+\mathbf B_u\boldsymbol\ell\bigr)$
\State $F_u\gets\vecop^{-1}\bigl(\mathbf h_u\vert_{T=1}\bigr)$
\EndFor
\State \Return $\sum_{u\in\Ex}\bZ^uF_u$
\end{algorithmic}
\end{algorithm}

Algorithm~\ref{alg:summary-output} introduces $T$ to keep the live equation solvable even when the loop does not terminate almost surely. The live series sums the quantum states at successive visits to summary locations, so a nonterminating component contributes repeatedly. If a counter increases on every continuing transition, these contributions occur at increasing powers of its indeterminate and each coefficient remains finite. Without such a counter, infinitely many contributions can accumulate in a single coefficient, and the ungraded equation $L=\iota+\Cmap(\bY)L$ may have no solution.

\begin{example}[A singular summary]\label{ex:singular}
Take the retry summary of Example~\ref{ex:retry-summary} without its counter ($d=0$), with $M_0=\proj0$ and $M_1=\proj1$. On input $\rho=\ket+\bra+$, the loop exits after the first iteration in state $\ket0$ with probability $\frac12$, and remains in $\ket1$ forever otherwise. Thus the output is $\frac12\proj0$. However, the live equation $L=\rho+M_1LM_1^\dagger$ gives $L_{11}=\frac12+L_{11}$, which has no finite solution.
\end{example}

The indeterminate $T$ acts as a \emph{ghost counter}: each continuing transition contributes a factor of $T$. The graded equation $\widehat L(T)=\iota+T\Cmap(\bY)\widehat L(T)$ has a unique solution over $\C(T,\bY)$, since $\det(I-T\mathbf M)$ has constant term $1$ in $T$. In Example~\ref{ex:singular}, this gives $\widehat L_{11}(T)=\frac12+T\widehat L_{11}(T)$ and hence $\widehat L_{11}(T)=\frac1{2(1-T)}$. The exit map discards this nonterminating component, leaving the output $\frac12\proj0$. Accordingly, the algorithm first forms the exit expressions $\delta_u+\Emap_u(\bY)\widehat L(T)$, cancels common polynomial factors in each entry, and then sets $T=1$. Only the exit expressions are specialized; the live series itself may have a pole at $T=1$.

This final substitution is valid because, for $0\le t<1$, setting $T=t$ weights each exiting run by $t^m$, where $m$ is its number of continuing transitions. The resulting terminal series has total mass at most one. This bound rules out a factor $T-1$ in any reduced exit denominator, so setting $T=1$ recovers the sum over all terminating runs (Appendix~\ref{app:T-specialization}). When every continuing transition increments a homogeneous variable, $\Cmap(\zero)=0$ and the ungraded live equation is already solvable, so $T$ can be omitted.

\paragraph{Further results.}
The rationality result extends to arbitrary loop summaries. Exact computation from source additionally assumes the syntactic condition of Section~\ref{sec:hb-def}, a given finite superset of $\Act$, evaluable expressions and guards, and algebraic program and input data. The appendices give details and extensions to nested loops and moments of homogeneous variables.

\subsection{Which Loops Are CQ-HB?}\label{sec:hb-scope}
CQ-HB loops are the quantum analogue of finite-state Markov chains with rewards. While the loop runs, they have finitely many classical control values, a fixed finite-dimensional quantum memory, and counters that only accumulate. Although there are continuously many quantum states, each is represented by a $D\times D$ matrix. The summary uses a matrix-valued generating series at each classical location, with the homogeneous variables recording the counters. Quantum operations act linearly on the matrix coefficients, so finitely many locations yield a finite linear system whose unknowns are scalar generating series, one per matrix entry at each location. No separate unknown is needed for each possible quantum state.

\paragraph{Loops in the class.}
A typical CQ-HB loop combines finite classical control with homogeneous counters that record iteration counts or resource use. The retry loop, repeat-until-success circuits (Example~\ref{ex:rus}), the Bernoulli-factory subroutine, and the absorbing walk (Section~\ref{sec:case-studies}) have this structure. Finite control can also record progress across iterations, as when a measurement is repeated until a designated outcome occurs twice in a row.

A counter read by the guard can instead be retained if the guard bounds its active values. Examples are a time-out, $\kw{while}\ b=1\wedge n<10$, and a readout between two fixed thresholds, $\kw{while}\ 0<x<2k$ for fixed $k$. In each case, the counter becomes part of the finite classical control. The set of active joint valuations of all retained variables must be finite, and the summary has one location per such valuation.

Thus the analysis chooses a partition satisfying Definition~\ref{def:hb}: homogeneous variables must satisfy the semantic shift property for the body and must not affect the guard, while retained variables must have finitely many active joint valuations.

\paragraph{Loops outside the class.}
Three kinds of loops fall outside the class.
\begin{itemize}
\item \emph{Unbounded classical control.} The guard reads a variable that can grow without bound, as in the one-sided readout of Section~\ref{sec:superinv-nonhb}. The output need not be rational. Superinvariants still bound it, and they can verify closed forms.
\item \emph{Unbounded non-homogeneous counters.} A counter that is not homogeneous for the body must be retained; it excludes CQ-HB only when the guard leaves its active values unbounded. For example, a retry loop whose guard depends only on a flag can double $n$ from $1$ on each attempt. With independent success probability $p\in(0,1)$, this gives the non-rational counter generating series $\sum_{m\ge1}p(1-p)^{m-1}N^{2^m}$.
\item \emph{Nested loops whose inner loop adds an unbounded amount to a counter.} One iteration of the outer loop then has infinitely many outcomes. Such loops can be analyzed with summaries that keep the locations of the inner loops.
\end{itemize}

\FloatBarrier
\section{Upper Bounds via Superinvariants}\label{sec:superinv}
The exact analysis of Section~\ref{sec:hb} requires the CQ-HB structure. This section develops upper bounds on loop outputs without that restriction, including loops with unbounded classical control. Section~\ref{sec:superinv-nonhb} illustrates the method on a sequential readout loop.

An OFPS upper bound provides bounds on several output properties at once: classical event probabilities, joint probabilities of classical events and quantum measurement outcomes, and submoments.

\subsection{The Rule}
We lift the superinvariant rule of \citet{Klinkenberg2021} to OFPSs. The rule retains the same form; its side condition now compares operator-valued outputs.

Fix $W=\kw{while}\ e\ \kw{do}\ S\ \kw{od}$. A candidate map $\chi$ assigns an output bound to every input. To be a \emph{superinvariant}, the candidate must dominate the sum of the output that exits immediately and the candidate applied after one body execution on the continuing part of the input.

\begin{definition}[Superinvariant]\label{def:superinvariant}
A \emph{superinvariant} of $W$ is a map $\chi:\OFPS{\Hil}{\bX}\to\Pos(\Hil)[[\bX]]$ such that, for every $F\in\OFPS{\Hil}{\bX}$,
\begin{equation}\label{eq:superinvariant}
 F|_{\neg e}+\chi\bigl(\sem S(F|_e)\bigr)\ \sqsubseteq\ \chi(F).
\end{equation}
\end{definition}

A superinvariant need not return an admissible OFPS: its total mass may exceed one, allowing more freedom in constructing an upper bound. Each coefficient must still be a finite positive semidefinite matrix.

\begin{theorem}[Superinvariant rule]\label{thm:superinvariant}
If $\chi$ is a superinvariant of $W$, then $\sem W(F)\sqsubseteq\chi(F)$ for every $F\in\OFPS{\Hil}{\bX}$.
\end{theorem}

\subsection{Checking a Superinvariant One Classical State at a Time}
Condition~\eqref{eq:superinvariant} quantifies over all inputs: infinitely many classical states and, at each classical state, a continuum of quantum states. In the probabilistic setting, linearity reduces it to inputs concentrated on a single classical state, one condition per classical state~\citep{Klinkenberg2021}. For OFPSs, linearity again reduces the condition to single classical states, but each classical state still carries all its quantum states. We therefore describe a candidate bound by \emph{bound maps}. The bound map $\mathcal V_{\tau,\sigma}$ says that if the loop starts in classical state $\sigma$ with state $\rho$, then at most $\mathcal V_{\tau,\sigma}(\rho)$ ends in classical state $\tau$. We require each bound map to send every input quantum state to a positive operator, so that its output is a valid coefficient of the candidate bound. Linearity allows us to assemble the candidate for a general input by summing the contributions from individual classical states.

\begin{definition}[Bound maps]\label{def:bound-maps}
A family of \emph{bound maps} is a family $\mathcal V=(\mathcal V_{\tau,\sigma})_{\sigma,\tau\in\Sigma}$ of positive linear maps on $\Lin(\Hil)$, indexed by a start $\sigma$ and an end $\tau$, such that $\sum_{\sigma}\mathcal V_{\tau,\sigma}(\rho_\sigma)$ converges for every $\tau\in\Sigma$ and every $F=\sum_\sigma \rho_\sigma\bX^\sigma\in\OFPS{\Hil}{\bX}$. Its \emph{candidate} is the map $\chi_{\mathcal V}:\OFPS{\Hil}{\bX}\to\Pos(\Hil)[[\bX]]$ with $[\bX^\tau]\chi_{\mathcal V}(F):=\sum_\sigma\mathcal V_{\tau,\sigma}(\rho_\sigma)$.
\end{definition}

\begin{definition}[Slack maps]\label{def:slack}
The \emph{slack map} of a family of bound maps from $\sigma$ to $\tau$ is
\begin{equation}\label{eq:residual}
 \Theta_{\tau,\sigma}=
 \begin{cases}
 \mathcal V_{\sigma,\sigma}-\id & \text{if }\sigma\not\models e\text{ and }\tau=\sigma\quad\text{(the input exits unchanged)},\\
 \mathcal V_{\tau,\sigma} & \text{if }\sigma\not\models e\text{ and }\tau\ne\sigma,\\
 \mathcal V_{\tau,\sigma}-\sum_{\mu}\mathcal V_{\tau,\mu}\circ\kappa^S_{\mu,\sigma} & \text{if }\sigma\models e\quad\text{(one iteration, then the bound)}.
 \end{cases}
\end{equation}
\end{definition}

For input $\rho\,\bX^\sigma$, the slack $\Theta_{\tau,\sigma}(\rho)$ is an operator difference: the candidate output at $\tau$ minus the output of one unfolding with the same candidate used for the remaining loop. The three cases in~\eqref{eq:residual} have the following meanings.
\begin{itemize}
\item If $\sigma\not\models e$ and $\tau=\sigma$, the loop exits immediately and returns $\rho$. The slack is therefore $\mathcal V_{\sigma,\sigma}(\rho)-\rho$.
\item If $\sigma\not\models e$ and $\tau\ne\sigma$, immediate exit contributes nothing at $\tau$. The slack is the candidate output $\mathcal V_{\tau,\sigma}(\rho)$ itself.
\item If $\sigma\models e$, one body execution produces $\kappa^S_{\mu,\sigma}(\rho)$ at each intermediate classical state $\mu$. Applying $\mathcal V_{\tau,\mu}$ to each contribution and summing over $\mu$ gives the candidate estimate for the remaining loop at $\tau$. The slack subtracts this sum from $\mathcal V_{\tau,\sigma}(\rho)$.
\end{itemize}
The second case is already positive by the definition of bound maps, so only the first and third require a positivity check.

\begin{theorem}[Positivity criterion for superinvariants]\label{thm:local-check}
Let $\mathcal V$ be a family of bound maps. Then $\chi_{\mathcal V}$ is a superinvariant of $W$ iff $\Theta_{\tau,\sigma}$ is a positive map for all $\sigma,\tau\in\Sigma$.
\end{theorem}
For input $\rho\,\bX^\sigma$, the difference between the right- and left-hand sides of~\eqref{eq:superinvariant} at coefficient $\bX^\tau$ is exactly $\Theta_{\tau,\sigma}(\rho)$. Thus checking~\eqref{eq:superinvariant} on these inputs is equivalent to checking positivity of every slack map. For a general input, the coefficientwise differences are sums of these positive operators, so the inequality still holds.

Theorem~\ref{thm:local-check} reduces the condition to positivity of linear maps, one per pair of classical states. When the candidate is described by finitely many symbolic families of maps, the slack maps form finitely many symbolic families. What remains is to show that a map is positive, that is, that it sends \emph{every} state to a positive operator. For $D\ge2$, no finite set of test states suffices; testing only the basis states $\ket0,\ket1,\ldots$, as a classical analysis would, can be misleading.

\begin{example}[Checking basis states is not enough]\label{ex:basis-not-enough}
Take the retry loop in \cref{prog:retry} with $M_0=\sqrt{0.8}\,I$ and $M_1=\sqrt{0.2}\,I$. The output terminating on the first attempt is $0.8\rho$. Consider bounding it by the dephased input $\mathcal D(\rho)=\mathrm{diag}(\rho_{00},\rho_{11})$, giving the slack $\Theta(\rho)=\mathcal D(\rho)-0.8\rho$.

For $\rho=\proj0$ or $\proj1$, the slack is $0.2\rho\succeq0$, so both basis-state tests pass. For $\rho=\proj+$, however, $\mathcal D(\rho)=I/2$ and
\[
 \bra+\Theta(\proj+)\ket+=0.5-0.8=-0.3.
\]
Thus the slack map is not positive, and the candidate fails on a coherent input despite passing the basis-state tests.
\end{example}

Positivity of a slack map can often be proved directly using operator inequalities. Alternatively, one can verify the stronger property of complete positivity using the Choi matrix defined in Section~\ref{sec:choi}. By Choi's theorem, a map is completely positive iff its Choi matrix is positive semidefinite~\citep{Choi1975,Watrous2018}. Thus, for a fixed slack map, checking that the Choi matrix has no negative eigenvalues certifies complete positivity and hence positivity.

The Choi test is sufficient but not necessary: a positive slack map may fail to be completely positive, so a failed Choi test does not by itself invalidate the candidate bound.\footnote{For maps on qubits, positivity itself also admits a semidefinite certificate~\citep{Stormer1963,Woronowicz1976}.} For a symbolic family of slack maps, the Choi condition must hold for every parameter value; checking a single member does not certify the whole family.

For example, a direct check shows that the retry loop of Section~\ref{sec:overview-retry} needs exactly $m$ attempts with probability at most $2(0.8)^{m-1}$, for every input state (Appendix~\ref{app:retry-cert}).

\subsection{Beyond Homogeneous-Bounded Loops}\label{sec:superinv-nonhb}
Superinvariants apply beyond CQ-HB loops, including loops with unbounded classical control. For probabilistic programs, \citet{Klinkenberg2021} use them to verify closed-form output expressions and prove failure of almost-sure termination. Here we analyze a quantum sequential readout loop with unbounded classical control: a superinvariant certifies a closed-form upper bound on its output and proves a positive probability of nontermination for some inputs.

Consider a readout procedure that repeatedly measures a qubit to collect evidence for $\ket1$. We use a noisy \emph{quantum non-demolition (QND) measurement}, which preserves the basis states $\ket0$ and $\ket1$ in each round. Repeated QND measurements are used in adaptive qubit readout~\citep{Hume2007}.

We use the measurement $M=\{M_0,M_1\}$ from \cref{prog:retry}. Outcome $0$ has probability $0.8$ on $\ket1$ and $0.2$ on $\ket0$, so we count it as a vote for $\ket1$; outcome $1$ counts as a vote for $\ket0$. The variable $x$ records how many more net votes for $\ket1$ are needed. A vote for $\ket1$ decreases $x$ by one, and a vote for $\ket0$ increases it by one. The counter $n$ records the number of measurement rounds. The loop stops when $x$ reaches zero:\footnote{This is a one-sided version of sequential hypothesis testing~\citep{Wald1945}. The stopping time is a first-passage time, as in analyses of adaptive readout~\citep{DAnjou2016}.}
\begin{equation}\label{prog:backlog}
\begin{aligned}
 \mathrm{Readout}\equiv{}&\kw{while}\ x>0\ \kw{do}\\
 &\quad n:=n+1;\ b\leftarrow\kw{meas}\ M[q];\\
 &\quad\kw{if}\ b=0\ \kw{then}\ x:=\max\{x-1,0\}\ \kw{else}\ x:=x+1\ \kw{fi}\\
 &\kw{od}.
\end{aligned}
\end{equation}

\begin{example}[One-sided sequential readout]\label{ex:backlog}
The program~\eqref{prog:backlog} is a \emph{one-sided} test: it stops only to declare $\ket1$. Such a test is useful when only one answer needs confirmation, such as heralding that a qubit has been prepared in $\ket1$. On $\ket0$ it may run forever. The guard reads the unbounded variable $x$, so the loop is not homogeneous-bounded, and its output is not rational.

\emph{The bound.} Because $M_0$ and $M_1$ are diagonal, each matrix entry of the qubit evolves on its own, with a weight $j$ per vote for $\ket1$ and $k$ per vote for $\ket0$: $(j,k)=(0.2,0.8)$ for the population of $\ket0$, $(0.8,0.2)$ for that of $\ket1$, and $(j,k)=(0.4,0.4)$ for the coherence. For one entry, gaining one net vote is a first-passage problem of a weighted random walk, whose formal generating series is $\lambda(N)=\bigl(1-\sqrt{1-4jkN^2}\bigr)/(2kN)$. The candidate multiplies each entry of the input state by $\lambda(N)^x$, with $\lambda$ chosen for that entry, and places the result at $x=0$. All its slack maps vanish, so it is a superinvariant.

\emph{What it shows.} The bounds are in fact attained, so they describe the exact behavior of the test. At $N=1$ the three multipliers are $\frac14$, $1$, and $\frac12$, so for a required margin $x$:
\begin{itemize}
\item On $\ket1$, the test halts almost surely, after $5x/3$ rounds on average. The mean is obtained by differentiating at $N=1$.
\item On $\ket0$, it halts, wrongly declaring $\ket1$, with probability $4^{-x}$. The false-alarm rate thus decays exponentially in the margin while the readout time grows only linearly, the trade-off that makes adaptive readout attractive~\citep{Hume2007}. With probability $1-4^{-x}$ the program never halts. The upper bound alone already proves that it diverges with at least this probability.
\item On $\ket+$ with $x=1$, the output is $\frac18\proj0+\frac12\proj1$ plus coherences $\frac14$. The test halts with probability $0.625$, and when it halts the qubit is \emph{not} collapsed to $\ket1$. It is left in the pure superposition $\sqrt{0.2}\ket0+\sqrt{0.8}\ket1$, which is what a subsequent operation in a dynamic circuit receives.
\end{itemize}
A probability-only analysis recovers the halting probabilities and readout times, but not the post-readout state. The distribution of the readout time is given by an algebraic, non-rational generating function.
\end{example}

\FloatBarrier
\section{Case Studies}\label{sec:case-studies}
We revisit the Bernoulli-factory subroutine and absorbing quantum walk studied by \citet{Liu2025} to show what a joint-output analysis adds to expected-runtime calculations. For the almost-surely terminating instances below, we recover their expected runtimes under the same cost model. The Bernoulli-factory example extracts the complete round-count distribution and higher cost moments, and establishes that the returned state is independent of the number of rounds. The walk example distinguishes distributions with the same mean and variance and exposes coherence between the absorbing boundaries. For each input, these properties are read from the same rational OFPS.

\subsection{A Bernoulli-Factory Subroutine}\label{sec:qbf}
A classical \emph{Bernoulli factory} uses independent flips of a coin with unknown bias $p$ to produce a coin with a prescribed bias $f(p)$. The quantum version receives copies of a \emph{quantum coin} $\ket p=\sqrt p\ket0+\sqrt{1-p}\ket1$. \citet{Dale2015} showed that quantum Bernoulli factories can realize strictly more functions $f$ than classical factories~\citep[see also][]{Patel2019}.

We analyze the state-preparation subroutine studied by \citet{Liu2025}. Its goal is to prepare $\ket{\phi_p}=(2p-1)\ket0+2\sqrt{p(1-p)}\ket1$: measuring this state in the computational basis returns $1$ with probability $4p(1-p)$. Each attempt uses two fresh copies of $\ket p$ and either returns the target state or reports failure. After failure, both qubits are reset and the procedure tries again. In \cref{prog:qbf}, we add a counter $n$ to record the number of attempts alongside the returned quantum state.

The fixed two-qubit unitary $U$, independent of $p$, combines the input coins according to
\begin{equation}\label{eq:qbf-action}
 U\ket p\ket p=\tfrac1{\sqrt2}\bigl(\ket{\phi_p}\ket0+\ket0\ket1\bigr).
\end{equation}
The second qubit flags success (outcome $0$) or failure (outcome $1$). Appendix~\ref{app:qbf} gives the full matrix of $U$.

\begin{equation}\label{prog:qbf}
\begin{aligned}
 \mathrm{QBF}\equiv{}&n:=0;\ q_1:=\ket0;\ q_1:=\NOT[q_1];\ q_2:=\ket0;\ q_2:=\NOT[q_2];\\
 &b\leftarrow\kw{meas}\ M_{\mathrm{comp}}[q_2];\\
 &\kw{while}\ b=1\ \kw{do}\\
 &\quad n:=n+1;\\
 &\quad q_1:=\ket0;\ q_1:=V_p[q_1];\\
 &\quad q_2:=\ket0;\ q_2:=V_p[q_2];\\
 &\quad q_1,q_2:=U[q_1,q_2];\\
 &\quad b\leftarrow\kw{meas}\ M_{\mathrm{comp}}[q_2]\\
 &\kw{od}.
\end{aligned}
\end{equation}

The initialization resets both qubits to $\ket0$ and flips them with $\NOT$, producing $\ket{11}$. Measuring the second qubit with $M_{\mathrm{comp}}$ in the basis $\ket0,\ket1$ therefore sets $b=1$ and enters the loop. Each iteration resets both qubits and applies $V_p$ to each, preparing the fresh input $\ket p\ket p$.
By~\eqref{eq:qbf-action}, measuring the second qubit after $U$ yields either outcome with probability $1/2$. Outcome $0$ sets $b=0$, exits the loop, and returns the pure two-qubit state $\ket{\phi_p}\ket0$. Outcome $1$ sets $b=1$, so the loop starts another attempt.

\paragraph{The summary and the output.}
The loop in~\cref{prog:qbf} is CQ-HB with $b$ retained and $n$ homogeneous. The guard allows only the active retained value $b=1$. The body increments $n$ by one, and no guard or quantum operation depends on $n$. From $(b,n)=(1,0)$, one iteration can end only in $(1,1)$ or $(0,1)$, for every quantum input. Thus the summary has one location, $b=1$, and one exit, $b=0$; both transitions increment the counter and carry a factor $N$.

Because each attempt resets both qubits, its branch maps depend on the input only through its trace. The weighted continuation and exit maps are
\[
 \Cmap(N)(\rho)=\frac N2\tr(\rho)\,\proj{01},
 \qquad
 \Emap_0(N)(\rho)=\frac N2\tr(\rho)\,\bigl(\proj{\phi_p}\otimes\proj0\bigr).
\]
To sum over repeated failed attempts, let $\ell(N)=\tr L(N)$ be the trace of the live series at the single location. For normalized input, initialization contributes unit mass, and each continuation contributes a factor $N/2$, so $\ell(N)=1+(N/2)\ell(N)$ and hence $\ell(N)=2/(2-N)$. The exit map then gives
\[
 F(N)=\frac{N}{2-N}\,\bigl(\proj{\phi_p}\otimes\proj0\bigr)
 =\sum_{m\ge1}2^{-m}\bigl(\proj{\phi_p}\otimes\proj0\bigr)\,N^m .
\]

\paragraph{What the output tells us.}
The loop terminates almost surely with geometric round count $\Prb(n=m)=2^{-m}$ for $m\ge1$. Under the cost model $\ops=5+6n$, we recover the mean runtime $17$ of \citet{Liu2025} and also obtain cost variance $72$ and higher moments (Appendix~\ref{app:qbf}).

The returned state $\ket{\phi_p}\ket0$ is independent of the round count, unlike the count-dependent retry output in Section~\ref{sec:overview-retry}. Thus a limit of $m$ rounds discards probability $2^{-m}$ while preserving the state of every successful output.

\subsection{Absorbing Quantum Walks}\label{sec:walk}
Quantum walks support algorithms for element distinctness~\citep{Ambainis2007} and oracle problems with exponential speedups~\citep{Childs2003}; their hitting times can also be exponentially shorter than those of classical walks~\citep{Kempe2005}. We consider a walk with absorbing boundaries on a finite line segment, a quantum version of gambler's ruin: it stops when a measurement detects the walker at either end~\citep{Ambainis2001,Bach2004,Kuklinski2018}.

\citet{Liu2025} analyzed the expected hitting time of this program. We use its joint output to show how two walks with the same mean and variance can have different hitting-time distributions, and how the quantum state returned by the Hadamard walk depends on when it reaches an endpoint.

\subsubsection{The Program and its Summary}\label{sec:walk-program}
The walker uses two quantum registers~\citep{Aharonov2001,Ambainis2001}. The \emph{coin} has basis states $\ket L,\ket R$, representing left and right directions; the \emph{position} register has basis states $\ket0,\ldots,\ket a$, for $a\ge2$. We abbreviate $\ket L\otimes\ket x$ as $\ket{L,x}$, and similarly for $R$. The walker may be in a superposition of these joint states.

Each step first applies a unitary $U_c$ to the coin register. The program allows any coin unitary. Our examples use the Hadamard gate $U_c=\frac{1}{\sqrt2}\bigl(\begin{smallmatrix}1&1\\1&-1\end{smallmatrix}\bigr)$ in the basis $(\ket L,\ket R)$. The shift $V$ then moves the position one step in the direction indicated by the coin, so that
\[
 V\ket{L,x}=\ket{L,x-1},\qquad V\ket{R,x}=\ket{R,x+1}.
\]
Position indices in this formula are taken modulo $a+1$.\footnote{The cyclic shift permutes the basis states, making $V$ unitary. The boundary test stops the walk before any step can wrap around.}

Initially and after each step, the program applies the boundary measurement $M_\partial=\{P_\partial,P_{\mathrm{int}}\}$, where
\[
 P_\partial=I_{\mathrm{coin}}\otimes(\proj0+\proj a),
 \qquad P_{\mathrm{int}}=I-P_\partial.
\]
Here $I_{\mathrm{coin}}$ acts on the coin alone and $I$ on both registers. Both endpoints share outcome $b=0$, which stops the loop. Outcome $b=1$ selects the interior and continues the loop. The classical counter $n$ counts the steps taken.
\begin{equation}\label{prog:walk}
\begin{aligned}
 \mathrm{Walk}\equiv{}&n:=0;\ b\leftarrow\kw{meas}\ M_\partial[\mathit{coin},\mathit{position}];\\
 &\kw{while}\ b=1\ \kw{do}\\
 &\quad n:=n+1;\ \mathit{coin}:=U_c[\mathit{coin}];\\
 &\quad\mathit{coin},\mathit{position}:=V[\mathit{coin},\mathit{position}];\\
 &\quad b\leftarrow\kw{meas}\ M_\partial[\mathit{coin},\mathit{position}]\\
 &\kw{od}.
\end{aligned}
\end{equation}
The program~\eqref{prog:walk} is CQ-HB, with $b$ retained and $n$ homogeneous, and its summary has one location. One iteration applies the step $U=V(U_c\otimes I)$ and then keeps either the interior part, $\Qop_{\mathrm{int}}(\rho)=P_{\mathrm{int}}U\rho U^\dagger P_{\mathrm{int}}$, which continues the loop, or the boundary part, $\Qop_\partial(\rho)=P_\partial U\rho U^\dagger P_\partial$, which exits. By Definition~\ref{def:live-terminal}, the output for an input density operator $\rho$ has the generating function
\begin{equation}\label{eq:walk-gf}
 f_\rho(z)=P_\partial\rho P_\partial+z\,\Qop_\partial\bigl(\id-z\,\Qop_{\mathrm{int}}\bigr)^{-1}\bigl(P_{\mathrm{int}}\rho P_{\mathrm{int}}\bigr).
\end{equation}
The constant term is the output of the initial boundary test. The coefficient of $z^m$ for $m\ge1$ sums the states that stay in the interior for $m-1$ steps and reach an endpoint at step $m$. Its trace is the probability of hitting time $n_{\mathrm{hit}}=m$. Appendix~\ref{app:walk-absorption} gives the general matrix formula.

\subsubsection{Same Mean and Variance, Different Behavior}\label{sec:walk-exact}
Take $a=6$ and start the Hadamard walk in $\ket{L,1}$. The first step absorbs half the mass at the left endpoint and leaves the other half in $\ket{R,2}$. Since each step changes the parity of the position and both endpoints are even, we group the remaining steps in pairs. The interior positions at the end of each pair are $2$ and $4$, giving a four-dimensional space including the coin. We solve the summary on this space, using $z^2$ to count each pair of steps, and add the first-step contribution.

In the basis $(\ket{L,0},\ket{R,6})$, the calculation gives the hitting-time generating function $g(z)=\tr f_\rho(z)$ and the total output state $f_\rho(1)$:
\[
 g(z)=\frac{z(4-4z^2+2z^4-z^6)}{8-10z^2+6z^4-5z^6+2z^8},
 \qquad
 f_\rho(1)=\frac1{58}\begin{pmatrix}41&-1\\-1&17\end{pmatrix}.
\]
Appendix~\ref{app:walk-certificate} gives the full matrix of rational functions. Table~\ref{tab:walk} compares the moments and the coefficient $[z^7]g(z)$ with those of a classical fair-coin walk from position $1$. The lower-right entry of $f_\rho(1)$ gives the probability of reaching the right endpoint.

\begin{table}[!htbp]
\caption{Absorbing walk on positions $0,\ldots,6$ started at position $1$: the Hadamard quantum walk of the program~\eqref{prog:walk} versus the classical fair-coin walk.}\label{tab:walk}
\small
\begin{tabular}{@{}lrr@{}}
\toprule
 & quantum walk & classical walk\\
\midrule
mean hitting time & $5$ & $5$\\
variance of the hitting time & $40$ & $40$\\
probability of absorption at the right end & $17/58\approx0.29$ & $1/6\approx0.17$\\
probability of absorption after exactly $7$ steps & $17/128\approx0.13$ & $9/128\approx0.07$\\
\bottomrule
\end{tabular}
\end{table}

For this starting state, the two walks agree in both mean and variance. The probability of absorption at step $7$ separates their hitting-time distributions, and the quantum walk reaches the far endpoint about $1.76$ times as often.

\FloatBarrier
\subsubsection{The Returned State at a Given Hitting Time}\label{sec:walk-coherence}
For the same walk started in $\ket{L,1}$, the coefficients of $f_\rho(z)$ give the quantum state at each hitting time. In the boundary basis $(\ket{L,0},\ket{R,6})$, the first-step and seventh-step coefficients are
\[
 [z]f_\rho(z)=\frac12\proj{L,0},\qquad
 [z^7]f_\rho(z)=\frac1{128}\begin{pmatrix}1&-4\\-4&16\end{pmatrix}.
\]
Dividing each coefficient by its trace gives the state conditioned on that hitting time. At step $1$ the walk returns $\ket{L,0}$. At step $7$ it returns the coherent superposition $(\ket{L,0}-4\ket{R,6})/\sqrt{17}$. The two diagonal entries give conditional endpoint probabilities $1/17$ and $16/17$. The off-diagonal entries retain the relative sign between the two amplitudes. Thus the same coefficient that gives the step-$7$ probability in Table~\ref{tab:walk} also gives the quantum state returned at that time.

\FloatBarrier
\section{Related Work}\label{sec:related}

\paragraph{Probabilistic programs.}
Probabilistic semantics describes output distributions~\citep{Kozen1981} or expectations~\citep{McIverMorgan2005}. Generating functions give symbolic representations of these distributions and support loop analysis~\citep{Klinkenberg2021}, distribution verification~\citep{Chen2022}, and Bayesian inference~\citep{Zaiser2023,Klinkenberg2024}. Related analyses derive moments and tail bounds~\citep{Kura2019,WangHoffmannReps2021,Moosbrugger2022}. One symbolic representation can then be used to recover probabilities and moments of the output. We extend this approach to joint classical-quantum outputs, where each classical outcome carries an unnormalized quantum state, and establish rationality for homogeneous-bounded loops without assuming termination.

\paragraph{Quantum programs.}
Quantum weakest preconditions and Hoare logics reason about program correctness~\citep{DHo2006,Feng2007,Ying2011,Ying2016,Zhou2019}, while relational logics compare programs~\citep{Unruh2019,Barthe2026}. Invariant generation, mechanized verification, and abstract interpretation provide complementary proof methods~\citep{YingYingWu2017,Chareton2021,Zhou2023,YuPalsberg2021}. Semantics and logics in infinite dimension also cover unbounded systems, conditioning, and continuous-variable computation~\citep{Cho2016,Unruh2024,Gehnen2025,Barthe2026,Yu2026cv}. They also support embeddings of classical-quantum programs into a purely quantum setting~\citep{Barthe2026}. We keep the quantum memory finite-dimensional and represent unbounded classical data explicitly; our superinvariants provide bounds on their joint outputs.

\paragraph{Classical-quantum programs.}
Languages and semantics for classical-quantum programs combine classical control with quantum data~\citep{Selinger2004,SelingerValiron2006,Chadha2006,FengYing2021,DengFeng2022,Jia2022}. Joint states and kernels of quantum operations describe the resulting correlations~\citep{FengYing2021,DengFeng2022,Barbosa2021}; OFPSs give these outputs an algebraic representation. Hybrid path sums support symbolic execution, equivalence checking, and invariant-based analysis of unbounded loops~\citep{Chareton2026hps,Chareton2026,Ke2026}. Other approaches model recursive hybrid computation and use graded execution series~\citep{Rice2026,Magnot2026}. 

\paragraph{Expected runtimes.}
Expected-runtime calculi extend probabilistic cost analysis~\citep{Kaminski2018} to quantum~\citep{OlmedoDiazCaro2019,Liu2025} and mixed~\citep{Avanzini2022} programs. Their main objective is to analyze resource use, such as iteration counts or the cost of executing commands. These methods include automated bounds for mixed programs~\citep{MoserSchaper2026} and analyses of infinite-dimensional programs with unbounded expectations~\citep{Gehnen2026}. Finite-dimensional matrix methods compute runtime quantities directly~\citep{Liu2025,Yu2023}. \citet{yu2026parameterizedweakestpreconditioncalculusfinegrained} use parameterized weakest preconditions to derive termination probabilities, conditional runtime moments, and divergence rates of infinite moments, including for infinite-dimensional systems. Our focus is exact rational representations of joint classical-quantum outputs for CQ-HB loops with finite-dimensional quantum memory, retaining the quantum state associated with each count.
Comparisons with expected runtimes require matching cost models and divergence conventions (Section~\ref{sec:ofps-queries}).

\paragraph{Quantum loops and quantum Markov chains.}
Analyses of quantum loops study termination and runtimes through quantum operations~\citep{YingFeng2010,Ying2013,YuYing2012,LiYing2018,Ying2016,AndresMartinez2022}, while quantum Markov chains describe systems by graphs with quantum transitions~\citep{FengYuYing2013}. Our summaries use a similar finite graph and attach monomials to record classical increments. Generating functions also appear in quantum recurrence~\citep{Grunbaum2013,Bourgain2014}, where they describe return amplitudes. Our coefficients instead describe unnormalized states at classical outcomes, and the semantics follows program composition.

\FloatBarrier
\section{Conclusion}\label{sec:conclusion}
We presented operator-valued formal power series as a compositional semantics for classical-quantum programs, preserving correlations between classical outcomes and quantum states. We proved that homogeneous-bounded loops admit rational output representations, which can be computed exactly by finite linear algebra under algebraic-data assumptions, without requiring almost-sure termination or user-supplied loop invariants. For more general loops, we established sound output bounds through superinvariants. Together, these results provide a forward approach to quantitative analysis, allowing termination probabilities, distributions and moments of accumulated costs over terminating executions, and outcome-conditioned quantum states to be derived from a single symbolic representation.
\FloatBarrier

\bibliographystyle{ACM-Reference-Format}
\bibliography{references}

\clearpage
\appendix
\section{Quantum and Analytic Preliminaries}\label{app:prelim}
This appendix collects standard facts used throughout the proofs: positive operator sums and monotone limits (Appendix~\ref{app:positive}), complete positivity and Choi matrices (Appendix~\ref{app:cp}), and rational power series (Appendix~\ref{app:rational-germ}). All Hilbert spaces are finite-dimensional. We write $\|A\|_1=\tr\sqrt{A^\dagger A}$ for the trace norm, and recall that $\|A\|_1=\tr A$ for $A\succeq0$ and $|\tr(OA)|\le\|O\|_\infty\|A\|_1$~\citep{Watrous2018,HornJohnson2012}.

\subsection{Positive Sums and Increasing Limits}\label{app:positive}
\begin{lemma}[Positive monotone convergence]\label{lem:positive}
\leavevmode
\begin{enumerate}
\item If $(A_i)_{i\in I}$ is a finite or countable family of positive operators with $\sum_i\tr A_i<\infty$, then $\sum_iA_i$ converges absolutely in trace norm, independently of the order of summation, and is the L\"owner supremum of its finite partial sums.
\item Every increasing sequence $A_1\preceq A_2\preceq\cdots$ of positive operators with bounded traces converges in trace norm to its L\"owner supremum.
\item If $A_{i,n}\succeq0$ is increasing in $n$ for every $i$ and $\sup_n\sum_i\tr A_{i,n}<\infty$, then $\sup_n\sum_iA_{i,n}=\sum_i\sup_nA_{i,n}$.
\end{enumerate}
\end{lemma}
\begin{proof}
(1) Since $\|A_i\|_1=\tr A_i$, the series is absolutely convergent. Its finite partial sums increase and are bounded by the limit, and every common upper bound of the partial sums bounds the limit because the positive cone is closed.

(2) For $m\ge n$, $\|A_m-A_n\|_1=\tr A_m-\tr A_n$. Since the traces form a bounded increasing sequence, $(A_n)$ is Cauchy in trace norm. Its limit is the least upper bound because the positive cone is closed.

(3) Both sides are the supremum, over $n$ and over finite $J\subseteq I$, of $\sum_{i\in J}A_{i,n}$: for a finite sum the supremum over $n$ commutes with the sum by (2), and the supremum over $J$ is given by (1). All families are bounded by the trace assumption.
\end{proof}

\begin{corollary}\label{cor:mass-sup}
For an increasing sequence $F_1\sqsubseteq F_2\sqsubseteq\cdots$ in $\OFPS{\Hil}{\bX}$, $\mass{\bigsqcup_nF_n}=\sup_n\mass{F_n}$.
\end{corollary}
\begin{proof}
Both sides are the supremum over $n$ and over finite sets of classical states of the sum of the traces of the corresponding coefficients.
\end{proof}

\subsection{Complete Positivity and Choi Matrices}\label{app:cp}
Recall the Choi matrix $\Choi(\Theta)=\sum_{i,j}\ket i\bra j\otimes\Theta(\ket i\bra j)$. The equivalence below is Choi's characterization of completely positive maps~\citep{Choi1975}; see also \citet{Watrous2018}.

\begin{lemma}[Choi's theorem]\label{lem:choi}
For a linear map $\Theta:\Lin(\Hil)\to\Lin(\Hil')$, the following are equivalent: (a)~$\Theta$ is CP; (b)~$\Choi(\Theta)\succeq0$; (c)~$\Theta(A)=\sum_iK_iAK_i^\dagger$ for some operators $K_i$.
\end{lemma}
\begin{proof}
(a)$\Rightarrow$(b): $\Choi(\Theta)=(\id\otimes\Theta)(\proj\Omega)$ with $\ket\Omega=\sum_i\ket i\ket i$, and $\proj\Omega\succeq0$.

(b)$\Rightarrow$(c): Write $\Choi(\Theta)=\sum_i\ket{v_i}\bra{v_i}$ and define $K_i$ by $\ket{v_i}=\sum_j\ket j\otimes K_i\ket j$. Comparing blocks gives $\Theta(\ket j\bra{j'})=\sum_iK_i\ket j\bra{j'}K_i^\dagger$.

(c)$\Rightarrow$(a): $(\id\otimes\Theta)(B)=\sum_i(I\otimes K_i)B(I\otimes K_i)^\dagger\succeq0$ for $B\succeq0$.
\end{proof}

\begin{lemma}[Closure of CP maps]\label{lem:cp-closure}
\leavevmode
\begin{enumerate}
\item If $(\Theta_i)_{i\in I}$ is a countable family of CP maps with $\sum_i\tr\Theta_i(I)<\infty$, then $\sum_i\Theta_i$ converges (in any norm on the finite-dimensional space of linear maps) to a CP map.
\item A limit of CP maps is CP.
\end{enumerate}
\end{lemma}
\begin{proof}
The map $\Theta\mapsto\Choi(\Theta)$ is a linear bijection. For CP $\Theta_i$, $\Choi(\Theta_i)\succeq0$ and $\tr\Choi(\Theta_i)=\tr\Theta_i(I)$, so (1) follows from Lemma~\ref{lem:positive}(1) and Lemma~\ref{lem:choi}. For (2), the cone of positive-semidefinite matrices is closed.
\end{proof}

\paragraph{Positivity versus complete positivity.}
A map is \emph{positive} if it maps $\Pos(\Hil)$ into $\Pos(\Hil')$. Every CP map is positive, but not conversely. The transpose map $A\mapsto A^{\mathsf T}$ on a qubit is positive, yet its Choi matrix is the swap operator, which has eigenvalue $-1$ on $\ket{01}-\ket{10}$. When $\dim\Hil\ge2$, no finite set of test states suffices to establish positivity of a map. The conic hull of finitely many states is a polyhedral cone, whereas $\Pos(\Hil)$ is not polyhedral for $\dim\Hil\ge2$. By the separation theorem there is a real linear functional $\ell$ on Hermitian operators that is nonnegative on the test states but negative on some positive operator, and the map $A\mapsto\ell(A)\,I$, extended complex-linearly, passes all the tests but is not positive. Choi's theorem replaces infinitely many positivity conditions by one semidefiniteness condition, at the price of checking the stronger property.

\subsection{Rational Power Series}\label{app:rational-germ}
\begin{lemma}\label{lem:rational-germ}
Let $P,Q\in\C[\bY]$ be coprime with $Q\ne0$, and let $G\in\C[[\bY]]$ satisfy $QG=P$ in $\C[[\bY]]$. Then $Q(\zero)\ne0$, so $G$ is the Taylor series of $P/Q$ and is a rational power series.
\end{lemma}
\begin{proof}
Let $R=\C[\bY]_{(\bY)}$ be the local ring at the origin. Its completion $\C[[\bY]]$ is faithfully flat over $R$~\citep[Theorem~8.14]{Matsumura1986}, so $Q\,\C[[\bY]]\cap R=QR$~\citep[Theorem~7.5]{Matsumura1986}. Since $P$ lies in the left-hand side, $P=Q\,A/S$ with $A,S\in\C[\bY]$ and $S(\zero)\ne0$, that is, $PS=QA$. As $P$ and $Q$ are coprime in the factorial ring $\C[\bY]$, $Q$ divides $S$, and therefore $Q(\zero)\ne0$.
\end{proof}

\begin{corollary}\label{cor:rational-analytic}
Let $G$ be analytic on a neighborhood of $\zero\in\C^d$, and let $p,q\in\C[\bY]$ with $q\ne0$ satisfy $q\,G=p$ near $\zero$. Then the Taylor series of $G$ is rational.
\end{corollary}
\begin{proof}
Write $p/q=P/Q$ in lowest terms. Then $q(QG-P)=0$ in the integral domain $\C[[\bY]]$, so $QG=P$, and Lemma~\ref{lem:rational-germ} applies.
\end{proof}

\FloatBarrier
\section{Proofs for the Semantics}\label{app:semantics}
This appendix establishes the domain and convergence properties of OFPSs, proves healthiness and unfolding, and derives the kernel representation in Proposition~\ref{prop:kernels}.

\subsection{The Domains}\label{app:domain}
Through the encoding, the domain properties of Proposition~\ref{prop:ofps-domain} correspond to those of cq-states~\citep{FengYing2021}; we give a direct proof.
\begin{proof}[Proof of Proposition~\ref{prop:ofps-domain}]
Let $F_1\sqsubseteq F_2\sqsubseteq\cdots$ in $\OFPS{\Hil}{\bX}$. For each classical state $\sigma$, the coefficients $[\bX^\sigma]F_n$ increase and have trace at most one, so by Lemma~\ref{lem:positive}(2) they converge in trace norm to their supremum $\rho_\sigma\succeq0$. For every finite $J\subseteq\Sigma$, $\sum_{\sigma\in J}\tr \rho_\sigma=\lim_n\sum_{\sigma\in J}\tr[\bX^\sigma]F_n\le1$. Hence $F=\sum_\sigma \rho_\sigma\bX^\sigma\in\OFPS{\Hil}{\bX}$, and $F$ is the least upper bound because suprema are taken coefficientwise. Applying the argument to the partial sums of an admissible family shows that the admissible sum is their supremum. The zero series is least, and mass commutes with suprema by Corollary~\ref{cor:mass-sup}.
\end{proof}

\begin{proposition}[Evaluation and extraction]\label{prop:gf}
For $F=\sum_\sigma \rho_\sigma\bX^\sigma\in\OFPS{\Hil}{\bX}$, the generating function $f_F(\bz)=\sum_\sigma \rho_\sigma\bz^\sigma$ converges absolutely and uniformly in trace norm on the closed unit polydisc and is analytic in its interior. Moreover, $[\bX^\sigma]F=\partial^\sigma f_F(\zero)/\sigma!$, $f_F(\one)=\sum_\sigma \rho_\sigma$, and $\tr f_F(\one)=\mass F$. For a variable $x_i$, the $r$-th factorial submoment equals $\lim_{z\to1^{-}}g_i^{(r)}(z)$, where $g_i(z)=\tr f_F(1,\ldots,z,\ldots,1)$ has $z$ in position $i$.
\end{proposition}
\begin{proof}
For $|z_i|\le1$, $\sum_\sigma\|\rho_\sigma\bz^\sigma\|_1\le\sum_\sigma\tr \rho_\sigma\le1$, because $\|A\|_1=\tr A$ for $A\succeq0$, which gives absolute and uniform convergence, analyticity, and the coefficient formulas. For the submoments, the $r$-th derivative of $g_i$ on $(0,1)$ is a power series with nonnegative coefficients $m^{\underline{r}}\,a_m$, where $a_m$ is the total trace of the coefficients with $\sigma_i=m$. By monotone convergence, its limit as $z\to1^{-}$ is $\sum_m m^{\underline{r}}\,a_m\in[0,\infty]$.
\end{proof}

Two consequences of healthiness are used repeatedly. If $\psi$ is linear on admissible combinations, then $\psi(0)=0$ (take $a=b=0$). It also preserves admissible countable sums: it preserves finite ones by linearity, and countable ones by continuity together with Proposition~\ref{prop:ofps-domain}. Moreover, if $F\sqsubseteq G$ are admissible, then $G-F$ is admissible, so $\psi(G)=\psi(F)+\psi(G-F)\sqsupseteq\psi(F)$.

\begin{lemma}[Healthy transformers]\label{lem:transformers}
$(\TF,\sqsubseteq)$, ordered pointwise, is a pointed $\omega$-cpo. Its least element is the zero transformer, and suprema of increasing sequences are pointwise.
\end{lemma}
\begin{proof}
Let $\psi_1\sqsubseteq\psi_2\sqsubseteq\cdots$ in $\TF$ and put $\psi(F)=\bigsqcup_n\psi_n(F)$, which exists by Proposition~\ref{prop:ofps-domain}. Coefficientwise limits preserve linearity on admissible combinations, and $\mass{\psi(F)}=\sup_n\mass{\psi_n(F)}\le\mass F$ by Corollary~\ref{cor:mass-sup}. For an increasing sequence $(F_m)$,
\[
 \psi\Bigl(\bigsqcup_mF_m\Bigr)=\bigsqcup_n\bigsqcup_m\psi_n(F_m)=\bigsqcup_m\bigsqcup_n\psi_n(F_m),
\]
since both iterated suprema equal $\bigsqcup_j\psi_j(F_j)$: each pair $(n,m)$ is dominated by $(j,j)$ with $j=\max(n,m)$. Hence $\psi\in\TF$, and it is the least pointwise upper bound.
\end{proof}

\subsection{Well-Definedness}\label{app:well-defined}
\paragraph{Primitive commands.}
In coefficient form, for $F=\sum_\sigma \rho_\sigma\bX^\sigma\in\OFPS{\Hil}{\bX}$,
\begin{align*}
 [\bX^\tau]\sem{x:=e}(F)&=\sum_{\sigma:\,\sigma[e(\sigma)/x]=\tau}\rho_\sigma,
 &[\bX^\tau]\sem{x\sample\mu}(F)&=\sum_{(\sigma,v):\,\sigma[v/x]=\tau}\mu(v)\rho_\sigma,\\
 [\bX^\tau]\sem{q:=\ket0}(F)&=\Rmap_q(\rho_\tau),
 &[\bX^\tau]\sem{\bar q:=U[\bar q]}(F)&=U\rho_\tau U^\dagger,\\
 [\bX^\tau]\sem{x\leftarrow\kw{meas}\ M[\bar q]}(F)&=\sum_{(\sigma,i):\,\sigma[i/x]=\tau}M_i\rho_\sigma M_i^\dagger.
\end{align*}
Every pair $(\sigma,v)$ or $(\sigma,i)$ contributes to exactly one output classical state. Summing traces over all output classical states therefore gives $\sum_\sigma\tr \rho_\sigma\sum_v\mu(v)$ for sampling and $\sum_\sigma\tr\bigl(\rho_\sigma\sum_iM_i^\dagger M_i\bigr)$ for measurement, and both equal $\mass F$. Reset and unitaries preserve the trace of each coefficient. The summands are positive and their traces are summable, so the sums converge (Lemma~\ref{lem:positive}). Linearity on admissible combinations follows by distributing the sums, and continuity follows from Lemma~\ref{lem:positive}(3). Hence every primitive command denotes a mass-preserving healthy transformer, including assignments that merge infinitely many classical states and measurements with countably many outcomes.

\begin{proof}[Proof of healthiness and unfolding]
We proceed by induction on $S$. The primitive commands were treated above.

\emph{Sequencing.} Composition preserves linearity on admissible combinations (the intermediate combination is admissible because mass does not increase), continuity, and the mass bound, and it preserves mass if both components do.

\emph{Conditional.} Restriction to a guard and to its negation is healthy, and $\mass{F|_e}+\mass{F|_{\neg e}}=\mass F$. Hence $\sem{S_1}(F|_e)+\sem{S_2}(F|_{\neg e})$ has mass at most $\mass F$ and is therefore admissible, with equality in the mass bound if both branches preserve mass.

\emph{Loop.} Let $W=\kw{while}\ e\ \kw{do}\ S\ \kw{od}$, where $\sem S\in\TF$ by induction. For $\psi\in\TF$,
\[
 \mass{F|_{\neg e}}+\mass{\psi(\sem S(F|_e))}\le\mass{F|_{\neg e}}+\mass{F|_e}=\mass F,
\]
so $\Phi_{e,S}(\psi)(F):=F|_{\neg e}+\psi(\sem S(F|_e))$ is an admissible sum and $\Phi_{e,S}(\psi)$ is mass-nonincreasing. Continuity in $F$ and linearity on admissible combinations follow from these properties of guard restriction, $\sem S$, $\psi$, and admissible addition. Thus $\Phi_{e,S}$ maps $\TF$ to $\TF$. It is monotone in $\psi$ and preserves pointwise suprema of increasing sequences, because the argument $\sem S(F|_e)$ does not depend on $\psi$. By Lemma~\ref{lem:transformers} and Kleene's theorem, $\psi^*=\bigsqcup_m\Phi^m_{e,S}(0)$ is a fixed point. For any fixed point $\chi\in\TF$, induction gives $\Phi^m_{e,S}(0)\sqsubseteq\chi$ for all $m$, so $\psi^*$ is the least fixed point, and $\sem W=\bigsqcup_m\Phi^m_{e,S}(0)$.

\emph{Unrollings.} For the program $\kw{abort}$, which is $\kw{while}\ \mathrm{true}\ \kw{do}\ \kw{skip}\ \kw{od}$, the unfolding functional is $\Phi(\psi)(F)=0+\psi(F)$, the identity on $\TF$, whose least fixed point is the zero transformer; hence $\sem{W_0}=0=\Phi^0_{e,S}(0)$. By the clauses for sequencing and conditionals,
\[
 \sem{W_{m+1}}(F)=\sem{W_m}\bigl(\sem S(F|_e)\bigr)+F|_{\neg e}=\Phi_{e,S}(\sem{W_m})(F).
\]
By induction, $\sem{W_m}=\Phi^m_{e,S}(0)$ for all $m$. Hence $\sem W=\bigsqcup_m\sem{W_m}$, and the unfolding identity is the fixed-point equation $\sem W=\Phi_{e,S}(\sem W)$.
\end{proof}

\subsection{Coefficient Kernels}\label{app:kernels}
A coefficient kernel is initially defined on partial density operators by $\kappa^S_{\tau,\sigma}(\rho)=[\bX^\tau]\sem S(\rho\bX^\sigma)$. EasyPQC establishes a representation by quantum operations indexed by classical states~\citep[Proposition~A.9]{Barbosa2021}. We give a direct proof for the OFPS presentation, including its countable coefficient sums. We first show that each kernel extends uniquely to a linear map on all operators.
\begin{proposition}[Kernel representation]\label{prop:kernels}
For every program $S$ and all $\sigma,\tau\in\Sigma$:
\begin{enumerate}
\item $\kappa^S_{\tau,\sigma}$ is completely positive;
\item $\sum_{\tau\in\Sigma}\tr\kappa^S_{\tau,\sigma}(A)\le\tr A$ for every $A\succeq0$;
\item $[\bX^\tau]\sem S(F)=\sum_{\sigma\in\Sigma}\kappa^S_{\tau,\sigma}\bigl([\bX^\sigma]F\bigr)$ for every $F\in\OFPS{\Hil}{\bX}$.
\end{enumerate}
\end{proposition}
\paragraph{Existence and uniqueness of the kernel.}
Fix $S,\sigma,\tau$. For $A\succeq0$, choose $r\ge\max(1,\tr A)$ and set $\kappa(A)=r\,[\bX^\tau]\sem S\bigl((A/r)\bX^\sigma\bigr)$. Linearity on admissible combinations makes this independent of $r$, positively homogeneous, and additive on $\Pos(\Hil)$. Indeed, for $A,B\succeq0$ choose $r\ge\max(1,\tr A+\tr B)$; then $(A/r)\bX^\sigma+(B/r)\bX^\sigma$ is an admissible sum. A Hermitian operator is a difference of positive operators, and if $A-B=A'-B'$ with all four positive, then $A+B'=A'+B$, so additivity makes $\kappa(A)-\kappa(B)=\kappa(A')-\kappa(B')$. This yields a real-linear map on Hermitian operators, which extends uniquely to a complex-linear map on $\Lin(\Hil)=\{H_1+iH_2\}$. Uniqueness holds because partial density operators span $\Lin(\Hil)$. Scaling the mass inequality gives Proposition~\ref{prop:kernels}(2).

\begin{proof}[Proof of Proposition~\ref{prop:kernels}]
Part~(2) was shown above. Part~(3) holds because $F$ is the admissible sum of its elementary terms $\rho_\sigma\bX^\sigma$ and $\sem S$ preserves admissible countable sums. We prove part~(1) by induction on $S$.

\emph{Primitive commands.} The coefficient formulas above give
\begin{align*}
 \kappa^{x:=e}_{\tau,\sigma}&=\iv{\tau=\sigma[e(\sigma)/x]}\id,
 &\kappa^{x\sample\mu}_{\tau,\sigma}&=\sum_{v:\sigma[v/x]=\tau}\mu(v)\id,\\
 \kappa^{q:=\ket0}_{\tau,\sigma}&=\iv{\tau=\sigma}\Rmap_q,
 &\kappa^{\bar q:=U[\bar q]}_{\tau,\sigma}&=\iv{\tau=\sigma}U(\cdot)U^\dagger,\\
 \kappa^{x\leftarrow\kw{meas}M[\bar q]}_{\tau,\sigma}&=\sum_{i:\sigma[i/x]=\tau}M_i(\cdot)M_i^\dagger.
\end{align*}
Each kernel is a sum of CP maps whose values at $I$ have summable traces, and is therefore CP by Lemma~\ref{lem:cp-closure}(1).

\emph{Sequencing.} Since $\sem{S_1;S_2}(\rho\bX^\sigma)=\sem{S_2}\bigl(\sum_\mu\kappa^{S_1}_{\mu,\sigma}(\rho)\bX^\mu\bigr)$,
\begin{gather*}
 \kappa^{S_1;S_2}_{\tau,\sigma}=\sum_{\mu\in\Sigma}\kappa^{S_2}_{\tau,\mu}\circ\kappa^{S_1}_{\mu,\sigma},\\
 \sum_\mu\tr\kappa^{S_2}_{\tau,\mu}\bigl(\kappa^{S_1}_{\mu,\sigma}(I)\bigr)\le\sum_\mu\tr\kappa^{S_1}_{\mu,\sigma}(I)\le D .
\end{gather*}
Compositions of CP maps are CP, and Lemma~\ref{lem:cp-closure}(1) applies to the sum.

\emph{Conditional.} $\kappa_{\tau,\sigma}$ equals $\kappa^{S_1}_{\tau,\sigma}$ if $\sigma\models e$ and $\kappa^{S_2}_{\tau,\sigma}$ otherwise.

\emph{Loop.} By the previous cases and induction on $m$, every unfolding $W_m$ has CP kernels. By the unfolding characterization proved above, $\kappa^{W_m}_{\tau,\sigma}(\rho)\to\kappa^W_{\tau,\sigma}(\rho)$ for every partial density operator $\rho$. Choose a finite spanning set of partial density operators, for example the projections onto $\ket u$, $(\ket u\pm\ket v)/\sqrt2$, and $(\ket u\pm i\ket v)/\sqrt2$ for basis vectors $u,v$, scaled to trace at most one. Convergence on this set implies convergence of the linear maps, so $\kappa^W_{\tau,\sigma}$ is CP by Lemma~\ref{lem:cp-closure}(2).
\end{proof}

\paragraph{Reference systems.}
Let $\mathcal R$ be a finite-dimensional reference space on which the program does not act, and interpret every primitive on $\Hil\otimes\mathcal R$ as its tensor product with $\id_{\Lin(\mathcal R)}$. The same induction, using composition and limits, shows that the kernels of $S$ on $\Hil\otimes\mathcal R$ are $\kappa^S_{\tau,\sigma}\otimes\id_{\Lin(\mathcal R)}$, as in the extension to untouched subsystems of \citet[Proposition~A.9]{Barbosa2021}. Complete positivity is exactly the statement that these are positive~\citep{Watrous2018}. It is a property of program denotations: an abstract element of $\TF$ need not have CP kernels.

\FloatBarrier
\section{Adequacy, Conservativity, Observables, and Costs}\label{app:correspondence}
This appendix proves adequacy, conservativity (Corollary~\ref{cor:conservative}), and the characterization by classical observations. It then relates OFPS semantics to weakest preconditions on bounded observables and expands on the interpretation of submoments.

\subsection{Adequacy}\label{app:adequacy}
\paragraph{The common fragment.}
\citet{FengYing2021} use a countable set of Boolean and integer classical variables, global classical states, and finite-dimensional quantum subsystems. \cqwhile{} programs belong to their language, with the variables of $\cvar$ assigned integer type and all other classical variables left untouched. A cq-state $\Delta$ on $\Sigma=\N^k$ is embedded into their state space by fixing the untouched variables and assigning the zero operator to classical states with a negative coordinate. Expressions have the same interpretation on states with natural-number values. Their values elsewhere play no role, because every nonzero branch from such a state again has natural-number values: assignments return naturals, sampling distributions are supported on $\N$, and measurement labels lie in $\N$. We write $\fy S$ for their denotational semantics restricted to these inputs, with the output viewed as a cq-state on $\Sigma$. In their framework, $\fy S(\Delta)$ is the sum of the cq-states reached by terminating computations, and the denotation of a loop is the supremum of the denotations of its finite unfoldings.

\begin{proof}[Proof of adequacy]
We prove $\sem S(\Enc\Delta)=\Enc(\fy S(\Delta))$ by induction on $S$.

\emph{Primitive commands.} For a classical state $\sigma$ and an unnormalized quantum state $\rho$, the operational rules of \citet{FengYing2021} update $(\sigma,\rho)$ to $(\sigma[e(\sigma)/x],\rho)$ for assignment, to $(\sigma[v/x],\mu(v)\rho)$ for sampling, to $(\sigma[i/x],M_i\rho M_i^\dagger)$ for measurement, and to $(\sigma,\Rmap_q(\rho))$ for reset. A unitary also preserves $\sigma$. Their denotation sums the resulting operators at each final classical state, which is precisely the coefficient formula of Appendix~\ref{app:well-defined}. Distinct outcome histories are summed even when they lead to the same classical state and operator, on both sides.

\emph{Sequencing and conditionals.} $\Enc$ preserves admissible sums and commutes with restriction to a guard, so the induction hypotheses for the components give the claim.

\emph{Loops.} Let $W=\kw{while}\ e\ \kw{do}\ S\ \kw{od}$. Using the induction hypothesis for $S$ and the conditional and sequencing cases, a second induction on $m$ gives $\sem{W_m}(\Enc\Delta)=\Enc(\fy{W_m}(\Delta))$ for all unfoldings, with both sides zero at $m=0$. The reference semantics of $W$ is the supremum of $\fy{W_m}(\Delta)$. Since $\Enc$ preserves suprema, the unfolding characterization in Appendix~\ref{app:well-defined} identifies its image with $\sem W(\Enc\Delta)$. Every terminating computation completes finitely many iterations of the outer body, whatever loops the body contains, so the unfoldings capture all of them.
\end{proof}

\subsection{Conservativity}\label{app:conservative}
\begin{corollary}[Conservativity]\label{cor:conservative}
Let $\Hil=\C$, so that $\OFPS{\C}{\bX}$ is the set of scalar formal power series encoding subdistributions on $\Sigma$, and write $\sem S_{\mathrm{K}}$ for the semantics of \citet{Klinkenberg2021}.
For every program $S$ built from $\kw{skip}$, assignments, sequencing, conditionals, and loops whose expressions and guards are interpreted identically in both languages, $\sem S(F)=\sem S_{\mathrm K}(F)$ for every $F\in\OFPS{\C}{\bX}$.
Moreover, for such programs $S_1,S_2$, a fresh variable $c$, and the distribution $\mu$ with $\mu(0)=p$ and $\mu(1)=1-p$, setting the indeterminate of $c$ to $1$ in the output of $c\sample\mu;\ \kw{if}\ c=0\ \kw{then}\ S_1\ \kw{else}\ S_2\ \kw{fi}$ gives their rule for the probabilistic choice $\{S_1\}[p]\{S_2\}$.
\end{corollary}
\begin{proof}[Proof of Corollary~\ref{cor:conservative}]
For $\Hil=\C$, positive operators are nonnegative reals and $\OFPS{\C}{\bX}$ is the set of subprobability series. On $\kw{skip}$, assignment, sequencing, and conditionals, the semantic clauses coincide with the rules of \citet{Klinkenberg2021}. For loops both semantics are least fixed points of the same unfolding functional, and both are the supremum of the same Kleene iterates: theirs in a complete lattice that contains our $\omega$-cpo, ours in that $\omega$-cpo. Each iterate maps subprobability series to subprobability series.

For probabilistic choice, fix a variable $c$ that occurs in neither $S_1$ nor $S_2$ and let $\mu(0)=p$, $\mu(1)=1-p$. For $F=\sum_{\sigma,v}a_{\sigma,v}\bX^\sigma C^v\in\OFPS{\C}{(\bX,C)}$, let $\pi(F)=F|_{C=1}=\sum_\sigma\bigl(\sum_va_{\sigma,v}\bigr)\bX^\sigma\in\OFPS{\C}{\bX}$ sum out $c$. The sampling command overwrites $c$, and neither branch reads or writes it, so
\[
 \pi\Bigl(\sem{c\sample\mu;\ \kw{if}\ c=0\ \kw{then}\ S_1\ \kw{else}\ S_2\ \kw{fi}}(F)\Bigr)
 =p\,\sem{S_1}(\pi F)+(1-p)\,\sem{S_2}(\pi F),
\]
which is their rule for $\{S_1\}[p]\{S_2\}$.
\end{proof}
\subsection{Classical Observations}\label{app:observations}
Throughout this subsection, $\ket0$ denotes the all-zero product state of $\Hil$.

\begin{proof}[Proof of the characterization by classical observations]
\emph{From left to right.} Each semantic clause defines the transformer of a compound program from the transformers of its immediate parts. By induction on the context $C$, the transformer $\sem{C[S]}$ is therefore determined by $\sem S$. If $\sem{S_1}=\sem{S_2}$, then $\sem{C[S_1]}=\sem{C[S_2]}$, and the observations agree.

\emph{From right to left.} Assume that $\operatorname{obs}(P;S_1;T)=\operatorname{obs}(P;S_2;T)$ for all loop-free programs $P$ and $T$. Let $\bar q$ be the tuple of all subsystems in $\qvar$. Fix classical states $\sigma,\tau\in\Sigma$ and unit vectors $\ket\psi,\ket\varphi\in\Hil$. Choose a unitary $U_\psi$ with $U_\psi\ket0=\ket\psi$, and let $M_\varphi$ be the measurement with outcomes $0$ and $1$ and operators $M_0=\proj\varphi$ and $M_1=I-\proj\varphi$. Write $x=\tau$ for the guard that only $\tau$ satisfies, and define the loop-free programs
\begin{align*}
 P&:=x_1:=\sigma_1;\ \ldots;\ x_k:=\sigma_k;\ \bar q:=U_\psi[\bar q],\\
 T&:=\kw{if}\ x=\tau\ \kw{then}\ x_1\leftarrow\kw{meas}\ M_\varphi[\bar q]\ \kw{else}\ x_1:=2\ \kw{fi}.
\end{align*}
By the semantic clauses, $\sem P(\proj0\bX^\zero)=\proj\psi\,\bX^\sigma$. For $j=1,2$ and $\mu\in\Sigma$, let $B^j_\mu:=\kappa^{S_j}_{\mu,\sigma}(\proj\psi)$. By Proposition~\ref{prop:kernels}(3), $\sem{S_j}(\proj\psi\,\bX^\sigma)=\sum_\mu B^j_\mu\bX^\mu$, and applying $T$ gives
\[
 \sem{P;S_j;T}(\proj0\bX^\zero)=M_0B^j_\tau M_0\,\bX^{\tau[0/x_1]}+M_1B^j_\tau M_1\,\bX^{\tau[1/x_1]}+\sum_{\mu\ne\tau}B^j_\mu\,\bX^{\mu[2/x_1]} .
\]
Only the first term has $x_1=0$. The coefficients of $\operatorname{obs}(P;S_j;T)$ at classical states with $x_1=0$ therefore add up to $\tr(M_0B^j_\tau M_0)=\bra\varphi B^j_\tau\ket\varphi$. By assumption, $\bra\varphi B^1_\tau\ket\varphi=\bra\varphi B^2_\tau\ket\varphi$ for every unit vector $\ket\varphi$, and by scaling for every vector. On a complex vector space, the values $\bra\varphi B\ket\varphi$ determine the operator $B$ by polarization, so $B^1_\tau=B^2_\tau$. Thus $\kappa^{S_1}_{\tau,\sigma}$ and $\kappa^{S_2}_{\tau,\sigma}$ agree on all pure states. Pure states span $\Lin(\Hil)$ (Appendix~\ref{app:kernels}), and kernels are linear, so $\kappa^{S_1}_{\tau,\sigma}=\kappa^{S_2}_{\tau,\sigma}$. Since $\sigma$ and $\tau$ were arbitrary, Proposition~\ref{prop:kernels}(3) gives $\sem{S_1}=\sem{S_2}$.

If $\qvar$ is empty, omit the unitary from $P$ and replace the measurement in $T$ by $x_1:=0$. The same argument applies with $\Hil=\C$.
\end{proof}
The proof uses only one number from each observation: the probability that $x_1$ ends at $0$. Without a classical variable, the observation is only the termination probability, and the proposition fails as soon as $\Hil\ne\C$. For example, the programs $\kw{skip}$ and $q:=U[q]$, for a unitary $U$ that is not a multiple of $I$, have different semantics but the same termination probability in every context.

\subsection{Bounded Observables}\label{app:duality}
The \emph{adjoint} of a linear map $\Qop$ on $\Lin(\Hil)$ is the map $\Qop^\dagger$ with $\tr(O\,\Qop(A))=\tr(\Qop^\dagger(O)\,A)$ for all $O,A$; for $\Qop(A)=\sum_iK_iAK_i^\dagger$ it is $\Qop^\dagger(O)=\sum_iK_i^\dagger OK_i$. Adjoints evolve observables backward, as quantum weakest preconditions do~\citep{DHo2006,FengYing2021}.

\begin{proposition}\label{prop:duality}
Let $O:\Sigma\to\Pos(\Hil)$ satisfy $O(\tau)\preceq cI$ for all $\tau$. Then $(\wp_SO)(\sigma)=\sum_\tau(\kappa^S_{\tau,\sigma})^\dagger(O(\tau))$ converges, satisfies $0\preceq\wp_SO\preceq cI$, and for every cq-state $\Delta$
\begin{align*}
 \sum_\tau\tr\bigl(O(\tau)\,[\bX^\tau]\sem S(\Enc\Delta)\bigr)
 &=\sum_\tau\tr\bigl(O(\tau)\,\fy S(\Delta)(\tau)\bigr)\\
 &=\sum_\sigma\tr\bigl((\wp_SO)(\sigma)\,\Delta(\sigma)\bigr).
\end{align*}
\end{proposition}
\begin{proof}
For a finite set $J$ of classical states and $A\succeq0$, Proposition~\ref{prop:kernels} gives
\[
 \tr\Bigl(A\sum_{\tau\in J}(\kappa^S_{\tau,\sigma})^\dagger(O(\tau))\Bigr)=\sum_{\tau\in J}\tr\bigl(O(\tau)\kappa^S_{\tau,\sigma}(A)\bigr)\le c\tr A .
\]
Testing against rank-one $A$ shows that the partial sums are positive and bounded by $cI$, so they converge by Lemma~\ref{lem:positive}. The first equality follows from adequacy (Appendix~\ref{app:adequacy}). For the second, expand the left-hand side using Proposition~\ref{prop:kernels}(3), apply trace duality, and interchange the nonnegative sums, whose total is at most $c\sum_\sigma\tr\Delta(\sigma)$.
\end{proof}
The kernel definition satisfies the usual backward clauses, for instance
\begin{gather*}
 (\wp_{x:=e}O)(\sigma)=O(\sigma[e(\sigma)/x]),\qquad
 \wp_{S_1;S_2}=\wp_{S_1}\circ\wp_{S_2},\\
 (\wp_{x\leftarrow\kw{meas}\,M[\bar q]}O)(\sigma)=\textstyle\sum_iM_i^\dagger O(\sigma[i/x])M_i .
\end{gather*}
For a loop, $\wp_WO=\bigsqcup_mH_m$ with $H_0=0$ and $H_{m+1}=\iv{\neg e}O+\iv{e}\wp_SH_m$, which follows by pairing the finite unfoldings with elementary inputs. On the assertions of \citet{FengYing2021}, these clauses give their weakest-precondition rules, and Proposition~\ref{prop:duality} gives the duality identity.

\subsection{Counters and Nontermination}\label{app:counts}
Let $n$ be a classical variable, $F=\sem S(\Enc\Delta)$, and $a_m=\sum_{\sigma:\sigma(n)=m}\tr[\bX^\sigma]F$, so that $\sum_ma_m\le\mass\Delta$. For $0<z<1$, the marginal $g_n(z)=\sum_ma_mz^m$ has $g_n'(z)=\sum_{m\ge1}ma_mz^{m-1}$. Every summand increases as $z\to1^{-}$, so $\lim_{z\to1^{-}}g_n'(z)=\sum_mma_m\in[0,\infty]$. Alternatively, apply Proposition~\ref{prop:duality} to the bounded observables $O_r(\sigma)=\min(\sigma(n),r)I$ and let $r\to\infty$.

For the probabilistic interpretation in Section~\ref{sec:ofps-queries}, assume $\mass\Delta=1$ and that $n$ starts at zero. The termination probability is $p_{\mathrm{term}}=g_n(1)$, and $\mu_{\mathrm{term}}:=\sum_mma_m=g_n'(1^-)$ is the first terminal submoment. If $p_{\mathrm{term}}>0$, normalization gives
\[
 \E[n\mid\mathrm{termination}]=\frac{\mu_{\mathrm{term}}}{p_{\mathrm{term}}}.
\]
When $n$ records elapsed steps, define time to termination $T$ as $n$ on terminating runs and $\infty$ otherwise. If $p_{\mathrm{term}}<1$, then $T$ is infinite with positive probability, so $\E[T]=\infty$. If $p_{\mathrm{term}}=1$, then $\E[T]=\mu_{\mathrm{term}}$, including the possibility that both are infinite.

For a general accumulated resource cost, the charging policy on divergent executions matters. A divergent execution can accumulate zero cost if all its steps have zero charge, even though its time to termination is infinite. Terminal-output semantics does not record that cost. Comparisons with expected-runtime or expected-cost calculi~\citep{Liu2025,Avanzini2022} therefore require matching both the cost model and the treatment of divergence.

\FloatBarrier
\section{Loop Summaries and Rationality}\label{app:rationality}
This appendix gives formal statements and proofs for the summaries and rational outputs of single loops. Nested loops are treated in Appendix~\ref{app:nested}, and moments in Appendix~\ref{app:moments}. In matrix coordinates, a linear map on $\Vsp=\bigoplus_{s\in\Lambda}\Lin(\Hil)$ is represented by a square matrix of size $|\Lambda|D^2$. We use $\vecop(E_{ij})=\ket i\otimes\ket j$, under which $A\mapsto KAK^\dagger$ has matrix $K\otimes\overline K$~\citep{Watrous2018}. With the Choi convention of Section~\ref{sec:choi}, the matrix of $\Qop$ is $\mathfrak R(\Choi(\Qop))$, where $[\mathfrak R(J)]_{Da+b,Di+j}=J_{Di+a,Dj+b}$ for $0\le a,b,i,j<D$. Maps with polynomial or formal power-series coefficients act on their arguments by convolution of coefficients.

\subsection{Formal Statements}\label{app:formal}
For a CQ-HB loop $W=\kw{while}\ e\ \kw{do}\ S\ \kw{od}$, fix a partition satisfying Definition~\ref{def:hb} and write $(s,\bh)\in\N^\ell\times\N^d$ for retained and homogeneous values. Homogeneity is required on all admissible inputs, as in Definition~\ref{def:homogeneous}. The finite active set $\Act=\{s:(s,\zero)\models e\}$ contains all joint retained valuations satisfying the guard, including unreachable ones. Uniform finite support requires, for each active $s$, a finite set $\Omega_s$ containing $\csupp\sem S(\rho\,\bZ^s\bY^\zero)$ for every partial density operator $\rho$.

\begin{lemma}[Loop homogeneity]\label{lem:loop-homogeneous}
If $x_i$ is homogeneous for $S$ and $e$ is independent of $x_i$, then $x_i$ is homogeneous for $\kw{while}\ e\ \kw{do}\ S\ \kw{od}$.
\end{lemma}
\begin{proof}
Restriction to $e$ commutes with multiplication by $X_i$, because $e$ is independent of $x_i$. By induction on $m$, each unrolling satisfies $\sem{W_{m+1}}(X_iF)=X_iF|_{\neg e}+\sem{W_m}(X_i\sem S(F|_e))=X_i\sem{W_{m+1}}(F)$. Multiplication by $X_i$ commutes with suprema, so taking the unfolding limit gives the claim.
\end{proof}

\begin{lemma}[Full-rank witness]\label{lem:witness}
The uniform finite-support condition holds if and only if, for every active $s$, the set $\csupp\sem S\bigl((I/D)\,\bZ^s\bY^\zero\bigr)$ is finite.
\end{lemma}
\begin{proof}
Every partial density operator $\rho$ satisfies $0\preceq \rho\preceq I=D\cdot(I/D)$. Kernels are positive (Proposition~\ref{prop:kernels}), so $0\preceq\kappa^S_{\tau,\sigma}(\rho)\preceq D\,\kappa^S_{\tau,\sigma}(I/D)$. If the coefficient at $\tau$ vanishes for the input $I/D$, it vanishes for every $\rho$.
\end{proof}
Checking only the actual input is not enough. Suppose the body measures $q$ and, on outcome $1$, runs an inner loop that increments $n$ a geometric number of times. One iteration from $\ket0$ then has finitely many outcomes, but one iteration from $\ket1$ has infinitely many, and the check on $I/D$ detects this.

\begin{proposition}[Syntactic criterion]\label{prop:syntactic-hb}
Let $S$ be loop-free, with finitely supported sampling distributions and finite measurement outcome sets. Suppose every homogeneous variable $h_i$ occurs only in assignments $h_i:=h_i+r$, where the expression $r$ reads only retained variables, and that neither $e$ nor any other expression of $S$ reads a homogeneous variable. Then the homogeneity, guard-independence, and uniform finite-support conditions hold. If, in addition, $\Act$ is finite, then $W$ is CQ-HB.
\end{proposition}
The proof is in Appendix~\ref{app:uniform-support}.

We collect the algebraic representation of the loop summaries introduced in Section~\ref{sec:summaries}.
\begin{definition}[Matrix form of a loop summary]\label{def:summary-formal}
Let $\Lambda$ be a finite set of \emph{locations}, $\Vsp=\bigoplus_{s\in\Lambda}\Lin(\Hil)$ the space of block vectors, and $\Ex$ a finite set of \emph{exit labels}. A \emph{loop summary} consists of polynomial maps
\[
 \Cmap(\bY)=\sum_{\br}\bY^\br\,\Cmap_\br:\Vsp\to\Vsp,
 \qquad
 \Emap_u(\bY)=\sum_{\br}\bY^\br\,\Emap_{u,\br}:\Vsp\to\Lin(\Hil)\quad(u\in\Ex),
\]
with finitely many nonzero coefficients, each \emph{blockwise completely positive} (every component map from one block to another is CP), such that
\begin{equation}\label{eq:subconservation}
 \tr\Cmap(\one)A+\sum_{u\in\Ex}\tr\Emap_u(\one)A\le\tr A\qquad(A\in\Vsp,\ A\succeq0).
\end{equation}
An \emph{input} consists of polynomials $\iota(\bY)\in\Vsp[\bY]$ and $\delta_u(\bY)\in\Lin(\Hil)[\bY]$ with positive coefficients and total mass $\tr\iota(\one)+\sum_u\tr\delta_u(\one)\le1$. The graded live series is $\widehat L(T):=\sum_{m\ge0}T^m\Cmap(\bY)^m\iota$, and the terminal series is $F_u:=\delta_u+\sum_{m\ge0}\Emap_u(\bY)\Cmap(\bY)^m\iota$. The ungraded live series $L=(L_s)_{s\in\Lambda}:=\sum_{m\ge0}\Cmap(\bY)^m\iota$ is defined only when every coefficient sum converges in trace norm.
\end{definition}

\begin{proposition}[Matrix summaries of CQ-HB loops]\label{prop:summary-formal}
Let $W=\kw{while}\ e\ \kw{do}\ S\ \kw{od}$ be CQ-HB, and let $F_{\mathrm{in}}=\sum_t\bZ^tR_t(\bY)\in\OFPS{\Hil}{\bX}$ be finitely supported. Put $\Lambda=\Act$ and
\begin{align*}
 \mathcal B_{t,s,\br}&:=\kappa^S_{(t,\br),(s,\zero)},\\
 (\Cmap(\bY)A)_t&=\sum_{s\in\Act,\,\br}\bY^\br\,\mathcal B_{t,s,\br}(A_s)\quad(t\in\Act),\\
 \Emap_u(\bY)A&=\sum_{s\in\Act,\,\br}\bY^\br\,\mathcal B_{u,s,\br}(A_s)\quad(u\notin\Act).
\end{align*}
Let $\Ex$ be the finite set of retained values $u\notin\Act$ that are retained components of classical states in some $\Omega_s$ or in $\csupp F_{\mathrm{in}}$, and put $\iota=(R_s)_{s\in\Act}$ and $\delta_u=R_u$. This is a loop summary with an input, and $\sem W(F_{\mathrm{in}})=\sum_{u\in\Ex}\bZ^u\,F_u(\bY)$.
\end{proposition}
The proof is in Appendix~\ref{app:path-expansion}.

\begin{theorem}[Rationality of loop summaries]\label{thm:rational}
For every loop summary with an input and every exit $u$, the terminal series $F_u$ is a rational OFPS in $\bY$. A rational representation is obtained as follows. The system $(\id-T\,\Cmap(\bY))\,\widehat L(T)=\iota(\bY)$ has a unique solution over the field $\C(T,\bY)$. If $p/q$ is a matrix entry of $\delta_u+\Emap_u(\bY)\widehat L(T)$ in lowest terms, then $q(1,\bY)\ne0$, and the corresponding entry of $F_u$ is represented by $p(1,\bY)/q(1,\bY)$. No termination assumption is made.
\end{theorem}

\begin{corollary}[Direct formula]\label{cor:direct}
If $\det(\id-\Cmap(\bY))$ is not the zero polynomial, then $F_u=\delta_u+\Emap_u(\bY)\,(\id-\Cmap(\bY))^{-1}\iota(\bY)$. This holds in particular when $1$ is not an eigenvalue of $\Cmap(\zero)$, for example when $\Cmap(\zero)=0$.
\end{corollary}
The theorem and corollary are proved in Appendix~\ref{app:T-specialization}.

\begin{theorem}[Exact computation]\label{thm:computation}
Let $W=\kw{while}\ e\ \kw{do}\ S\ \kw{od}$ satisfy Proposition~\ref{prop:syntactic-hb}, and assume:
\begin{enumerate}
\item a finite set containing $\Act$ is given, and the expressions and the guard can be evaluated on given classical states;
\item all sampling probabilities and all entries of the unitaries and measurement operators of $S$ are algebraic numbers;
\item the input $F_{\mathrm{in}}$ is finitely supported with algebraic coefficients.
\end{enumerate}
Then a rational representation of $\sem W(F_{\mathrm{in}})$ is computable. It is obtained by solving a linear system with $|\Act|D^2$ unknowns over $\mathbb K(T,\bY)$, where $\mathbb K$ is the number field generated by the data and their complex conjugates.
\end{theorem}
The proof is in Appendix~\ref{app:computation}.

\begin{theorem}[Nested loops]\label{thm:nested}
Let $S$ be a \cqwhile{} program with arbitrarily nested loops, and let $F_{\mathrm{in}}\in\OFPS{\Hil}{\bX}$ be finitely supported. Suppose the variables of $S$ are partitioned into retained variables, with values $s$, and \emph{additive} variables $\bh$, such that
\begin{enumerate}
\item every update of an additive variable has the form $h_i:=h_i+r(s)$ with $r(s)\in\N$, and no other expression or guard reads $\bh$;
\item all sampling distributions have finite support, and all measurement outcome sets are finite;
\item finite value sets for the retained variables are given that contain the input support and are closed under all updates reachable from it;
\item all numerical data and the coefficients of $F_{\mathrm{in}}$ are algebraic, and expressions can be evaluated on given classical states.
\end{enumerate}
Then $\sem S(F_{\mathrm{in}})$ is a rational OFPS, and a rational representation is computable. Nontermination and cycles without increments are allowed. Closure in condition~(3) can be checked by finite exploration.
\end{theorem}
The proof is in Appendix~\ref{app:nested}.

\begin{proposition}[Moments]\label{prop:moments}
Consider a loop summary with an input.
\begin{enumerate}
\item Each terminal generating function $f_{F_u}(\by)$ is analytic on a neighborhood of the closed unit polydisc. Consequently all mixed factorial submoments of the homogeneous variables over terminating runs are finite, and they are the corresponding derivatives of the rational function of Theorem~\ref{thm:rational} at $\by=\one$.
\item Suppose~\eqref{eq:subconservation} holds with equality and the loop terminates almost surely from $\rho:=\iota(\one)$. Let $\Cone:=\Cmap(\one)$, and let $\Vsp_\rho\subseteq\Vsp$ be the smallest block space of the form $\bigoplus_s\Lin(\Hil_s)$ that contains every $\Cone^m\rho$. Then $\Cone$ maps $\Vsp_\rho$ into itself with spectral radius less than one. Let $n_{\mathrm c}$ be the number of continuing transitions before exit, and let $\E[X;u]$ denote the expectation of $X$ restricted to runs that exit through $u$. Then $\E[n_{\mathrm c}^{\underline{r}}\,;\,u]=\tr\Emap_u(\one)(L_r)$ for $r\ge1$, where $L_0,L_1,\ldots\in\Vsp_\rho$ solve $(\id-\Cone)L_0=\rho$ and $(\id-\Cone)L_r=r\,\Cone(L_{r-1})$ in $\Vsp_\rho$.
\end{enumerate}
\end{proposition}
The proofs are in Appendices~\ref{app:terminal-regularity} and~\ref{app:moment-recurrences}, respectively. Part~(1) follows from geometric decay of the terminal coefficients in the number of steps. Part~(2) is the operator form of the classical fundamental-matrix method~\citep{KemenySnell1976} and is the moment method used for quantum expected runtimes~\citep{Ying2013,Liu2025}. The restriction to $\Vsp_\rho$ is needed: on the full operator space, $\id-\Cone$ can be singular when an unreachable component never terminates. For a run that enters and later exits a CQ-HB loop, the number of loop iterations is $n_{\mathrm c}+1$.

\subsection{Homogeneity and Uniform Support}\label{app:uniform-support}
\begin{lemma}[Shifted body]\label{lem:shifted-body}
Let every $h_i$ be homogeneous for $S$. For all retained $s,t$, homogeneous values $\bh,\bh'$, and partial density operators $\rho$,
\[
 \kappa^S_{(t,\bh'),(s,\bh)}=
 \begin{cases}\kappa^S_{(t,\bh'-\bh),(s,\zero)}&\text{if }\bh'\ge\bh\text{ componentwise},\\0&\text{otherwise,}\end{cases}
\]
and hence $\sem S(\rho\,\bZ^s\bY^\bh)=\sum_{t,\br}\mathcal B_{t,s,\br}(\rho)\,\bZ^t\bY^{\bh+\br}$, where $\mathcal B_{t,s,\br}=\kappa^S_{(t,\br),(s,\zero)}$.
\end{lemma}
\begin{proof}
Applying homogeneity $h_i$ times in each coordinate gives $\sem S(\rho\bZ^s\bY^\bh)=\bY^\bh\sem S(\rho\bZ^s\bY^\zero)$. Comparing coefficients gives the kernel identity on partial density operators, and linearity extends it to all operators.
\end{proof}

\begin{proof}[Proof of Proposition~\ref{prop:syntactic-hb}]
\emph{Homogeneity.} We show by induction on loop-free $S$ that $\sem S(X_iF)=X_i\sem S(F)$ for every homogeneous $h_i$. For $h_i:=h_i+r$, both sides add $r(s)$ to $h_i$, since $r$ reads only retained variables. All other primitive commands neither read nor write $h_i$, so they commute with the shift. For sequencing, apply both induction hypotheses. For a conditional, the guard does not read $h_i$, so $(X_iF)|_e=X_i(F|_e)$.

\emph{Uniform finite support.} A loop-free program with finitely supported sampling and finite measurement outcome sets has a finite execution tree from every classical state, so $\sem S(\rho\bX^\sigma)$ is supported on the finitely many classical states at the leaves of that tree, for every partial density operator $\rho$.
\end{proof}

\subsection{Summaries of CQ-HB Loops}\label{app:path-expansion}
\begin{proof}[Proof of Proposition~\ref{prop:summary-formal}]
\emph{The data form a loop summary.} By uniform finite support, $\mathcal B_{t,s,\br}=0$ unless $(t,\br)\in\Omega_s$, so only finitely many maps are nonzero, and $\Ex$ is finite. Each $\mathcal B_{t,s,\br}$ is CP by Proposition~\ref{prop:kernels}(1), so the coefficients of $\Cmap$ and $\Emap_u$ are blockwise CP. For $A\in\Vsp$ positive, Proposition~\ref{prop:kernels}(2) gives $\sum_{t,\br}\tr\mathcal B_{t,s,\br}(A_s)\le\tr A_s$ for each $s$, and summing over $s$ gives~\eqref{eq:subconservation}. The input polynomials have positive coefficients and total mass $\mass{F_{\mathrm{in}}}\le1$.

\emph{Path expansion.} Write $G_{\mathrm{act}}(\iota)=\sum_{t\in\Act}\bZ^t\iota_t(\bY)$ for the OFPS of a block vector of polynomials. By Lemma~\ref{lem:shifted-body}, for a positive polynomial block vector $\iota$,
\[
 \sem S\bigl(G_{\mathrm{act}}(\iota)\bigr)=G_{\mathrm{act}}\bigl(\Cmap(\bY)\iota\bigr)+\sum_{u\in\Ex}\bZ^u\,\Emap_u(\bY)\iota .
\]
By unfolding, $\sem{W_{j+1}}(F)=F|_{\neg e}+\sem{W_j}(\sem S(F|_e))$, and $F_{\mathrm{in}}|_e=G_{\mathrm{act}}(\iota)$ by guard independence. Induction on $j$, for arbitrary positive polynomial inputs $(\iota,\delta)$, gives for $j\ge1$
\[
 \sem{W_j}(F_{\mathrm{in}})=\sum_{u\in\Ex}\bZ^u\Bigl(\delta_u+\sum_{m=0}^{j-2}\Emap_u(\bY)\Cmap(\bY)^m\iota\Bigr),
\]
with an empty sum for $j=1$. Here the $m$-th term collects runs in which $m$ body executions return to an active state and the next body execution exits. The partial sums increase coefficientwise, and their limit is $\sem W(F_{\mathrm{in}})$ by the same unfolding characterization.
\end{proof}

\subsection{Terminal Specialization}\label{app:T-specialization}
Fix a loop summary with an input as in Definition~\ref{def:summary-formal}, and put $\mu_{\mathrm{in}}=\tr\iota(\one)+\sum_u\tr\delta_u(\one)\le1$. Let $T$ be a fresh indeterminate. The \emph{graded terminal series} is
\[
 H_u=\delta_u+\sum_{m\ge0}T^m\,\Emap_u(\bY)\Cmap(\bY)^m\iota(\bY)\ \in\ \Lin(\Hil)[\bY][[T]],
 \qquad \rho_{u,m,\bh}:=[T^m\bY^\bh]H_u\succeq0 .
\]
The coefficient of $T^m$ collects the runs that exit through $u$ after exactly $m$ continuing transitions. It is a polynomial in $\bY$, so no convergence question arises at this stage.

\begin{lemma}[The graded live series]\label{lem:T-solution}
In $\Vsp[\bY][[T]]$, the equation $\widehat L(T)=\iota+T\,\Cmap(\bY)\widehat L(T)$ has the unique solution $\widehat L(T)=\sum_mT^m\Cmap(\bY)^m\iota$, and $H_u=\delta_u+\Emap_u(\bY)\widehat L(T)$. In matrix coordinates, $\mathfrak d:=\det(\id-T\Cmap(\bY))\in\C[T,\bY]$ satisfies $\mathfrak d(0,\bY)=1$. Hence $\widehat L(T)$ is also the unique solution over $\C(T,\bY)$, and each matrix entry of $H_u$ equals $\mathfrak n/\mathfrak d$ for some $\mathfrak n\in\C[T,\bY]$, as formal power series in $(T,\bY)$.
\end{lemma}
\begin{proof}
Comparing coefficients gives $[T^0]\widehat L(T)=\iota$ and $[T^{m+1}]\widehat L(T)=\Cmap\bigl([T^m]\widehat L(T)\bigr)$, which proves existence and uniqueness in $\Vsp[\bY][[T]]$. Since $\mathfrak d(0,\bY)=1$, $\mathfrak d$ is a unit in $\C[[T,\bY]]$, and Cramer's rule gives the entries of $(\id-T\Cmap)^{-1}$ as polynomials divided by $\mathfrak d$; their expansions agree with $\sum_mT^m\Cmap^m$.
\end{proof}

Lemma~\ref{lem:T-solution} gives a rational representation with $T$ recording the number of continuing transitions. Recovering the terminal output requires summing over these counts, but the live series may have a pole at $T=1$. The next lemma shows that the total mass of the terminal coefficients is bounded and that this bound makes specialization at $T=1$ valid for the reduced entries of $H_u$.

\begin{lemma}[Terminal specialization]\label{lem:T-specialization}
For the loop summary and input fixed above, let $H_u$ be the graded terminal series with coefficients $\rho_{u,m,\bh}$ and input mass $\mu_{\mathrm{in}}$. The following hold without assuming almost-sure termination.
\begin{enumerate}
\item $\sum_{u,m,\bh}\tr \rho_{u,m,\bh}\le\mu_{\mathrm{in}}$. Consequently each coefficient $[\bY^\bh]F_u=\sum_m\rho_{u,m,\bh}$ of the terminal series is a trace-norm convergent sum of positive operators, and $\sum_u\mass{F_u}\le\mu_{\mathrm{in}}$.
\item Let $p/q$ be a matrix entry of $H_u$, written in lowest terms in $\C[T,\bY]$ (with $0/1$ for zero). Then $q(1,\bY)$ is a nonzero polynomial, and the corresponding entry of $F_u$ is a rational power series represented by $p(1,\bY)/q(1,\bY)$. Thus the terminal output is obtained by reducing each entry of $H_u$ and then setting $T=1$.
\end{enumerate}
\end{lemma}
\begin{proof}
(1) Put $\Cone=\Cmap(\one)$, $\Xone_u=\Emap_u(\one)$, and $\rho=\iota(\one)\succeq0$. Evaluating the polynomial $[T^m]H_u$ at $\bY=\one$ sums its coefficients, so $\sum_\bh \rho_{u,m,\bh}=\Xone_u\Cone^m\rho$ for $m\ge0$, plus $\delta_u(\one)$ when $m=0$. Applying~\eqref{eq:subconservation} to $\Cone^m\rho\succeq0$ gives $\sum_u\tr\Xone_u\Cone^m\rho\le\tr\Cone^m\rho-\tr\Cone^{m+1}\rho$. Summing over $m<n$ telescopes to at most $\tr\rho-\tr\Cone^n\rho\le\tr\rho$. Adding $\sum_u\tr\delta_u(\one)$ and letting $n\to\infty$ proves the bound. The remaining claims follow from Lemma~\ref{lem:positive}.

(2) Fix an entry and let $a_{m,\bh}$ be the corresponding entries of $\rho_{u,m,\bh}$, so that $|a_{m,\bh}|\le\tr \rho_{u,m,\bh}$. By (1), the series $h(t,\by)=\sum_{m,\bh}a_{m,\bh}t^m\by^\bh$ converges absolutely and uniformly on the closed unit polydisc $\overline{\mathbb D}^{1+d}$, with $|h|\le\mu_{\mathrm{in}}$. It is therefore continuous there and analytic inside. Near the origin, $h=\mathfrak n/\mathfrak d$ by Lemma~\ref{lem:T-solution}, so the analytic function $\mathfrak dh-\mathfrak n$ vanishes near the origin and hence on the connected open polydisc. By continuity, $\mathfrak dh=\mathfrak n$ on the closed polydisc. From $\mathfrak nq=p\mathfrak d$ we get $\mathfrak d(qh-p)=0$. Since $\mathfrak d$ is a nonzero polynomial, its zero set has empty interior, so $qh=p$ on the closed polydisc by continuity.

Suppose that $q(1,\bY)=0$. Then $T-1$ divides $q$, and since $p$ and $q$ are coprime and $T-1$ is irreducible, $p(1,\bY)\ne0$. A nonzero polynomial cannot vanish on the open box $(0,1)^d$, so choose $\by\in(0,1)^d$ with $p(1,\by)\ne0$; if $d=0$, there is nothing to choose. Then $0=q(1,\by)h(1,\by)=p(1,\by)\ne0$, a contradiction. Hence $q(1,\bY)\ne0$.

Let $G(\by)=h(1,\by)=\sum_\bh\bigl(\sum_ma_{m,\bh}\bigr)\by^\bh$, which is the corresponding entry of the generating function of $F_u$. It is analytic on the open unit polydisc and satisfies $q(1,\by)G(\by)=p(1,\by)$ there. By Corollary~\ref{cor:rational-analytic}, its Taylor series, the entry of $F_u$, is rational, with representation $p(1,\bY)/q(1,\bY)$.
\end{proof}

\begin{proof}[Proof of Theorem~\ref{thm:rational}]
By Lemma~\ref{lem:T-solution}, the system $(\id-T\Cmap)\widehat L(T)=\iota$ has a unique solution over $\C(T,\bY)$. Substituting this solution into $\delta_u+\Emap_u\widehat L(T)$ gives the graded terminal series $H_u$. By Lemma~\ref{lem:T-specialization}, the entries of $F_u=\sum_m\rho_{u,m,\cdot}$ are rational with representations $p(1,\bY)/q(1,\bY)$.
\end{proof}

\begin{proof}[Proof of rationality for CQ-HB loops]
Combine Proposition~\ref{prop:summary-formal} with Theorem~\ref{thm:rational}; the finitely many retained factors $\bZ^u$ preserve rationality.
\end{proof}

\begin{proof}[Proof of Corollary~\ref{cor:direct}]
By Lemma~\ref{lem:T-solution}, each entry of $H_u$ has the form $\mathfrak n/\mathfrak d$ with $\mathfrak n\in\C[T,\bY]$ and $\mathfrak d=\det(\id-T\Cmap(\bY))$. From $\mathfrak nq=p\mathfrak d$ we get $\mathfrak n(1,\cdot)\,q(1,\cdot)=p(1,\cdot)\,\mathfrak d(1,\cdot)$. Both $q(1,\cdot)$ and $\mathfrak d(1,\cdot)$ are nonzero, so $\mathfrak n(1,\cdot)/\mathfrak d(1,\cdot)=p(1,\cdot)/q(1,\cdot)$. If $1$ is not an eigenvalue of $\Cmap(\zero)$, then $\mathfrak d(1,\zero)=\det(\id-\Cmap(\zero))\ne0$.
\end{proof}

\paragraph{Specialization as a ring homomorphism.}
The rational functions $p/q$ whose reduced denominator is not divisible by $T-1$ form the local ring $\C[T,\bY]_{(T-1)}$, and $T\mapsto1$ is a ring homomorphism from it onto $\C(\bY)$. Lemma~\ref{lem:T-specialization} says that the entries of $H_u$ lie in this ring and that the homomorphism sends them to rational representations of the terminal series. Because the homomorphism is linear, specialization commutes with linear combinations of inputs. This is used in Appendix~\ref{app:full-transformer}.

\subsection{The Nilpotent Case}\label{app:nilpotent}
When $\Cmap_\zero=\Cmap(\zero)$ is nilpotent, the live series itself is a formal power series, and Theorem~\ref{thm:rational} has an elementary proof that avoids $T$.

\begin{proposition}\label{prop:nilpotent}
If $\Cmap_\zero^\nu=0$, then for each total degree $r$ only the powers $\Cmap(\bY)^m$ with $m<\nu(r+1)$ contribute to coefficients of total degree at most $r$. Hence $L=\sum_m\Cmap(\bY)^m\iota$ converges in $\Vsp[[\bY]]$, $(\id-\Cmap(\bY))L=\iota$, and $F_u=\delta_u+\Emap_u(\id-\Cmap)^{-1}\iota$ with $\det(\id-\Cmap(\zero))=1$.
\end{proposition}
\begin{proof}
Write $\Cmap=\Cmap_\zero+\Cmap_+$, where $\Cmap_+$ has only terms of positive degree. Expand $\Cmap^m$ into words in $\Cmap_\zero$ and $\Cmap_+$. A word that contributes in total degree at most $r$ contains at most $r$ letters $\Cmap_+$. They separate at most $r+1$ runs of $\Cmap_\zero$, and a run of length at least $\nu$ vanishes. The word therefore has length at most $r+(r+1)(\nu-1)=\nu(r+1)-1$. Thus each coefficient of $L$ is a finite sum, and multiplying by $\id-\Cmap$ telescopes coefficientwise. All eigenvalues of $\Cmap_\zero$ are zero, so $\det(\id-\Cmap_\zero)=1$.
\end{proof}
This argument gives the direct formula for the preceding examples that increment a counter in every round, for which $\Cmap_\zero=0$. Corollary~\ref{cor:direct} extends the formula to every summary with $\det(\id-\Cmap(\bY))\ne0$, in which case $L$ itself need not converge formally.

\subsection{Full Transformers and Transposed Systems}\label{app:full-transformer}
 The loop kernels are obtained from the formal series of linear maps $\mathcal W_u(T,\bY)=\sum_mT^m\Emap_u(\bY)\Cmap(\bY)^m:\Vsp\to\Lin(\Hil)[\bY][[T]]$. Splitting a run after its first continuing transition gives the first-step equation
\[
 \mathcal W_u=\Emap_u+T\,\mathcal W_u\,\Cmap,\qquad\text{so}\qquad\mathcal W_u=\Emap_u(\id-T\Cmap)^{-1}.
\]
The continuation acts first: $\mathcal W_u\Cmap$ means ``continue, then run the loop''. In matrix coordinates, with $\mathbf M$ representing $\Cmap$, $\mathbf B_u$ representing $\Emap_u$, and $\Psi_u$ representing $\mathcal W_u$, the forward system for a fixed input and the system for the transformer are
\[
 (I-T\mathbf M)\,\boldsymbol\ell=\vecop\iota,\qquad \Psi_u\,(I-T\mathbf M)=\mathbf B_u .
\]
Each row of the second system is solved with the transposed coefficient matrix $(I-T\mathbf M)^{\mathsf T}$. This is an ordinary coordinate transpose, not a quantum adjoint.

A matrix unit $\ket i\bra j$ placed at location $a$ need not be a state, but it is a linear combination of at most four partial density operators. The output of $\mathcal W_u$ on each of these is a graded terminal series, so specialization applies to each of them. By linearity of the homomorphism above, it also applies to $\mathcal W_u$. Setting $T=1$ in the reduced entries of $\mathcal W_u$ thus yields the terminal loop kernels. For CQ-HB loops, an input at homogeneous value $\bh$ multiplies them by $\bY^\bh$ (Lemma~\ref{lem:loop-homogeneous}), and an input outside $\Act$ exits unchanged. The retry loop's rational output is an instance of this construction.

\subsection{Exact Computation}\label{app:computation}
\begin{proof}[Proof of Theorem~\ref{thm:computation}]
Let $\mathbb K$ be the number field generated by the given algebraic numbers and their complex conjugates. Arithmetic in $\mathbb K$, and in polynomial rings and rational-function fields over $\mathbb K$, including greatest common divisors, is effective with standard exact encodings of algebraic numbers~\citep{Basu2006}.

\emph{Active set.} Evaluate the guard on the given finite superset of $\Act$, with homogeneous values zero.

\emph{Body maps.} For each $s\in\Act$, enumerate the execution tree of the loop-free body $S$ from $(s,\zero)$, evaluating expressions and guards on concrete classical states. Construct the superoperator matrix along each path directly over $\mathbb K$. A unitary or measurement outcome with operator $K$ contributes $K\otimes\overline K$, whose entries lie in $\mathbb K$ since the field contains the data and their complex conjugates. A reset has an integer matrix in the matrix-unit basis, and a sampling outcome of probability $p$ contributes $pI$. Multiply these matrices in execution order and sum over leaves with equal final classical state $(t,\br)$ to obtain the matrix of $\mathcal B_{t,s,\br}$ over $\mathbb K$. The tree is finite because $S$ is loop-free and all branchings are finite.

\emph{Solution.} Assemble $\Cmap(\bY)$, $\Emap_u(\bY)$, $\iota$, and $\delta_u$ as in Proposition~\ref{prop:summary-formal}. Solve $(I-T\mathbf M)\boldsymbol\ell=\vecop\iota$, a system of $|\Act|D^2$ linear equations over $\mathbb K(T,\bY)$, by Gaussian elimination. Form the entries of $\delta_u+\Emap_u\boldsymbol\ell$, reduce each to lowest terms, and substitute $T=1$. By Theorem~\ref{thm:rational} and Proposition~\ref{prop:summary-formal}, the results give rational representations of the components $F_u$, and $\sem W(F_{\mathrm{in}})=\sum_u\bZ^uF_u$. If $\det(I-T\mathbf M)|_{T=1}\ne0$, which can be tested exactly, Corollary~\ref{cor:direct} allows solving directly at $T=1$.
\end{proof}

\FloatBarrier
\section{Nested Loops via Control-Flow Summaries}\label{app:nested}
This appendix proves Theorem~\ref{thm:nested}. Instead of summarizing a loop body by its output, we keep every control location of the program, paired with a valuation of the retained variables. This follows the quantum flowchart and superoperator-valued transition-system view~\citep{Selinger2004,YingYingWu2017,Liu2025}, with monomials added to track classical increments. It gives a loop summary even when inner loops produce infinitely many outputs or diverge.

\paragraph{Assumptions.}
Let $S$ satisfy the hypotheses of Theorem~\ref{thm:nested}. The variables are partitioned into retained variables, with values $s$, and additive variables $\bh$. Every update of $h_i$ has the form $h_i:=h_i+r(s)$ with $r(s)\in\N$, and no other expression reads $\bh$. Sampling distributions have finite support, and measurement outcome sets are finite. Finite value sets for the retained variables are given, and the input has finite classical support and positive operator coefficients with algebraic entries and total trace at most one. Branch decisions depend only on retained values, while additive values may grow without bound.

\subsection{The Control-Flow Summary}
\paragraph{Construction.}
For a command $S'$ and a continuation location $k$, the fragment $\mathrm{Build}(S',k)$ is defined as follows.
\begin{enumerate}
\item A primitive command receives a fresh entry location with one edge to $k$ for each listed outcome: one edge for deterministic commands, one per value $v$ in the given list of sampling outcomes, and one per label for measurement.
\item For $S_1;S_2$, build $S_2$ with continuation $k$, then $S_1$ with continuation the entry of $S_2$.
\item For a conditional, build both branches with continuation $k$, and add a fresh guard location with a true edge to the first entry and a false edge to the second.
\item For a loop, create a fresh guard location $g$, build the body with continuation $g$, and add a true edge from $g$ to the body entry and a false edge from $g$ to $k$.
\end{enumerate}
The top-level continuation is a fresh terminal location. Next, pair locations with retained valuations. Starting from the retained valuations in the input support, follow every listed outcome, including outcomes of probability zero, evaluating retained updates and guards. The given value sets are accepted only if every reachable retained valuation lies in them; this closure check terminates because there are finitely many pairs. Let $\Lambda$ be the reachable pairs with a nonterminal location, and $\Ex$ the retained valuations reachable at the terminal location.

\paragraph{Edge labels.}
Each edge $e$ carries an additive increment $\br_e\in\N^d$ and a quantum operation $\Qop_e$:
\begin{itemize}
\item an additive assignment carries its evaluated increment and $\id$;
\item retained assignments and guard edges carry $\zero$ and $\id$;
\item the outcome $v$ of a sampling command carries $\zero$ and $\rho\mapsto\mu(v)\rho$;
\item the outcome $i$ of a measurement carries $\zero$ and $\rho\mapsto M_i\rho M_i^\dagger$;
\item a reset gives one edge labelled $(\zero,\Rmap_q)$, and a unitary gives one labelled $(\zero,U(\cdot)U^\dagger)$.
\end{itemize}
Distinct outcomes remain distinct edges even when they share a target. Edges between pairs in $\Lambda$ form the continuation, $\Cmap(\bY)=\sum_e\bY^{\br_e}\Qop_e$ placed in the corresponding block, and edges into the terminal location with retained valuation $u$ form the exit $\Emap_u(\bY)$. The initial input places each coefficient at the entry pair of its retained valuation, multiplied by its \emph{initial} additive monomial, and we take $\delta_u=0$. At each nonterminal pair, the operations on the enabled outgoing edges together preserve the trace, so
\begin{equation}\label{eq:graph-conservation}
 \tr\Cmap(\one)A+\sum_u\tr\Emap_u(\one)A=\tr A\qquad(A\in\Vsp,\ A\succeq0),
\end{equation}
and the data form a loop summary in the sense of Definition~\ref{def:summary-formal}. For a path $\pi=e_1\cdots e_m$, put $\br(\pi)=\sum_i\br_{e_i}$ and $\Qop_\pi=\Qop_{e_m}\circ\cdots\circ\Qop_{e_1}$.

\begin{lemma}[Path correspondence]\label{lem:cfg}
For every fragment $\mathrm{Build}(S',k)$, every retained valuation $s$, every additive value $\bh$, and every partial density operator $\rho$, the terminating executions of $S'$ from $(s,\bh,\rho)$ correspond bijectively to the graph paths from the entry of the fragment to their first arrival at $k$. Executions and paths whose quantum output is zero are omitted on both sides. The correspondence preserves the final retained valuation, the final additive value $\bh+\br(\pi)$, the quantum output $\Qop_\pi(\rho)$, and the sequence of observed outcomes. In particular, the top-level terminating executions from the input support correspond to the terminal paths of the control-flow summary.
\end{lemma}
\begin{proof}
We proceed by induction on $S'$, using the graph before restriction to reachable pairs.

\emph{Primitive commands.} Each listed outcome is one edge carrying exactly its classical state update and quantum operation. Its increment does not depend on $\bh$, since increments read only retained values.

\emph{Sequencing.} A terminating execution of $S_1;S_2$ splits uniquely into one of $S_1$ followed by one of $S_2$. By induction, these correspond to a path to the first arrival at the entry $k_1$ of $S_2$, followed by a path to the first arrival at $k$. Concatenation adds increments and composes the second operation after the first. Conversely, every such path splits uniquely at its first arrival at $k_1$.

\emph{Conditional.} The guard reads only retained values, so it takes the same value in the program and in the graph, and the enabled guard edge selects the same branch. The guard edge carries $\zero$ and $\id$.

\emph{Loop.} Proceed by a second induction on the number of completed body executions before the guard fails. With zero executions, the false edge leads directly to $k$. Otherwise, the true edge enters the body, whose returns to $g$, including those that pass through nested loops, are matched by the structural hypothesis, and the remaining executions are matched by the second induction hypothesis. Conversely, a path to $k$ splits uniquely at its successive first returns to $g$ and ends with the false edge, and it has finitely many returns. A body execution that never returns gives neither a terminating execution nor a path reaching $k$.

All steps hold for an arbitrary additive offset, and distinct outcome sequences stay distinct. Once a quantum output is zero, all later outputs are zero, so omitting such histories changes no sum. Finally, restriction to reachable pairs removes no path from the input support, because exploration follows every listed outcome and closure holds.
\end{proof}

\begin{proof}[Proof of Theorem~\ref{thm:nested}]
Summing the contributions matched by Lemma~\ref{lem:cfg} gives the OFPS semantics. For primitive commands, this follows from their semantic clauses; sequencing and conditionals use admissible sums and restriction; and loops take suprema of finite unfoldings. Hence the terminal series of the control-flow summary, with $u$ restored as $\bZ^u$, equals $\sem S(F_{\mathrm{in}})$. Its $m$-th term collects paths with $m$ continuing edges. The summary satisfies~\eqref{eq:graph-conservation}, so Theorem~\ref{thm:rational} gives rationality. Computability follows as in Appendix~\ref{app:computation}: the summary has entries in the number field of the data, and the system has $|\Lambda|D^2$ unknowns.
\end{proof}

Solving for a basis of quantum inputs at each entry pair yields the full terminal maps of $S$ on the explored retained valuations, exactly as in Appendix~\ref{app:full-transformer}. Further classical inputs must be included in the exploration before the closure check.

\paragraph{Why internal locations are kept.}
Replacing an inner loop by its terminating output would remove its divergent mass from the summary, and~\eqref{eq:graph-conservation} would become a strict inequality. Theorem~\ref{thm:rational} would still apply, but the summary would no longer be finite whenever the inner loop has infinitely many outputs. Keeping the locations makes the summary finite at the price of zero-shift cycles, which Theorem~\ref{thm:rational} handles.

\subsection{Example: a Diverging Inner Loop}
Let $\mu(0)=\mu(1)=\frac12$ and consider the program with one variable $b$:
\begin{equation}\label{prog:zero-shift}
\begin{aligned}
 \mathrm{Nested}\equiv{}&b\sample\mu;\\
 &\kw{while}\ b=0\ \kw{do}\\
 &\quad\kw{while}\ \mathrm{true}\ \kw{do}\ \kw{skip}\ \kw{od}\\
 &\kw{od}.
\end{aligned}
\end{equation}
From a density operator $\rho$, half of the mass exits with $b=1$, and the other half never leaves the inner loop. After contracting identity steps, the control-flow summary has two locations, the outer guard with $b=1$ and the inner loop. Its data are
\[
 \Cmap=\begin{pmatrix}0&0\\0&\id\end{pmatrix},\qquad
 \Emap_1=\begin{pmatrix}\id&0\end{pmatrix},\qquad
 \iota=(\rho/2,\rho/2).
\]
The live solution has second block $\rho/(2(1-T))$, which diverges at $T=1$, and $\id-\Cmap$ is singular. The graded terminal series is $H_1=\rho/2$, and specialization gives the correct output $B\rho/2$.

\FloatBarrier
\section{Termination and Moments}\label{app:moments}
This appendix proves Proposition~\ref{prop:moments} by studying evaluation at $\bY=\one$. The results require only finite-dimensional CP maps; their entries need not be algebraic. Throughout, fix a loop summary with an input, and put $\Cone=\Cmap(\one)$, $\Xone_u=\Emap_u(\one)$, and $\rho=\iota(\one)\succeq0$. For a block observable $O\in\Vsp$, write $\langle O,A\rangle=\sum_s\tr(O_sA_s)$.

\subsection{Survival and the Reachable Space}\label{app:transience}
Spectral methods for quantum termination are developed by \citet{YingFeng2010,Ying2013,Liu2025}. Here we apply them to the reachable block space of a loop summary.
We consider the runs that start from $\rho$; the direct exits $\delta_u$ play no role here. Let $p_{u,m}=\tr\Xone_u\Cone^m\rho$ be the probability of exiting through $u$ after exactly $m$ continuing transitions, and let $\varsigma_m=\tr\Cone^m\rho$. Suppose, in this subsection, that~\eqref{eq:subconservation} holds with equality, as it does for mass-preserving loop-free bodies and for control-flow summaries~\eqref{eq:graph-conservation}, and that $\tr\rho=1$. Then $\varsigma_m-\varsigma_{m+1}=\sum_up_{u,m}$, and the nontermination probability is $p_\infty=\lim_m\varsigma_m$. The number $n_{\mathrm c}$ of continuing transitions, set to $\infty$ on divergence, satisfies
\begin{equation}\label{eq:tail}
 \Prb(n_{\mathrm c}=m)=\sum_up_{u,m},\qquad\Prb(n_{\mathrm c}\ge m)=\varsigma_m .
\end{equation}
For a linear map $\mathcal M$ on a finite-dimensional space, $\spr(\mathcal M)$ denotes its spectral radius. The \emph{reachable space} of $\rho$ is
\begin{equation}\label{eq:reachability}
 \Hil_{\rho,s}=\spanop\bigcup_{m\ge0}\supp\bigl((\Cone^m\rho)_s\bigr),
 \qquad
 \Vsp_\rho=\bigoplus_{s\in\Lambda}\Lin(\Hil_{\rho,s}),
\end{equation}
with operators extended by zero. Each continuing transition maps $\Vsp_\rho$ into itself. Indeed, if $K$ is a Kraus operator of a transition from location $s$ to location $t$, then $K(\Cone^m\rho)_sK^\dagger\preceq(\Cone^{m+1}\rho)_t$, so $K$ maps $\supp(\Cone^m\rho)_s$ into $\supp(\Cone^{m+1}\rho)_t$. Taking spans shows that $K\Hil_{\rho,s}\subseteq\Hil_{\rho,t}$, and hence $K\Lin(\Hil_{\rho,s})K^\dagger\subseteq\Lin(\Hil_{\rho,t})$.

\begin{theorem}[Transience on the reachable space]\label{thm:transience}
Assume equality in~\eqref{eq:subconservation} and $\tr\rho=1$. The following are equivalent: (1)~the summary terminates almost surely from $\rho$; (2)~$\varsigma_m\to0$; (3)~$\spr(\Cone|_{\Vsp_\rho})<1$. In that case there are $c<\infty$ and $\lambda<1$ with $\varsigma_m\le c\lambda^m$, and $(\id-t\Cone)^{-1}\rho$ is analytic near $t=1$. More generally, $(\id-\Cmap(\by))^{-1}$ restricted to $\Vsp_\rho$ is analytic near $\by=\one$ and equals the weighted path sum there.
\end{theorem}
\begin{proof}
The equivalence (1)$\Leftrightarrow$(2) follows from~\eqref{eq:tail}.

For (2)$\Rightarrow$(3), finitely many supports span each finite-dimensional $\Hil_{\rho,s}$, so for some $r$ the operator $\omega=\sum_{i\le r}\Cone^i\rho$ is positive definite on every nonzero block of $\Vsp_\rho$. Every positive $A\in\Vsp_\rho$ satisfies $A\preceq c_A\omega$ for some $c_A$, hence $\tr\Cone^mA\le c_A\sum_{i\le r}\varsigma_{m+i}\to0$. Positive operators span $\Vsp_\rho$, so $\Cone^m|_{\Vsp_\rho}\to0$, which for a finite-dimensional map is equivalent to $\spr<1$.

For (3)$\Rightarrow$(2), the spectral-radius condition gives $\Cone^m\rho\to0$. For any $\lambda$ with $\spr(\Cone|_{\Vsp_\rho})<\lambda<1$, there is a constant $c$ with $\|\Cone^m|_{\Vsp_\rho}\|\le c\lambda^m$~\citep{HornJohnson2012}. This gives the geometric bound and analyticity of the resolvent near $t=1$.

For the weighted statement, every weighted transition preserves $\Vsp_\rho$, and $\det(\id-\Cmap(\by))|_{\Vsp_\rho}\ne0$ near $\by=\one$. Let $a(\by)$ be the maximum of $1$ and of $\prod_i\max(1,|y_i|)^{(\br_e)_i}$ over the finitely many continuing transitions $e$. Expanding $\Cmap(\by)^m\rho$ along branches gives the trace-norm majorant $a(\by)^m\varsigma_m$. Since $a(\one)=1$, the series $\sum_m\Cmap(\by)^m\rho$ converges locally uniformly near $\one$ and equals the resolvent there.
\end{proof}

A component that is unreachable from $\rho$ may carry a nonterminating evolution and make $\id-\Cone$ singular on all of $\Vsp$, without affecting the input $\rho$. If an inner loop is summarized by its terminating output, the decrease in live mass can include mass lost to divergence inside that loop. The control-flow summary in Appendix~\ref{app:nested} retains the inner locations.

\subsection{Terminal Moments without Almost-Sure Termination}\label{app:terminal-regularity}
\begin{proposition}[Terminal decay]\label{prop:terminal-regularity}
Assume only~\eqref{eq:subconservation}. The function $f_u(t)=\sum_{m\ge0}t^m\Xone_u\Cone^m\rho$ is analytic on $|t|<R$ for some $R>1$, and $\tr\Xone_u\Cone^m\rho\le c\lambda^m$ for some $c<\infty$ and $\lambda<1$.
\end{proposition}
\begin{proof}
By telescoping (Lemma~\ref{lem:T-specialization}(1)), $\sum_{m,u}\tr\Xone_u\Cone^m\rho\le\tr\rho$, so each matrix entry of $f_u$ is bounded on the unit disc and its coefficients are absolutely summable. By Cramer's rule, $f_u(t)=\Xone_u(\id-t\Cone)^{-1}\rho$ is a rational function of $t$. After cancellation, a pole inside the open disc would contradict analyticity there, and a pole on the unit circle would make the entry unbounded along the radius approaching it, contradicting the uniform bound. Hence all poles lie outside the closed disc, and Cauchy's estimates give geometric decay.
\end{proof}

\begin{proof}[Proof of Proposition~\ref{prop:moments}(1)]
Let $c_{\max}$ bound the total increment $\sum_i(\br_e)_i$ of every transition, and let $h_0$ bound the total degree of the input polynomials. A run with $m$ continuing transitions and one exit has a final homogeneous value of total size at most $h_0+(m+1)c_{\max}$. For $|y_i|\le\varrho$ with $\varrho\ge1$, the generating function of $F_u$ is therefore majorized in trace norm by $\sum_m\varrho^{h_0+(m+1)c_{\max}}\,\tr\Xone_u\Cone^m\rho+\varrho^{h_0}\tr\delta_u(\one)$. By Proposition~\ref{prop:terminal-regularity}, this converges whenever $\varrho^{c_{\max}}\lambda<1$. Hence $f_{F_u}(\by)$ is analytic on a polydisc of radius greater than one, it agrees there with the rational function of Theorem~\ref{thm:rational}, and its derivatives at $\one$ are the factorial submoments.
\end{proof}

Proposition~\ref{prop:moments}(1) does not make the extended count $n_{\mathrm c}$ of~\eqref{eq:tail} integrable when $p_\infty>0$. For example, with $\Cone(\rho)=\proj1\rho\proj1$, $\Xone_0(\rho)=\proj0\rho\proj0$, and $\rho=\proj+$, the terminal function is $\frac12\proj0$ and every positive-order terminal submoment is zero, yet $\Prb(n_{\mathrm c}=\infty)=\frac12$ and $\E n_{\mathrm c}=\infty$.

\subsection{Moment Recurrences}\label{app:moment-recurrences}
\begin{proof}[Proof of Proposition~\ref{prop:moments}(2)]
For $\rho=0$, the reachable space is zero and every contribution from it vanishes. For $\rho\ne0$, normalize $\rho$ by its trace to apply Theorem~\ref{thm:transience}. This leaves $\Vsp_\rho$ unchanged and gives $\spr(\Cone|_{\Vsp_\rho})<1$. Use the original $\rho$ in $\ell(t)=(\id-t\Cone)^{-1}\rho$ on $\Vsp_\rho$ and set $L_r=\ell^{(r)}(1)$. Differentiating $\ell=\rho+t\Cone\ell$ gives
\begin{equation}\label{eq:factorial-rec}
 (\id-\Cone)L_0=\rho,\qquad(\id-\Cone)L_r=r\,\Cone(L_{r-1})\quad(r\ge1).
\end{equation}
Since $\sum_mt^m\Xone_u\Cone^m\rho=\Xone_u\ell(t)$, the $r$-th factorial submoment at exit $u$ is $\tr\Xone_u(L_r)$.
\end{proof}
Conditional moments follow by dividing by $p_u=\tr\Xone_u(L_0)$, provided $p_u>0$. For instance,
\[
 \E[n_{\mathrm c}\mid u]=\frac{\tr\Xone_u(L_1)}{p_u},\qquad
 \Var(n_{\mathrm c}\mid u)=\frac{\tr\Xone_u(L_2+L_1)}{p_u}-\Bigl(\frac{\tr\Xone_u(L_1)}{p_u}\Bigr)^2 .
\]
All systems share the matrix $\id-\Cone$, so the same factorization can be reused for every order.

For several homogeneous variables, let $\Cone_{[\mathbf j]}=(\partial^{\mathbf j}\Cmap)(\one)$, $\Xone_{u,[\mathbf j]}=(\partial^{\mathbf j}\Emap_u)(\one)$, $\rho_{\mathbf i}=(\partial^{\mathbf i}\iota)(\one)$, $\delta_{u,\mathbf i}=(\partial^{\mathbf i}\delta_u)(\one)$, and $V_{\mathbf i}=(\partial^{\mathbf i}f_L)(\one)$, where $f_L(\by)=(\id-\Cmap(\by))^{-1}\iota(\by)$ is the generating function of the live series on $\Vsp_\rho$, analytic near $\one$ by Theorem~\ref{thm:transience}. Multi-indices $\mathbf i,\mathbf j\in\N^d$ are compared componentwise, $\binom{\mathbf i}{\mathbf j}=\prod_k\binom{i_k}{j_k}$, and $\Cone_{[\zero]}=\Cmap(\one)=\Cone$ is the unweighted continuation. The Leibniz rule applied to $f_L=\iota+\Cmap f_L$ and to $f_{F_u}=\delta_u+\Emap_u f_L$ gives
\begin{align}
 (\id-\Cone)V_{\mathbf i}&=\rho_{\mathbf i}+\sum_{\zero\ne\mathbf j\le\mathbf i}\binom{\mathbf i}{\mathbf j}\Cone_{[\mathbf j]}(V_{\mathbf i-\mathbf j}),\label{eq:mixed-live}\\
 (\partial^{\mathbf i}f_{F_u})(\one)&=\delta_{u,\mathbf i}+\sum_{\mathbf j\le\mathbf i}\binom{\mathbf i}{\mathbf j}\Xone_{u,[\mathbf j]}(V_{\mathbf i-\mathbf j}).\label{eq:mixed-terminal}
\end{align}
These equations account for initial offsets (through $\rho_{\mathbf i}$ and $\delta_{u,\mathbf i}$) and for increments on exit edges (through $\Xone_{u,[\mathbf j]}$). Without almost-sure termination, Proposition~\ref{prop:moments}(1) still allows differentiating the \emph{terminal} rational function, but the live system~\eqref{eq:mixed-live} may be singular.

\subsection{Adjoint Equations and Certificates}
On the reachable space, let $I$ denote the block identity and define $R(t)=\sum_mt^m(\Cone^\dagger)^m(I)$, so that $h(t):=\tr\ell(t)=\langle R(t),\rho\rangle$. With $g(t)=\sum_u\tr f_u(t)$, conservation (equality in~\eqref{eq:subconservation}) gives the identity
\[
 t\,g(t)=1+(t-1)\,h(t)
\]
for $\tr\rho=1$, obtained by multiplying $\sum_up_{u,m}=\varsigma_m-\varsigma_{m+1}$ by $t^{m+1}$ and summing. Under transience, $R_0=R(1)$ solves $(\id-\Cone^\dagger)R_0=I$, and $R_r=R^{(r)}(1)$ solves $(\id-\Cone^\dagger)R_r=r\Cone^\dagger(R_{r-1})$. Hence
\[
 \E n_{\mathrm c}=\langle R_0,\rho\rangle-1,\qquad\E n_{\mathrm c}^2=2\langle R_1,\rho\rangle-\langle R_0,\rho\rangle+1 .
\]
These are the adjoint equations for the same path sums as~\eqref{eq:factorial-rec}, and they are the runtime-observable equations of \citet{Liu2025} specialized to loop summaries.

Without transience, $\Cone^\dagger(I)\preceq I$, and the decreasing sequence $(\Cone^\dagger)^m(I)$ converges to an effect $Q_\infty$ with $\Cone^\dagger(Q_\infty)=Q_\infty$. Under conservation, $\langle Q_\infty,\rho\rangle$ is the nontermination probability. Every fixed effect lies below every iterate, so $Q_\infty$ is the greatest fixed effect. A positive certificate $\Xi\succeq\rho+\Cone(\Xi)$ bounds every partial live sum. It therefore implies $\sum_m\varsigma_m<\infty$ and, under conservation, almost-sure termination. Such certificates avoid an explicit spectral computation.

\FloatBarrier
\section{Proofs for Superinvariants}\label{app:superinvariants}

\subsection{The Superinvariant Rule}\label{app:superinv-rule}
The proof is the fixed-point induction argument behind the probabilistic superinvariant rule~\citep{Klinkenberg2021}; its order-theoretic basis is standard~\citep{AbramskyJung1994}.
\begin{proof}[Proof of Theorem~\ref{thm:superinvariant}]
Let $\psi_m=\sem{W_m}=\Phi^m_{e,S}(0)$ (Appendix~\ref{app:well-defined}). Then $\psi_0(F)=0\sqsubseteq\chi(F)$, because $\chi(F)$ has positive coefficients. If $\psi_m(G)\sqsubseteq\chi(G)$ for every $G\in\OFPS{\Hil}{\bX}$, then, applying this to $G=\sem S(F|_e)$ and using the superinvariant condition,
\[
 \psi_{m+1}(F)=F|_{\neg e}+\psi_m\bigl(\sem S(F|_e)\bigr)\sqsubseteq F|_{\neg e}+\chi\bigl(\sem S(F|_e)\bigr)\sqsubseteq\chi(F).
\]
Each coefficient of $\sem W(F)$ is the limit of the increasing coefficients of $\psi_m(F)$, all bounded by the corresponding coefficient of $\chi(F)$, so $\sem W(F)\sqsubseteq\chi(F)$. The argument uses neither linearity nor continuity of $\chi$.
\end{proof}

\subsection{The Positivity Criterion}\label{app:superinv}
\begin{proof}[Proof of Theorem~\ref{thm:local-check}]
\emph{The slack maps are well defined.} Fix $\sigma\models e$ and $\tau$. For a partial density operator $\rho$, the series $\sem S(\rho\bX^\sigma)=\sum_\mu\kappa^S_{\mu,\sigma}(\rho)\bX^\mu$ is admissible. By the convergence requirement in Definition~\ref{def:bound-maps}, $\sum_\mu\mathcal V_{\tau,\mu}(\kappa^S_{\mu,\sigma}(\rho))$ converges, and it is a sum of positive operators. A finite spanning family of partial density operators (Appendix~\ref{app:kernels}) shows that the partial sums of $\sum_\mu\mathcal V_{\tau,\mu}\circ\kappa^S_{\mu,\sigma}$ converge as linear maps. Hence $\Theta_{\tau,\sigma}$ is a well-defined linear map.

\emph{Elementary inputs.} For $F=\rho\bX^\sigma$ with $\rho$ a partial density operator, the coefficient of the left-hand side of~\eqref{eq:superinvariant} at $\bX^\tau$ is
\[
 \iv{\sigma\not\models e}\iv{\tau=\sigma}\rho
 +\iv{\sigma\models e}\sum_\mu\mathcal V_{\tau,\mu}\bigl(\kappa^S_{\mu,\sigma}(\rho)\bigr).
\]
The right-hand coefficient is $\mathcal V_{\tau,\sigma}(\rho)$, so their difference is exactly $\Theta_{\tau,\sigma}(\rho)$. By scaling, the inequality on all elementary inputs is equivalent to positivity of every slack map.

\emph{General inputs.} Let $F=\sum_\sigma \rho_\sigma\bX^\sigma\in\OFPS{\Hil}{\bX}$. The coefficient at $\tau$ on each side of the superinvariant inequality is a sum, over $\sigma$, of the corresponding elementary contributions. For the left-hand side, this follows from Proposition~\ref{prop:kernels}(3) by interchanging the sums of positive operators over $\sigma$ and $\mu$. The result is finite by the convergence requirement applied to the admissible input $\sem S(F|_e)$. Summing the elementary inequalities gives the general inequality. Conversely, the general inequality applies in particular to elementary inputs.

\end{proof}

\subsection{The Retry Certificate}\label{app:retry-cert}
We prove the uniform retry bound quoted in Section~\ref{sec:superinv}. For a parameterized version of~\cref{prog:retry}, the operators are $J=\mathrm{diag}(\sqrt{1-\alpha},\sqrt{1-\beta})$ (exit) and $K=\mathrm{diag}(\sqrt\alpha,\sqrt\beta)$ (continuation), with $0\le\alpha,\beta\le1$.
Classical states are pairs $(b,n)$, with $b=1$ indicating continuation. Fix $r$ with $\max(\alpha,\beta)\le r<1$. Let $\Upsilon(\rho)=\tr(\rho)I$, $\mathcal J(\rho)=J\rho J^\dagger$, and $\mathcal K(\rho)=K\rho K^\dagger$. The body maps an active input $\rho\,BN^h$ to $\mathcal J(\rho)N^{h+1}+\mathcal K(\rho)BN^{h+1}$, so $\kappa^S_{(0,h+1),(1,h)}=\mathcal J$ and $\kappa^S_{(1,h+1),(1,h)}=\mathcal K$, and all other kernels from active states vanish. For a constant $c\ge1$, consider the candidate
\[
 \mathcal V_{(0,h+m),(1,h)}=c\,r^{m-1}\Upsilon\quad(m\ge1),\qquad
 \mathcal V_{\sigma,\sigma}=\id\quad(\sigma=(b,n),\ b\ne1),
\]
and zero otherwise. For an output state $(0,n')$, only finitely many active inputs $(1,h)$ with $h<n'$ and the single inactive input $(0,n')$ contribute, so the convergence requirement of Definition~\ref{def:bound-maps} holds. On an active input of mass one, the candidate has total trace $2c/(1-r)$, which may exceed one.

\paragraph{Slack maps.}
For an inactive input $\sigma$, the slack is $\iv{\tau=\sigma}(\id-\id)=0$. For an active input $(1,h)$, the slack is zero except at the output states $(0,h+m)$:
\[
 \Theta_{(0,h+1),(1,h)}=c\,\Upsilon-\mathcal J,
 \qquad
 \Theta_{(0,h+m),(1,h)}=c\,r^{m-2}\bigl(r\Upsilon-\Upsilon\circ\mathcal K\bigr)\quad(m\ge2).
\]
When $r=0$, we have $K=0$ and use the convention $0^0=1$.

\paragraph{Direct positivity proof ($c=1$).}
For $\rho\succeq0$, the operator $J\rho J^\dagger$ is positive, so its largest eigenvalue is at most its trace, and $J\rho J^\dagger\preceq\tr(J\rho J^\dagger)I=\tr(J^\dagger J\rho)I\preceq\tr(\rho)I$ because $J^\dagger J\preceq I$. Hence $\Upsilon-\mathcal J$ is positive. Moreover, $(r\Upsilon-\Upsilon\circ\mathcal K)(\rho)=\tr\bigl((rI-K^\dagger K)\rho\bigr)I\succeq0$ because $K^\dagger K=\mathrm{diag}(\alpha,\beta)\preceq rI$. Both families are positive for all $h$ and $m$, so the positivity criterion and superinvariant rule give $\sem W(\rho\,BN^0)\sqsubseteq\sum_{m\ge1}r^{m-1}\tr(\rho)I\,N^m$. Taking traces, the loop needs exactly $m$ attempts with probability at most $2r^{m-1}\tr \rho$.

For $\alpha=0.8$, $\beta=0.2$, $r=0.8$, and normalized input, this gives the probability bound $2(0.8)^{m-1}$ quoted in the main text.

\paragraph{Choi certificates ($c=2$).}
Since $J$ is diagonal, the Choi matrix of $\mathcal J$ is $\vecop(J)\vecop(J)^\dagger$, and that of $\Upsilon$ is $I_4$. Hence $\Choi(c\Upsilon-\mathcal J)=cI_4-\vecop(J)\vecop(J)^\dagger$, with eigenvalues $c,c,c$ and $c-\|\vecop J\|^2=c-(2-\alpha-\beta)$. For $c=1$ the last eigenvalue is $\alpha+\beta-1$, which is negative when $\alpha+\beta<1$: the first slack map is then positive but not completely positive, and the Choi check fails although the candidate is a superinvariant. For $c=2$ the eigenvalue is $\alpha+\beta\ge0$. For the second family, $\Choi(r\Upsilon-\Upsilon\circ\mathcal K)=(rI-K^\dagger K)^{\mathsf T}\otimes I=\operatorname{diag}(r-\alpha,r-\beta)\otimes I\succeq0$. With $c=2$, both symbolic families have positive semidefinite Choi matrices for all $h$ and $m$, which mechanically certifies the weaker bound $2\sum_{m\ge1}r^{m-1}\tr(\rho)I\,N^m$.

\subsection{The Sequential Readout Test}\label{app:backlog}
We verify the candidate for~\cref{prog:backlog} and justify the exact values in Example~\ref{ex:backlog}. Use the diagonal operators $J$ and $K$ from Appendix~\ref{app:retry-cert}, and write $\mathcal J(\rho)=J\rho J^\dagger$ and $\mathcal K(\rho)=K\rho K^\dagger$. The maps $\mathcal J$ and $\mathcal K$ multiply entry $(a,b)$ by $j_{ab}$ and $k_{ab}$, respectively:
\[
 \begin{aligned}
 (j_{00},k_{00})&=(1-\alpha,\alpha),\qquad
 (j_{11},k_{11})=(1-\beta,\beta),\\
 (j_{01},k_{01})&=(j_{10},k_{10})
 =\bigl(\sqrt{(1-\alpha)(1-\beta)},\sqrt{\alpha\beta}\bigr).
 \end{aligned}
\]

\paragraph{The candidate.}
Let
\[
 \lambda_{ab}(N)=\sum_{m\ge0}C_m\,j_{ab}^{m+1}k_{ab}^m N^{2m+1},
\]
where $C_m$ are the Catalan numbers~\citep{Stanley2015}. This is the unique formal series satisfying $\lambda_{ab}=j_{ab}N+k_{ab}N\lambda_{ab}^2$. Its closed form is
\[
 \lambda_{ab}(N)=\frac{1-\sqrt{1-4j_{ab}k_{ab}N^2}}{2k_{ab}N}
\]
when $k_{ab}>0$, and $\lambda_{ab}(N)=j_{ab}N$ when $k_{ab}=0$.

To include every source variable, write classical states in the order $(x,b,n)$. For an active source $(x,v,h)$, define
\[
 \mathcal V_{(0,0,h+m),\,(x,v,h)}(\rho)
 =\bigl([N^m]\lambda_{ab}(N)^x\,\rho_{ab}\bigr)_{a,b=0}^1
 \qquad(x\ge1,\ m\ge0).
\]
For a source with $x=0$, set $\mathcal V_{\sigma,\sigma}=\id$, and set all other maps to zero. Thus initial counter values shift the output by $N^h$, the initial measurement flag is overwritten on active inputs, and every terminating active execution ends with $x=b=0$.

\paragraph{Positivity, convergence, and exactness.}
A terminating execution from $x\ge1$ is a path of unit up and down steps from $x$ to $0$, staying positive until the final step. Each step contributes $N$ and applies $\mathcal J$ (down) or $\mathcal K$ (up). Because $J$ and $K$ are diagonal, a path with $d$ down steps and $u$ up steps multiplies entry $(a,b)$ by $j_{ab}^dk_{ab}^uN^{d+u}$. Catalan counting gives $\lambda_{ab}$ for a first passage down one level; concatenating $x$ such passages gives $\lambda_{ab}^x$.

Consequently, each bound map above is the sum of the CP maps of the paths reaching its target. The total trace of all terminal paths is at most the input trace. For any admissible input, summing these contributions at a fixed target therefore converges, as required by Definition~\ref{def:bound-maps}. Thus the candidate maps each admissible input $F$ to $\sem W(F)$.

\paragraph{The superinvariant check.}
On $x=0$, the candidate returns the input unchanged, so the slack is zero. For $x\ge1$, one body execution followed by the candidate multiplies entry $(a,b)$ by
\[
 j_{ab}N\lambda_{ab}^{x-1}+k_{ab}N\lambda_{ab}^{x+1}
 =\lambda_{ab}^{x-1}(j_{ab}N+k_{ab}N\lambda_{ab}^2)
 =\lambda_{ab}^x.
\]
When $x=1$, the down step exits with $b=0$, and the candidate is the identity there, consistent with $\lambda_{ab}^0=1$. Hence every slack map vanishes. The positivity criterion and superinvariant rule certify the output upper bound; equality is supplied separately by the path argument above.

\paragraph{Values used in the main text.}
For $\alpha=0.8$ and $\beta=0.2$,
\[
 \lambda_{00}(1)=\tfrac14,\qquad
 \lambda_{11}(1)=1,\qquad
 \lambda_{01}(1)=\lambda_{10}(1)=\tfrac12,\qquad
 \lambda_{11}'(1)=\tfrac53.
\]
Thus the termination probabilities from $\ket0$ and $\ket1$ at margin $x$ are $4^{-x}$ and $1$, respectively, and the mean number of rounds on $\ket1$ is $5x/3$. The upper bound alone proves a nontermination probability of at least $1-4^{-x}$ on $\ket0$; exactness makes this an equality. For input $\proj+$ at $x=1$ and $n=0$, summing over the final counter gives
\[
 \begin{pmatrix}1/8&1/4\\1/4&1/2\end{pmatrix}
 =\frac58\proj\psi,\qquad
 \ket\psi=\sqrt{0.2}\ket0+\sqrt{0.8}\ket1.
\]
Finally, $\lambda_{00}(N)$ is not rational: its closed form contains $\sqrt{1-0.64N^2}$, which cannot be rational because $1-0.64N^2$ has simple roots. The output series for this input is therefore algebraic but not rational.

\FloatBarrier
\section{Calculations for the Examples}\label{app:examples}
This appendix calculates the Bernoulli-factory and quantum-walk examples.

\subsection{The Bernoulli-Factory Subroutine}\label{app:qbf}
In the ordered basis $(\ket{00},\ket{01},\ket{10},\ket{11})$ of $q_1q_2$, let
\[
 U=\frac1{\sqrt2}\begin{pmatrix}1&0&0&-1\\1&0&0&1\\0&1&1&0\\0&1&-1&0\end{pmatrix},
 \qquad
 V_p=\begin{pmatrix}\sqrt p&-\sqrt{1-p}\\\sqrt{1-p}&\sqrt p\end{pmatrix}.
\]
The rows of $U$ are orthonormal, and $V_p\ket0=\ket p$. Direct multiplication gives
\[
 U\ket p\ket p=\tfrac{2p-1}{\sqrt2}\ket{00}+\tfrac1{\sqrt2}\ket{01}+\sqrt{2p(1-p)}\ket{10}
 =\tfrac1{\sqrt2}\bigl(\ket{\phi_p}\ket0+\ket0\ket1\bigr),
\]
and $\|\phi_p\|^2=(2p-1)^2+4p(1-p)=1$ for all $p\in[0,1]$. Because the body resets both qubits, on any partial density operator $\rho$ its branch maps are
\[
 \Emap_0(\rho)=\tfrac12\tr(\rho)\,\bigl(\proj{\phi_p}\otimes\proj0\bigr),\qquad\Cmap(\rho)=\tfrac12\tr(\rho)\proj{01},
\]
each multiplied by $N$ in the summary. The initialization produces $\proj{11}$ with $b=1$ and $n=0$. Corollary~\ref{cor:direct} gives $F(N)=N/(2-N)\cdot\bigl(\proj{\phi_p}\otimes\proj0\bigr)$. The scalar generating function $g(z)=z/(2-z)$ has factorial moments $g^{(r)}(1)=2\,r!$ for $r\ge1$. Expanding powers into falling factorials gives the ordinary moments in Table~\ref{tab:qbf-moments}. Initialization performs two resets, two $\NOT$ gates, and one measurement; each iteration performs two resets, three unitaries, and one measurement. Hence $\ops=5+6n$, and its expectation $17$ coincides with the expected runtime reported by \citet{Liu2025}. Charging one more unit for each guard test gives $6+7n$ instead. Changing the cost model changes the moments of the cost, but not the law of $n$.

\begin{table}[!htbp]
\caption{Ordinary moments of the iteration count $n$ and the operation count $\ops=5+6n$ of the Bernoulli-factory subroutine.}\label{tab:qbf-moments}
\centering\small
\setlength{\tabcolsep}{8pt}
\begin{tabular}{lrrrrr}
\toprule
 & $\E[\cdot]$ & $\E[(\cdot)^2]$ & $\E[(\cdot)^3]$ & $\E[(\cdot)^4]$ & $\Var(\cdot)$\\
\midrule
$n$ & $2$ & $6$ & $26$ & $150$ & $2$\\
$\ops$ & $17$ & $361$ & $9881$ & $345745$ & $72$\\
\bottomrule
\end{tabular}
\end{table}

\FloatBarrier
\subsection{Absorption for Quantum Walks}\label{app:walk-absorption}
On positions $0,\ldots,a$ ($a\ge2$), let $U=V(U_c\otimes I)$, where $U_c$ is the coin unitary and $V\ket{L,x}=\ket{L,x-1}$, $V\ket{R,x}=\ket{R,x+1}$, modulo $a+1$. Let $P_\partial$ project onto positions $0,a$, $P_{\mathrm{int}}=I-P_\partial$, and $\Gamma=UP_{\mathrm{int}}$. Since $\Gamma^\dagger\Gamma=P_{\mathrm{int}}$, $\Gamma$ is a contraction. A run passes its $m$-th guard test exactly when the state $\Gamma^{m-1}\rho\,\Gamma^{\dagger(m-1)}$ is found in the interior, so the probability of performing at least $m$ steps is $\tr(\Gamma^m\rho\,\Gamma^{\dagger m})$.

\paragraph{Two equivalent forms.}
Expanding the terminal path sum, the coefficient of $z^m$ for $m\ge1$ is
\[
 P_\partial U(P_{\mathrm{int}}U)^{m-1}P_{\mathrm{int}}\rho P_{\mathrm{int}}(U^\dagger P_{\mathrm{int}})^{m-1}U^\dagger P_\partial=P_\partial\Gamma^m\rho\,\Gamma^{\dagger m}P_\partial .
\]
Hence $f_\rho(z)=\sum_{m\ge0}z^mP_\partial\Gamma^m\rho\,\Gamma^{\dagger m}P_\partial$. Here $\Gamma$ records each step before the next boundary measurement. This matches the loop structure of \citet{Liu2025}, in which the boundary measurement is the loop guard.

\paragraph{Rational solution for a general walk.}
Write $A=P_{\mathrm{int}}UP_{\mathrm{int}}$ and $B=P_\partial UP_{\mathrm{int}}$ as matrices from the interior to the interior and boundary spaces, respectively. With $\rho_{\mathrm{int}}=P_{\mathrm{int}}\rho P_{\mathrm{int}}$, the coefficient of $z^m$ for $m\ge1$ is $BA^{m-1}\rho_{\mathrm{int}}(A^\dagger)^{m-1}B^\dagger$. Set $d=2(a-1)$ and $K=A\otimes\overline A$. Using the matrix-entry coordinates of Section~\ref{sec:rationality} to sum the powers of $K$ gives
\begin{equation}\label{eq:walk-matrix-solution}
 \vecop(f_\rho(z))
 =\vecop(P_\partial\rho P_\partial)
 +\frac{z(B\otimes\overline B)\operatorname{adj}(I_{d^2}-zK)
             \vecop(\rho_{\mathrm{int}})}{\det(I_{d^2}-zK)} .
\end{equation}
Here $\operatorname{adj}$ is the adjugate matrix. The determinant gives a common denominator for all boundary entries and has constant term $1$. This formula applies to any coin unitary and any initial density operator, including inputs for which the walk has positive probability of never reaching an endpoint.

\begin{theorem}[Absorption]\label{thm:absorption}
Consider the absorbing walk on positions $0,\ldots,a$ with coin $U_c=\bigl(\begin{smallmatrix}c_{LL}&c_{LR}\\c_{RL}&c_{RR}\end{smallmatrix}\bigr)$, whose diagonal entries are the amplitudes for keeping the direction of motion.
\begin{enumerate}
\item If $a=2$, every run performs at most one iteration.
\item If $a\ge3$, the walk terminates with probability $1$ from every input density operator if and only if $c_{LL}\ne0$. In that case there are $c<\infty$ and $\lambda<1$ such that $\Prb(n_{\mathrm{hit}}\ge m)\le c\lambda^m$ for every input and every $m$.
\end{enumerate}
\end{theorem}

\begin{proof}[Proof of Theorem~\ref{thm:absorption}]
Write $U_c=\bigl(\begin{smallmatrix}c_{LL}&c_{LR}\\c_{RL}&c_{RR}\end{smallmatrix}\bigr)$, so that $U_c\ket L=c_{LL}\ket L+c_{RL}\ket R$ and $U_c\ket R=c_{LR}\ket L+c_{RR}\ket R$.

\emph{$a=2$.} The interior is position $1$, and one step moves $L$ to $0$ and $R$ to $2$. Hence $P_{\mathrm{int}}\Gamma=0$ and every run stops after at most one step.

\emph{$c_{LL}\ne0$.} Suppose $\Gamma\ket\psi=\lambda\ket\psi$ with $|\lambda|=1$. Then $\|P_{\mathrm{int}}\psi\|=\|\Gamma\psi\|=\|\psi\|$, so $\psi$ is supported in the interior, and $\Gamma\psi=\lambda\psi$ has no boundary component. Let $\alpha_j=\langle L,j|\psi\rangle$ and $\beta_j=\langle R,j|\psi\rangle$, with $\alpha_0=\beta_0=\alpha_a=\beta_a=0$. The components of $\Gamma\psi$ are $(\Gamma\psi)_{L,j}=c_{LL}\alpha_{j+1}+c_{LR}\beta_{j+1}$ and $(\Gamma\psi)_{R,j}=c_{RL}\alpha_{j-1}+c_{RR}\beta_{j-1}$. The boundary component $(\Gamma\psi)_{L,0}=c_{LL}\alpha_1+c_{LR}\beta_1$ vanishes, and $\lambda\beta_1=(\Gamma\psi)_{R,1}=c_{RL}\alpha_0+c_{RR}\beta_0=0$, so $\beta_1=0$ and then $\alpha_1=0$ since $c_{LL}\ne0$. If $\alpha_j=\beta_j=0$ for some $1\le j\le a-2$, then $\lambda\beta_{j+1}=c_{RL}\alpha_j+c_{RR}\beta_j=0$ and $\lambda\alpha_j=c_{LL}\alpha_{j+1}+c_{LR}\beta_{j+1}$, which give $\beta_{j+1}=0$ and $\alpha_{j+1}=0$. By induction $\psi=0$. Hence $\Gamma$ has no eigenvalue on the unit circle, and, being a contraction, $\spr(\Gamma)<1$. There are $c<\infty$ and $\lambda'<1$ with $\|\Gamma^m\|_\infty\le c\lambda'^m$, so the survival probability $\tr(\Gamma^m\rho\Gamma^{\dagger m})\le c^2\lambda'^{2m}$ decays exponentially, and all moments of $n_{\mathrm{hit}}$ are finite.

\emph{$c_{LL}=0$, $a\ge3$.} Unitarity then implies $c_{RR}=0$ and $|c_{LR}|=|c_{RL}|=1$. A step maps $\ket{L,j}$ to $c_{RL}\ket{R,j+1}$ and $\ket{R,j+1}$ to $c_{LR}\ket{L,j}$. For $1\le j\le a-2$, both $j$ and $j+1$ are interior, so the span $\mathcal D$ of $\{\ket{L,j},\ket{R,j+1}:1\le j\le a-2\}$ is nonzero and invariant under $\Gamma$. The restriction of $\Gamma$ to $\mathcal D$ is unitary, so states in $\mathcal D$ are never absorbed.
\end{proof}

\paragraph{The anti-diagonal case.}
For $c_{LL}=0$, the interior basis states $\ket{R,1}$ and $\ket{L,a-1}$, which lie outside $\mathcal D$, are absorbed after one step, and boundary states are absorbed immediately. The invariant decomposition into $\mathcal D$, these two states, and the boundary gives, for every density operator $\rho$,
\[
 p_\infty=\tr(P_{\mathcal D}\rho),\qquad
 g_\rho(z)=\tr(P_\partial\rho)+z\bigl(\langle R,1|\rho|R,1\rangle+\langle L,a-1|\rho|L,a-1\rangle\bigr),
\]
where $P_{\mathcal D}$ is the projector onto $\mathcal D$. Coherences between orthogonal invariant parts do not contribute to traces.

\paragraph{Adjoint form.}
Let $R(z)=\sum_mz^m(\Gamma^\dagger)^m\Gamma^m$, which satisfies $R(z)=I+z\Gamma^\dagger R(z)\Gamma$. When $\spr(\Gamma)<1$, the equation at $z=1$ has a unique solution, and the survival identity of Appendix~\ref{app:moments} gives
\[
 \E_\rho n_{\mathrm{hit}}=\tr(R(1)\rho)-1,\qquad\E_\rho(3n_{\mathrm{hit}}+1)=\tr\bigl((3R(1)-2I)\rho\bigr).
\]
This is the Stein-equation form of the first moment, as used in runtime analysis~\citep{Liu2025}. The OFPS additionally retains the operator coefficients.

\subsection{The Seven-Position Walk}\label{app:walk-certificate}
Take $a=6$, the Hadamard coin, and $\rho=\proj{L,1}$, and use the boundary basis $(\ket{L,0},\ket{R,6})$. After one step the walker is either absorbed as $\ket{L,0}/\sqrt2$ or survives as $\ket{R,2}/\sqrt2$. From an even interior position, absorption needs an even number of further steps. In the basis $\ket{L,2},\ket{L,4},\ket{R,2},\ket{R,4}$, composing two steps and keeping the surviving and the absorbed amplitudes gives
\[
 \Gamma_2=\frac12\begin{pmatrix}1&1&-1&1\\0&1&0&-1\\1&0&1&0\\-1&1&1&1\end{pmatrix},\qquad
 \Pi_2=\frac12\begin{pmatrix}1&0&1&0\\0&-1&0&1\end{pmatrix},
\]
with $\Gamma_2^{\mathsf T}\Gamma_2+\Pi_2^{\mathsf T}\Pi_2=I_4$. With $e_3=(0,0,1,0)^{\mathsf T}$ and $\Cmap_2(\rho)=\Gamma_2\rho\Gamma_2^{\mathsf T}$, let $\ell_2(y)$ be the solution analytic near $y=0$ of $\ell_2(y)=e_3e_3^{\mathsf T}+y\,\Cmap_2(\ell_2(y))$. Then
\begin{equation}\label{eq:walk-small}
 f(z)=\frac z2\begin{pmatrix}1&0\\0&0\end{pmatrix}+\frac{z^3}2\,\Pi_2\ell_2(z^2)\Pi_2^{\mathsf T}.
\end{equation}
The reduction exploits the parity of this input. Solving the one-step summary on the full $14$-dimensional space gives the same result.

\paragraph{An all-degree certificate.}
Let
\[
 Q_{\mathrm w}(y)=8-10y+6y^2-5y^3+2y^4,\qquad
 D_{\mathrm w}(y)=(y^2-2y+2)(4y^4+y^2+8y+16),
\]
and let $\widehat Q_{\mathrm w}(x)=x^4Q_{\mathrm w}(1/x)$ and $\widehat D_{\mathrm w}(x)=x^6D_{\mathrm w}(1/x)$ be the reversed polynomials. Let $O_{\mathrm t}=I_2$, $O_{\mathrm d}=\operatorname{diag}(1,-1)$, and $O_{\mathrm c}=\frac12\bigl(\begin{smallmatrix}0&1\\1&0\end{smallmatrix}\bigr)$, and put $\widetilde O_\nu=\Pi_2^{\mathsf T}O_\nu\Pi_2$ for $\nu\in\{\mathrm t,\mathrm d,\mathrm c\}$. For the adjoint $\Cmap_2^\dagger(O)=\Gamma_2^{\mathsf T}O\Gamma_2$, finite matrix multiplication verifies
\begin{equation}\label{eq:walk-recurrence}
 \widehat Q_{\mathrm w}(\Cmap_2^\dagger)(\widetilde O_{\mathrm t})=0,\qquad
 \widehat D_{\mathrm w}(\Cmap_2^\dagger)(\widetilde O_{\mathrm d})=0,\qquad
 \widehat D_{\mathrm w}(\Cmap_2^\dagger)(\widetilde O_{\mathrm c})=0 .
\end{equation}
Hence the sequences $a_{\nu,m}=e_3^{\mathsf T}(\Cmap_2^\dagger)^m(\widetilde O_\nu)e_3$ satisfy linear recurrences at every index, and they are determined by the initial values in Table~\ref{tab:walk-initial}.

\begin{table}[!htbp]
\caption{Initial values of the walk recurrences.}\label{tab:walk-initial}
\centering\small
\setlength{\tabcolsep}{8pt}
\begin{tabular}{crrrrrr}
\toprule
$m$ & $0$ & $1$ & $2$ & $3$ & $4$ & $5$\\
\midrule
$a_{\mathrm t,m}$ & $1/4$ & $1/16$ & $17/64$ & $49/256$ & $17/1024$ & $113/4096$\\
$a_{\mathrm d,m}$ & $1/4$ & $-1/16$ & $-15/64$ & $-49/256$ & $-15/1024$ & $15/4096$\\
$a_{\mathrm c,m}$ & $0$ & $0$ & $-1/16$ & $0$ & $1/256$ & $7/512$\\
\bottomrule
\end{tabular}
\end{table}

\begin{proposition}[Output generating function]\label{prop:walk-law}
For this walk and input,
\begin{gather*}
 f(z)=\frac12\begin{pmatrix}g(z)+d(z)&2c(z)\\2c(z)&g(z)-d(z)\end{pmatrix},
 \qquad
 g(z)=\frac{z(4-4z^2+2z^4-z^6)}{Q_{\mathrm w}(z^2)},\\
 d(z)=\frac{-2z(z^2-2)(z^8+z^2+4)}{D_{\mathrm w}(z^2)},\qquad
 c(z)=\frac{z^7(z^2-2)}{2D_{\mathrm w}(z^2)} .
\end{gather*}
\end{proposition}
\begin{proof}
The equation for $\ell_2$ determines its Taylor coefficients uniquely, and all matrices are real, so the output coefficients are real symmetric. The operators $O_{\mathrm t}$, $O_{\mathrm d}$, and $O_{\mathrm c}$ extract the trace, the diagonal difference, and the off-diagonal entry. By~\eqref{eq:walk-recurrence} and Table~\ref{tab:walk-initial}, each of the three scalar series satisfies a recurrence with given initial values. Multiplying by $Q_{\mathrm w}$ or $D_{\mathrm w}$, substituting $y=z^2$, and including the first term of~\eqref{eq:walk-small} gives the stated rational functions by equality of their Taylor series at $z=0$.
\end{proof}

Expanding the generating function gives
\[
 g(z)=\frac z2+\frac{z^3}8+\frac{z^5}{32}+\frac{17z^7}{128}+\frac{49z^9}{512}+\cdots.
\]
The coefficient of $z^7$ in $f$ is $\frac1{128}\bigl(\begin{smallmatrix}1&-4\\-4&16\end{smallmatrix}\bigr)$, and $f(1)=\frac1{58}\bigl(\begin{smallmatrix}41&-1\\-1&17\end{smallmatrix}\bigr)$. The expansion
\[
 g(1+h)=1+5h+30h^2+210h^3+1614h^4+O(h^5)
\]
gives the factorial moments $5,60,1260,38736$ and the ordinary moments $5,65,1445,46721$; in particular, $\Var(n_{\mathrm{hit}})=40$. The diagonal entries give $p_L=41/58$, $p_R=17/58$, $\E[n_{\mathrm{hit}}\mid L]=93/29$, and $\E[n_{\mathrm{hit}}\mid R]=4597/493$. The cost $3n_{\mathrm{hit}}+1$ has mean $16$ and variance $360$.

For the effect $O_+$ that tests for $\frac1{\sqrt2}(\ket{L,0}+\ket{R,6})$, the acceptance probability is $14/29$ for $f(1)$ and $1/2$ for its boundary-dephased version. Conditioned on $n_{\mathrm{hit}}=7$, it is $(9/256)/(17/128)=9/34$.

\FloatBarrier
\subsection{The Symmetric Start}\label{app:walk-symmetric}
\begin{proposition}[Symmetric start]\label{prop:walk-symmetric}
For the seven-position Hadamard walk started at position $3$ with coin $\frac1{\sqrt2}(\ket L+i\ket R)$, the output has generating function $f(z)=g_{\mathrm{mid}}(z)\proj\phi$ with $\ket\phi=\frac1{\sqrt2}(\ket{L,0}-i\ket{R,6})$, where $g_{\mathrm{mid}}$ has nonnegative coefficients and $g_{\mathrm{mid}}(1)=1$.
\end{proposition}
\begin{proof}
Let $R$ be the unitary with $R\ket{L,x}=\ket{R,6-x}$ and $R\ket{R,x}=-\ket{L,6-x}$. It commutes with the shift $V$, since a left move at $x$ becomes a right move at $6-x$, and with $P_{\mathrm{int}}$ and $P_\partial$. On the coin it acts as $\sigma=\ket R\bra L-\ket L\bra R$, and $\sigma U_c=-U_c\sigma$ for the Hadamard coin. Hence $RU=-UR$ and $R\Gamma=-\Gamma R$ for $\Gamma=UP_{\mathrm{int}}$.

The input vector $\psi=\frac1{\sqrt2}(\ket L+i\ket R)\otimes\ket3$ satisfies $R\psi=-i\psi$. The absorbed amplitude after $m$ steps is $\phi_m=P_\partial\Gamma^m\psi=\alpha_m\ket{L,0}+\beta_m\ket{R,6}$, and it satisfies $R\phi_m=(-1)^m(-i)\phi_m$. Since $R\phi_m=\alpha_m\ket{R,6}-\beta_m\ket{L,0}$, comparing the components at $\ket{L,0}$ gives $\beta_m=(-1)^mi\alpha_m$.

A walker that starts at position $3$ can reach a boundary only after an odd number of steps, so $\phi_m=0$ for even $m$ and $\beta_m=-i\alpha_m$ for odd $m$. Hence $\phi_m=\sqrt2\,\alpha_m\ket\phi$, and by Appendix~\ref{app:walk-absorption} the coefficient of $z^m$ in $f(z)$ is $\phi_m\phi_m^\dagger=2|\alpha_m|^2\proj\phi$. Since $c_{LL}\ne0$, Theorem~\ref{thm:absorption} gives almost-sure termination, so $g_{\mathrm{mid}}(1)=\sum_m2|\alpha_m|^2=1$.
\end{proof}
Each end is therefore reached with probability $1/2$.

\FloatBarrier
\end{document}